\def\rfr#1{eq. (\ref{#1})}

\def\Rfr#1{Eq. (\ref{#1})}

\def\cf#1#2{\dot\Omega^{\rm #2}_{.#1}}

\def\derp#1#2{\rp{\partial{#1}}{\partial{#2}}}
\def\dert#1#2{\frac{{{d}}{#1}}{{{d}}{#2}}}              

\def\virg#1{``#1''}

\def\bar{\begin{eqnarray}}
\def\ear{\end{eqnarray}}
\def\bb{\bibitem}
\def\eqi{\begin{equation}}
\def\eqf{\end{equation}}
\def\eqia{\begin{eqnarray}}
\def\eqfa{\end{eqnarray}}
\def\rp#1#2{{#1\over#2}}

\def\lb#1{\label{#1}}

\def\oc2{$\mathcal{O}(c^{-2})$}

\def\bds#1{\vec{\it{#1}}}

\documentclass{aastex}
\usepackage{spr-astr-addons}
\usepackage{url}
\usepackage{amsmath,amsthm,amscd,amssymb}
\usepackage{latexsym}
\usepackage{graphicx,epsfig}

\RequirePackage{color}

\begin{document}

\title{Phenomenology of the Lense-Thirring effect in the Solar System }
\shorttitle{Phenomenology of the Lense-Thirring effect in the Solar System}
\shortauthors{L. Iorio, H.I.M. Lichtenegger, M.L. Ruggero, C. Corda}

\author{Lorenzo Iorio\altaffilmark{1} }
\affil{Ministero dell'Istruzione, dell'Universit\`{a} e della Ricerca (M.I.U.R.), Fellow of the Royal Astronomical Society (F.R.A.S.)  Viale Unit\`{a} di Italia 68, 70125, Bari (BA), Italy.}
\email{lorenzo.iorio@libero.it}

\author{Herbert I. M. Lichtenegger\altaffilmark{2}}
\affil{Institut f\"ur Weltraumforschung, \"Osterreichische Akademie der Wissenschaften. Schmiedlstrasse 12, 8042 Graz, Austria}

\author{Matteo Luca Ruggiero\altaffilmark{3}}
\affil{Dipartimento di Fisica, Politecnico di Torino, and INFN-Sezione di Torino. Corso Duca Degli Abruzzi 24, 10129, Torino (TO), Italy.}

\author{Christian Corda\altaffilmark{4}}
\affil{Associazione Scientifica Galileo Galilei. Via Pier Cironi 16, I-59100 Prato Italy. Institute for Basic
Research, P. O. Box 1577, Palm Harbor, FL 34682, USA}

\begin{abstract}
Recent years have seen increasing efforts to directly measure some aspects of the general relativistic gravitomagnetic
interaction in several astronomical scenarios
in the solar system. After briefly overviewing the concept of gravitomagnetism from a theoretical point of view,
we review the performed or proposed attempts to detect the Lense-Thirring effect affecting the orbital motions of  natural and artificial bodies in the gravitational fields
of the Sun, Earth, Mars and Jupiter. In particular, we will focus on
the evaluation of the impact of several sources of systematic uncertainties of dynamical origin
to realistically elucidate the present and future perspectives in directly measuring such an elusive relativistic effect.
\end{abstract}


\section{Introduction}\lb{introduzione}

The analogy between Newton's law of gravitation and Coulomb's law of electricity has been largely investigated since the nineteenth century, focusing on the possibility that the motion of masses
could produce a magnetic-like field of gravitational origin. For instance, \citet{Holz} and \citet{Tiss1,Tiss2}, taking into account the modification of the Coulomb law for the electrical
charges by \citet{Web1}, proposed  to modify Newton's law in a similar way, introducing in the radial
component of the force law a term depending on the relative velocity of the
two attracting particles, as described by \citet{north} and \citet{whittaker}. Moreover, \citet{heavi} investigated the analogy between gravitation and electromagnetism;
in particular, he explained the propagation of energy in a gravitational
field in terms of an electromagnetic-type Poynting vector.

Actually, today the term \virg{gravitomagnetism} (GM) \citep{Thorne,Rin,MashNOVA}  commonly indicates the collection of those gravitational phenomena regarding orbiting test
particles, precessing gyroscopes, moving clocks and
atoms and propagating electromagnetic waves \citep{Dymn,Rug,Scia04,Scia09} which, in the framework of the Einstein's
General Theory of Relativity (GTR), arise from non-static distributions of matter and energy.  In the weak-field
and slow motion approximation, the Einstein field equations  of GTR, which is a highly non-linear Lorentz-covariant tensor theory of gravitation, get
linearized, thus looking like  the Maxwellian equations of electromagntism. As a consequence, a \virg{gravitomagnetic}
field $\bds{B}_g$, induced by the off-diagonal components $g_{0i},\ i = 1,2,3$ of the spacetime metric
tensor related to mass-energy currents, does arise.
Indeed, bringing
together Newtonian gravitation and Lorentz invariance in a consistent field-theoretic framework necessarily requires the introduction of a
\virg{magnetic}-type gravitational field of some form \citep{Kha,Bed,Kolb}.

%
%
%
%
%
%
%
In general, GM is used to deal with aspects of GTR by means of an electromagnetic analogy.
However, it is important to point out that even though the linearization of the Einstein's field
equations produces the Maxwell-like equations (the so called `` linear perturbation approach'' to
GM, see e.g. \citet{MashNOVA}), often written in the literature including time dependent terms,
they are, in that case, just formal (i.e., a different notation to write linearized Einstein equations),
as the 3-vectors $\vec{E}_g$ and $\vec{B}_g$ (the ``gravito-electromagnetic fields'') showing up therein do
not have a clear physical meaning. A consistent physical analogy involving these objects is restricted to stationary
phenomena only \citep{Clark,Costa1,Costa2},
that is, actually, the case treated here. One may check, for instance, that from the geodesics
equation the corresponding Lorentz force is recovered -- to first order in $v/c$ -- only for stationary fields
 \citep{Costa1,bini08}. Moreover, the Maxwell-like equations obtained by linearizing GTR have
limitations\footnote{See, e.g., \citet{MTW} Section 7.1, box 7.1 and Section 18.3, and \citet{Ruff}, Chapter 3.}, since they are self-consistent at linear order only, which is what we are concerned with in this paper; in fact, inconsistencies arise when this fact is neglected\footnote{See, e.g., \citet{tartaglia04}.}.

Far from a localized rotating body with angular
momentum $\bds{S}$ the gravitomagnetic field can be written as \citep{Thorne0,Thorne,Mashetal01a}
\eqi
 \bds{B}_g(\bds{r}) = -\rp{G}{cr^3}\left[\bds{S} -3\left(\bds{S}\cdot\hat{r}\right)\hat{r} \right],\lb{gmfield}
\eqf
where $G$ is the Newtonian gravitational constant\textrm{, $c$ the speed of light in vacuum and $\hat{r}$ a unit vector along $\bds{r}$}.
\textrm{Eq. (\ref{gmfield}) equals the field of a magnetic dipole with a moment
$\vec{\mu}_g'=G\bds{S}/c$ and affects,}
e.g., a test particle
moving with velocity $\bds{v}$ with a non-central acceleration \citep{Mashetal01b,Lichten06}
\eqi \bds{A}_{\rm GM}=-2\left(\rp{\bds{v}}{c}\right)\times \bds{B}_g,\lb{accpertgm}\eqf which is the cause
of three of the most famous and empirically investigated GM features with which we will deal here: the \citet{LT} effect,
 the gyroscope precession \citep{Pu,Sci1}, and the gravitomagnetic clock effect \citep{Mashetal01a}. Corrections of higher order in $v/c$ to orbital motions of pointlike objects have been  studied by \citet{Capoz}; they may become relevant in stronger gravitational fields like those occurring in astrophysical scenarios (pulsars and black holes). For recent reviews of the Lense-Thirring effect in the astrophysical context see, e.g., \citet{Stella,Scia09}.

GM manifests itself in the gravitational tidal forces as well. Actually, a
gravito-electromagnetic analogy relying on exact and covariant equations stems from the
tidal dynamics of both theories \citep{Costa1}. In both theories it is possible to define electric and magnetic-type tidal tensors playing analogous physical roles. The electric tidal tensors of electromagnetism are gradients of the electric field, and play in the worldline deviation for
two neighboring test particles (with the same ratio charge to mass) the same role as the
so called \virg{electric part of the Riemann tensor} in the geodesic deviation equation. The
magnetic-type tidal tensors are, in the electromagnetic case, gradients of the magnetic field,
and their gravitational counterpart is the so-called magnetic part of the Riemann tensor.
\textrm{Physically, the latter} manifests itself, for instance, in the deviation from geodesic motion of
a spinning test \textrm{particle due} to a net gravitational force acting on it, in analogy with
the electromagnetic force exerted on a magnetic dipole. In the framework of the tidal tensor
formalism, the exact gravitational  force on a gyroscope \citep{Mathi,Papa,Pira}
is described by an equation formally identical to the electromagnetic force on a magnetic dipole.
Moreover, both Maxwell and part of the Einstein equations
(the time-time and time-space projections) may be expressed exactly as equations for tidal
tensors and sources, and such equations exhibit a striking analogy. In particular, these
equations show explicitly that mass currents generate the gravitomagnetic tidal tensor just
like currents of charge \textrm{involve} the magnetic tidal tensor of electromagnetism \citep{Costa1}.
At this point it is important to make a distinction between the \virg{gravitomagnetic field} causing, e.g., the precession of a gyroscope (see Section \ref{giroscopio}), and the gravitomagnetic tidal field causing e.g. the
net force on a the gyroscope. The gravitomagnetic field itself has no physical existence; it
is a pure coordinate artifact that can be gauged away by moving to a freely falling (non-rotating)
frame. For instance, it is well known that the spin 4-vector of a gyroscope undergoes
Fermi-Walker transport, with no real torques applied on it; thus
the gyroscope \virg{precession}  is a non-covariant notion
attached to a specific coordinate system, which is the reference frame of the distant stars,
i.e., it does not have a local existence, and can only be measured by locking to the distant
stars by means of a telescope \citep{Polna}. The gravitomagnetic
tidal tensor, instead, describes physical forces, which can be locally measured \citep{Polna}.

There have been several proposals to detect the GM tidal forces \textrm{originating} by the proper angular momentum $\vec{S}$ of a central body
like \textrm{the Earth.  
 In principle, they can be detected by an orbiting tidal} force sensor, or \virg{gravity gradiometer}, as suggested by \citet{GRADIO}.
 In the following years such a concept was further investigated \citep{gradius}, also from the
 point of view of a practical implementation in terms of a set of orbiting superconducting gravity gradiometers (SGG)
 \citep{gradio1,gradiop}. \citet{gradio2}
 recently reviewed the
 the feasibility of the gravity gradiometer experiment in view of the \textrm{latest} advancements in the field. According to \citet{gradio2}, the GM
 field of the Earth could be successfully detected by using an orbiting SGG. Such
a mission would benefit from the technologies already developed for GP-B \textrm{(see Section \ref{giroscopio})}. The SGG
would  be launched in a superfluid helium dewar, and the helium boil-off gas would be used
for drag-free control of the satellite. \textrm{GP-B, like gyros or telescopes,} may be used for
attitude control of the spacecraft.

Other GM effects which have recently received attention from the phenomenological point of view are those caused
by the orbital motion of the Earth-Moon system around the
Sun. They have nothing to do with
the proper angular momenta $\bds S$ of the Earth or the Sun causing the \virg{intrinsic} GM effects like those previously mentioned; in the case of
translational mass-energy currents it is customarily to speak
about \virg{extrinsic} GM effects. According to \citet{Nord1,Nord2}, the \textrm{extrinsic GM interaction has already been observed with a relative accuracy of 1 part to 1000 in} comprehensive fits of the motions of  several astronomical and astrophysical bodies like satellites, binary pulsars and the Moon.
In fact, a debate arose \textrm{about} the ability of the Lunar Laser Ranging (LLR) technique \citep{llr} of detecting genuine GM effects
\citep{Nord3,Kop1,llr1,llr2,llr3}. Concerning the Lense-Thirring precessions of the lunar orbit induced by the Earth's spin, they may remain still undetectable
also in \textrm{the foreseeable} future because of overwhelming systematic uncertainties \citep{IorLLR}, in spite
of the expected mm-level improvements in Earth-Moon ranging with LLR \citep{IMPRO}.

Several Earth-based laboratory experiments aimed to test the influence of the \virg{intrinsic} terrestrial GM field on classical and quantum objects
and electromagnetic waves have been proposed so far \citep{terra1,terra2,terra3,terra4,terraX,terra5,terra6,terra7,terra8}, but they have never
been implemented because of several technological difficulties in meeting the stringent requirements in terms of sensitivity and/or accuracy.

As gravitomagnetic effects \textrm{are, in general, analyzed} in the framework of the linearized theory of GTR, they have also an important connection with Gravitational Waves (GWs)  \citep{IorCor09,IorCor10,Grishchuk,CordaI}.
We recall that the data analysis of interferometric  detectors has nowadays been started, and the scientific community
\textrm{awaits a first direct detection of GWs in the near future}.
For the current status of \textrm{GW} interferometers \textrm{we refer to} \citet{Giazotto}. Thus, the indirect evidence \textrm{for} the existence of GWs by \citet{Pulsar} \textrm{might soon be confirmed}.
Detectors for GWs will be important for a better knowledge of the Universe \citep{IorCor10} and also because the interferometric GWs detection will be the ultimate test for GTR or, alternatively, a strong endorsement for Extended Theories of Gravity \citep{Essay}. In fact, if advanced projects on the detection of GWs improve their sensitivity, allowing the scientific community to perform a GW astronomy, accurate angle- and frequency-dependent response functions of interferometers for GWs arising from various theories of gravity will permit to discriminate among GTR and extended theories of gravity. This  ultimate test will work because standard GTR admits only two polarizations for GWs, while in all extended theories \textrm{there exist at least three polarizations states}; see \citet{Essay} for details.
On the other hand, the discovery of GW emission by the compact binary system composed \textrm{of} two Neutron Stars PSR1913+16 \citep{Pulsar} represents, for  scientists working in this research field, the definitive thrust allowing to reach the extremely sophisticated technology needed for investigating in this field of research \citep{IorCor10}.
GWs are a consequence of Einstein's GTR \citep{Albertino}, which presupposes GWs to be ripples in the spacetime curvature traveling at light speed \citep{AlbertinoII,AlbertinoIII,IorCor10}.
The importance of gravitomagnetic effects in the field of a GW  has been emphasized by \citet{Grishchuk}. For a complete review of such a topic, see \citet{IorCor10}. Recently, the analysis has been extended
\textrm{to} gravitomagnetic effects in the field of GWs arising by Scalar Tensor Gravity too \citep{Cafaro, IorCor10}.

\subsection{The Lense-Thirring effect}\lb{LETIZ}
After the birth  of the Einstein's Special Theory of Relativity (STR) in 1905,
the problem of a \virg{magnetic}-type component of the
gravitational field of non-static mass distributions was tackled in the framework of
the search for a consistent relativistic theory of gravitation \citep{ein}.

 With a preliminary and still incorrect version of GTR, Einstein and Besso in 1913 \citep{colle} calculated the node precession of planets in the field of
 the rotating Sun; the figures they obtained for Mercury and Venus were incorrect also because they used a wrong value for the solar mass.
 Soon after GTR was put forth by Einstein in 1915, \citet{DeS} \textrm{worked out the corresponding shifts of the planet's perihelia
 for ecliptic orbits due to the rotation of the Sun; however,} his result for Mercury ($-0.01$ \textrm{arcseconds per century}) was too
 large by one order of magnitude because he assumed a
homogenous and uniformly rotating Sun. In 1918 \citet{Thirr1} analyzed in a short article the formal analogies between the Maxwell
equations and the linearized Einstein equations. Later, \citet{Thirr1,Thirr2,Thirr3} computed the centrifugal and Coriolis-like gravitomagnetic
forces occurring inside  a rotating massive shell. \citet{LT}\footnote{However, in August 1917 Einstein \citep{coll2} wrote to Thirring that
he calculated the Coriolis-type field of the rotating Earth and Sun, and its influence on the orbital elements of planets (and moons). A detailed
history of the formulation of the so-called Lense-Thirring effect has recently been outlined by \citet{Pfi07}; according to him, it would be
more fair to speak about an Einstein-Thirring-Lense effect.}  worked out the gravitomagnetic effects on the orbital motions of
test particles outside a slowly rotating mass; in particular, they computed the gravitomagnetic rates \textrm{for the two} satellites of Mars (Phobos and Deimos),
and \textrm{for} some of the moons of the giant gaseous planets.
They found for the longitude of the ascending node $\Omega$ \textrm{and the argument of pericenter $\omega$ a pro- and retrograde precession,
respectively, according to}\footnote{In fact, \citet{LT} considered the longitude of pericenter $\varpi\doteq\Omega+\omega$.}
\eqi \dot\Omega_{\rm LT}=\rp{2GS}{c^2 a^3 (1-e^2)^{3/2}},\ \dot\omega_{\rm LT}=-\rp{6GS\cos I}{c^2 a^3 (1-e^2)^{3/2}},\lb{sblenda}\eqf where $a,e,I$ are
the semimajor axis, the eccentricity and the inclination of the test particle's orbital plane to the central body's equator, respectively. Later,
the Lense-Thirring
effect was re-derived in a number of different approaches; see, e.g., \citet{Bogo59,Zel71,Bark,Landau,Ash,IorNCB,Lamm,Chia}.
\textrm{In the following we give some more physical insights about the derivation and the characteristics of \rfr{gmfield}-\rfr{accpertgm} and, consequently, of \rfr{sblenda}. In the standard parameterized post-Newtonian\footnote{The post-Newtonian formalism allows to approximate the non-linear Einstein field equations for weak fields and slow motions in terms of the lowest order deviations from the Newton's theory. In the framework of the parameterized post-Newtonian formalism \citep{eddi,NI,Norda,Nordb,Will}, such departures from classical gravity are expressed in terms of a set of ten parameters to discriminate between competing metric theories of relativistic gravity. The modern notation is due to \citet{WilNo}. For a comprehensive review, see \citet{Willone}.} (PPN) framework for an isolated, weakly gravitating and slowly rotating body, the PPN spacetime metric coefficients  involving its angular momentum are the off-diagonal \textrm{terms}
\eqi g_{0i}=-\rp{(\gamma+1)}{c^2}\mathcal{V}_i,\;\; i=1,2,3,\lb{goi}\eqf
with \eqi \vec{\mathcal{V}}\doteq \rp{G}{c}\rp{\vec{S}\times\vec{r}}{r^3}.\eqf In \rfr{goi} $\gamma$ is the PPN parameter accounting for the spatial curvature caused by a unit \textrm{mass; in GTR $\gamma=1$. Thus, in the PPN approximation the resulting Christoffel symbols including $\mathcal{V}_i,\ i=1,2,3$ and entering the geodesic equation of motion are \citep{Sof}
\eqi\Gamma_{0j}^i=-\rp{(\gamma+1)}{c^2}\left(\derp{\mathcal{V}_i}{x^j}-\derp{\mathcal{V}_j}{x^i}\right),\ i,j=1,2,3;\eqf
they give rise to \rfr{gmfield}--\rfr{accpertgm} for $\gamma=1$.
It is interesting and useful to emphasize} the complete analogy of \rfr{accpertgm} with the Lorentz force of electromagnetism provided that the factor $-2$ entering \rfr{accpertgm} \textrm{can be} formally thought as a gravitomagnetic charge-to-mass ratio $q_{g}/m$.
As originally done by \citet{LT}, the orbital precessions of the node $\Omega$ and the pericenter $\omega$ of \rfr{sblenda} can be straightforwardly worked out with, e.g., the Gauss equations for the variation of the Keplerian orbital elements \citep{Sof,Roy} by treating \textrm{the gravitomagnetic force} \rfr{accpertgm} as a small perturbation of the Newtonian monopole. The Gauss equations for $\Omega$ and $\omega$ are \citep{Sof,Roy}
\eqi\dert\Omega t  =  \rp{1}{na\sin I\sqrt{1-e^2}}A_{\nu}\left(\rp{r}{a}\right)\sin u,\lb{gaus_O}\eqf
\eqi
\begin{array}{lll}
\dert\omega t  &=& \rp{\sqrt{1-e^2}}{nae}\left[-A_{r}\cos f + A_{\tau}\left(1+\rp{r}{p}\right)\sin f\right]-\\ \\
&-&\cos I\dert\Omega t,\lb{gaus_o}
\end{array}
\eqf
where  $n\doteq\sqrt{GM/a^3}$ is the Keplerian mean motion, $u\doteq\omega+f$ is the argument of latitude in which $f$ is the true anomaly, and $A_r,A_{\tau},A_{\nu}$ are the projections of the perturbing acceleration onto the radial, transverse and normal components of an orthonormal frame co-moving with the test particle. In the case of \rfr{accpertgm} \textrm{these projections correspond to} \citep{Sof}
\eqi
A_r^{\rm LT}=\eta\cos I(1+e\cos f),\lb{agmr}
\eqf
\eqi
A_{\tau}^{\rm LT}= -\eta e\cos I\sin f,\lb{agmt}
\eqf
\eqi
A_{\nu}^{\rm LT}=\eta\sin I(1+e\cos f)\left[2\sin u + e\left(\rp{\sin f\cos u}{1+e\cos f}\right)\right],\lb{agmn}
\eqf
with
 \eqi\eta\doteq\rp{\chi n}{a^2(1-e^2)^{7/2}}(1+e\cos f)^3,\eqf and \eqi \chi\doteq \rp{2 G S}{c^2}.\eqf
\textrm{It is immediately seen from \rfr{agmr}--\rfr{agmn} that for polar orbits (i.e. $I=90$ deg), both the radial and the transverse components} vanish, contrary to the out-of-plane component. This fact can easily be inferred also from \textrm{the dipolar field structur and the Lorentz-type force, i.e.} \rfr{gmfield}-\rfr{accpertgm}. Indeed, if $\vec{S}$ lies in the orbital plane, the latter one contains $\vec{B}_g$ as well, so that, according to \rfr{accpertgm}, $\vec{A}_{\rm GM}$ is entirely out-of-plane.
\textrm{In that case there is also no pericenter precession, as indicated by \rfr{sblenda} since all terms in \rfr{gaus_o} vanish for $I=90$ deg.
By analogous reasoning it is easy to figure out that for equatorial orbits (i.e. $I=0$ deg) $A_{\rm GM}$  completely lies in the orbital plane which is in agreement with \rfr{agmr}-\rfr{agmn}. If the orbit is circular in addition, $A_{\rm GM}$, which is perpendicular to $\vec{S}$ and $\vec{v}$, becomes entirely radial, as confirmed by \rfr{agmr}--\rfr{agmt}. Finally, since the factor $\sin I$ cancels in \rfr{gaus_O} and \rfr{agmn} the Lense-Thirring node precession is independent of $I$, as stated by \rfr{sblenda}.}
}
\subsection{The gyroscope precession}\lb{giroscopio}
Another well known GM effect consists of the precession of a
gyroscope moving in the field of a slowly rotating body. It was worked out in 1959 by \citet{Pu} and in 1960 by \citet{Sci1,Sci2,Sci3} on the basis of the \citet{Mathi} and \citet{Papa} equation \textrm{and became known as the Schiff effect}.
More recent derivations, based on a quantum mechanical approach to gravitation, can be found in \citet{Bark70,Bark72}.

\textrm{In order to yield a direct, physical insight of such a phenomenon, let us recall the following basic \textrm{facts of Maxwellian} electromagnetism. A charged spinning particle with electric charge $q$, mass $m$ and spin $\vec{\sigma}$ has a magnetic moment
\eqi\vec{\mu}=\left(\rp{q}{2mc}\right)\vec{\sigma},\lb{mu}\eqf
so that it precesses in an external magnetic field $\vec{B}$ according to
\eqi\dert{\vec{\sigma}}{t}=\vec{\mu}\times \vec{B}=-\left(\rp{q}{2mc}\right)\vec{B}\times\vec{\sigma}\doteq\vec{\mathfrak{O}}\times\vec{\sigma}\eqf
\textrm{with the precessional frequency}
\eqi \vec{\mathfrak{O}}\doteq -\left(\rp{q}{2mc}\right)\vec{B}.\eqf
Moving \textrm{now to the weak-field and slow-motion approximation of} linearized gravitomagnetism and recalling that the gravitomagnetic charge-to-mass ratio is $q_g/m\doteq -2$, \textrm{according to \rfr{mu}} a spinning particle like a gyroscope is endowed with a gravitomagnetic dipole moment
\eqi\vec{\mu}_g\doteq -\rp{\vec{\sigma}}{c},\eqf so that it undergoes a precession during its motion in an external gravitomagnetic field $\vec{B}_g$
with frequency
\eqi\vec{\mathfrak{O}_g}\doteq \rp{\vec{B}_g}{c}.\eqf
\textrm{Since the GM-field of a distant rotating astronomical body is given by \rfr{gmfield}, the precession frequency becomes}
\eqi \vec{\mathfrak{O}_g}=-\rp{G}{c^2 r^3}\left[\bds{S} -3\left(\bds{S}\cdot\hat{r}\right)\hat{r} \right].\eqf
If the gyroscope moves along an equatorial orbit\footnote{For simplicity, we will assume it circular.},
then
\eqi \dert{\bds{\sigma}}{t}=-\rp{G}{c^2 r^3}\bds{S}\times \bds{\sigma}\eqf
\textrm{and there is no precession of the orbit in case the two spins are aligned.}
If the orbit of the gyroscope is polar, then, by choosing a plane $\{xz\}$ reference frame with, say, \eqi\vec{S}=S\ \hat{z}\eqf and \eqi\hat{r}=\cos nt\ \hat{x} + \sin nt\ \hat{z}\eqf it is easy to show that the averaged precession vanishes if $\vec{\sigma}$ and $\vec{S}$ are parallel or antiparallel, i.e. for \eqi \vec{\sigma}=\pm \sigma\ \hat{z}.\eqf
\textrm{Finally, if} $\vec{\sigma}$ is orthogonal to $\vec{S}$ and to the orbital plane, i.e. \eqi \vec{\sigma}=\pm\sigma\ \hat{y}, \eqf then there is a net spin precession \textrm{given by}
\eqi\left\langle\dert{\vec{s}}{t}\right\rangle=\rp{G}{2c^2 r^3}\vec{S}\times\vec{\sigma}.\lb{bingo}\eqf
}

Soon after the formulation of the Schiff effect, in 1961 \citet{Fair61} submitted to NASA a proposal for a dedicated space-based
 project aimed to directly measure \textrm{the precession of \rfr{bingo}} in a \textrm{dedicated,} controlled experiment. Such an extraordinary and extremely sophisticated  mission, later named Gravity Probe B (GP-B) \citep{gpb1,gpb2}, consisted of a
drag-free, liquid helium-cooled spacecraft moving in a polar, low\textrm{\footnote{Its altitude was 642 km.}} orbit around the Earth and carrying onboard four superconducting gyroscopes whose
GM precessions \textrm{of 39 milliarcseconds per year (mas yr$^{-1}$ in the following)} should have been detected by Superconducting Quantum Interference Devices (SQUID) with an expected accuracy of $1\%$ or better. It
took 43 years to be implemented \textrm{before} GP-B was finally launched on 20 April 2004; the science data collection  lasted from 27 August 2004 to 29 September
2005, while the data analysis is still ongoing \citep{ANA1,ANA2}. It seems that the final accuracy obtainable \textrm{will be less than initially expected}  because
of the occurrence of unexpected systematic errors \citep{merdacce1,merdacce2,merdacce3}.
At present, \textrm{the} GP-B team reports\footnote{See on the WEB: \url{http://einstein.stanford.edu/}} evidence of the gravitomagnetic spin precessions as predicted by GTR, with a statistical error of $\approx 14\%$ and systematic uncertainty
of $\approx 10\%$.

In 1975 \citet{Haas} proposed to measure the angular momenta of the Sun and Jupiter by exploiting the Schiff effect with dedicated
spacecraft-based missions, but such a proposal was not carried out so far.

\subsection{The gravitomagnetic clock effect}\lb{gclock}
\citet{Zel65} discovered that the gravitational field of a rotating mass of radius $R$ and angular momentum $S$  splits the line emitted by an atom
with frequency $\mathfrak{f}_0$ into two components with opposite circular polarizations and frequencies $\mathfrak{f}_0\pm \Delta \mathfrak{f}$, where  \eqi\Delta \mathfrak{f}= \rp{2GS}{c^2 R^3}\eqf
for an electromagnetic signal emitted on the body's surface at the pole.
It is a gravitational analog of the electromagnetic Zeeman effect and, as such, it does not depend on the \textrm{specific} properties of the system emitting \textrm{the}
electromagnetic radiation, \textrm{thus} being the same for an atom \textrm{or} a molecule and \textrm{independent from the emitted frequency.}
A similar effect for \textrm{the orbital motion} of test particles around a spinning body of mass  $M$ and angular momentum $S$ was discovered later by
\citet{Vlad87}. They noted that in the equatorial plane of a rotating mass the gravitomagnetic force becomes purely radial (if the motion occurs
at a constant distance $r$), and acts as a supplement to the centripetal Newtonian monopole\textrm{; it is clearly elucidated by \rfr{agmr}-\rfr{agmt}}. Such an additional term, may be \textrm{either positive,}
i.e. directed outward so that it weakens the overall gravitational force, if the particle's motion coincides with the direction of rotation of the
central body, \textrm{or} negative, thus enhancing the force of gravity, if the motion is \textrm{opposite} to the rotation. As a consequence, the orbital period
of a particle moving along a circular and equatorial orbit in the Lense-Thirring metric is longer in the first case and shorter in the
second\footnote{Remember that the period of  a pendulum's oscillations decreases as the force acting on it increases}. \textrm{\textrm{
This fact contradicts the idea of frame-dragging where it is conceived that a moving object is \virg{dragged} by spacetime which in turn is \virg{twisted} by the rotation of the central mass. If this were the case, the \virg{dragged} test particle should move faster when co-rotating with the central body and should thus have a shorter period.}}
\textrm{Interestingly, such a change in the period of the particle's orbit is a universal and structurally very simple quantity, given by
\eqi
T_{\rm GM}=\pm 2\pi\frac{S}{Mc^2}\lb{vladi}.
\eqf
}
It is amazing that \rfr{vladi} is independent of both the Newtonian gravitational
constant $G$ and the orbital radius. \citet{Coh93} suggested to consider the difference between the orbital periods of two counter-orbiting
clocks moving around a rotating astronomical body along circular and equatorial orbits because it \textrm{cancels the} common Keplerian terms and
\textrm{enhances the gravitomagnetic ones by adding them up}. This is the so-called
gravitomagnetic clock effect \citep{Mashetal99,Mashetal01a}. \citet{Theiss} worked out the case of circular orbits with arbitrary inclinations showing
that the time difference decreases with increasing inclination;
for a polar orbit the effect vanishes as it is expected because of symmetry. The general case for arbitrary values of the eccentricity and the
inclination was treated by \citet{Mashetal01b}
and \citet{LIorio07}. \textrm{If one considers two satellites orbiting a slowly rotating
mass $M$ in opposite directions, their common initial position in the orbital plane is given by the argument of latitude $u_0\doteq \omega_0+f_0$.
In taking the difference of the sidereal periods of the satellites, the gravitoelectric perturbations cancel, leaving
\begin{equation}\label{dtsid2}
\Delta T^{\rm sid}=
 \frac{4\pi S\cos I}{c^2M}\biggl[-\frac{3}{\sqrt{1-e^2}}+\frac{2\left(2-\tan^2 I\cos^2u_0\right)}{(1+e\cos f_0)^2}\biggr]
\end{equation}
which reduces to the difference of the GM corrections $T_{\rm GM}$ for counter-orbiting particles of (\ref{vladi}) in case of $e=I=0$. It is interesting to note that this clock effect can reveal a relatively large
value by a careful choice of the initial parameters \citep{LIorio07}.}

\textrm{Based on various approaches (e.g., analogies with electromagnetism, spacetime geometric properties),
derivations of the gravitomagnetic clock effect for the circular and equatorial cases} can be found in \citep{You,Ioretal,Tart00a}.
\citet{Gron97} proposed to detect the gravitomagnetic clock effect with a space-based mission \textrm{-- dubbed Gravity Probe C -- }
in the gravitational
field of the Earth where such an effect would be as large as $10^{-7}$ s. In view of the challenging difficulties of implementing such a
demanding experiment, a number of studies were performed to mainly investigate the impact of several competing dynamical effects acting as
sources of insidious systematical uncertainty \citep{Lichten00,Ior01a,Ior01b,IorLichten05,Lichten06}. \citet{Tart00b} preliminarily looked at the
possibilities offered by other Solar System scenarios
finding that, in principle, the less unfavorable situation occurs for the Sun and Jupiter.

\textrm{
\subsection{Motivations for attempting to directly measure the Lense-Thirring effect}}

\textrm{
GTR is a basic pillar of our knowledge of Nature since \textrm{it currently represents our best theory of gravitation,}
which is one of the four fundamental interactions governing the physical world. The simplicity, the internal coherence and the mathematical elegance of GTR are remarkable, but the level of its empirical corroboration, although certainly satisfactorily up to now \citep{Willpalla}, is not comparable to that of the other theories describing the remaining fundamental interactions.
\textrm{This applies both to the number of successfully tested predictions of GTR as well as to the level of accuracy reached.
Notably the level of empirical corroboration of GM, which is a constitutive, fundamental  aspect of GTR is to date extraordinarily poor.
It is therefore desirable to expand and strengthen the empirical basis of the theory by testing as many diverse aspects and predictions as possible using different methods and techniques. It is also necessary, however, to devise observational/experimental tests in those extreme regimes in which the theory in its currently accepted form is believed to experience failures. This is particularly important in order to gain possible hints on a future quantum gravity theory which should combine both gravitational and quantum phenomena.\newline
Concerning the GP-B mission, aimed to test a well defined GM prediction using the gravitational field of the Earth, although different teams may have the possibility to repeat the analysis of the currently available data record with various approaches and techniques, the results of the mission
are likely doomed to remain unique since it will be impossible to replicate the entire experiment in any foreseeable future. For the moment,
its level of accuracy is}
more or less comparable to or better than that reached in the non-dedicated\footnote{LAGEOS and LAGEOS II were originally launched for other purposes.} tests of the Lense-Thirring effect with the terrestrial LAGEOS satellites.
Thus, it becomes of the utmost importance not only to reliably assess the total accuracy in such attempts but also to look for other possibilities offered by different astronomical scenarios by exploiting future planned/proposed spacecraft-based missions and expected improvements in ranging techniques. Such an effort has the merit, among other things, to realistically establish the limits which may likely be reached in such an endeavor.
Moreover, as a non-negligible by-product, the knowledge of several classical effects, regarded in the present context as sources of unwanted systematic biases but interesting if considered from \textrm{different} points of view, will turn out to be greatly improved by the tireless efforts towards the measurement of \textrm{such a} tiny relativistic feature of motion. Last but not least, it \textrm{should be recalled that reaching a satisfying level of knowledge of GM has important} consequences in the study of extreme astrophysical scenarios in which, as we know, GM may play a very important role \citep{Thorne0,Thorne,Williams1,Williams2,latest}.\newline
\textrm{Another source of motivation to exploit gravitomagnetic effects is their possible relation with Mach's principle.
While GTR as a whole -- despite its name --  appears not to fulfill Machian expectations of a description of motion with only relative concepts,
some special GM phenomena seem to be in accordance with Machian ideas.
In particular dragging effects are considered to be the most direct manifestations of Mach's principle in general relativity because they indicate
that local inertial frames are at least partially determined by the distribution and currents of mass-energy in the universe. Thus the study of
gravitomagnetism may provide a deeper insight into the presumed intimate connection between inertial properties and matter.}
}

\section{Measuring the Lense-Thirring effect in the gravitational field of the Earth}\lb{terra}
Soon after the dawn of the space age with the launch of Sputnik in 1957 \textrm{Soviet scientists proposed} to directly test the
general relativistic Lense-Thirring effect with artificial satellites orbiting the Earth. In particular, \textrm{\citet{Ginz57,Ginz59,Ginz62}}
proposed to use the perigee of a terrestrial spacecraft in \textrm{a} highly elliptic orbit, while \textrm{\citet{Bogo59}} considered also the node.
\textrm{\citet{Yil59}}, aware of the aliasing effect of the much larger classical precessions induced by the non-sphericity of the Earth,
proposed to launch a satellite in a polar orbit to cancel them. About twenty years later,\citet{vpe76a,vpe76b}
suggested to use a pair of drag-free, counter-orbiting terrestrial spacecraft in nearly polar orbits to detect their combined Lense-Thirring
node precessions. Almost contemporaneously, \citet{Cug77,Cug78} suggested to use the passive geodetic satellite LAGEOS, in orbit
around the Earth since 1976 and tracked with the Satellite Laser Ranging (SLR) technique \citep{slr}, along with the other existing laser-ranged targets
to measure the Lense-Thirring node precession. About ten years later, \citet{Ciu86} proposed a somewhat simpler version of the van Patten-Everitt
mission consisting of looking at the sum of the nodes of LAGEOS and of another SLR satellite to be launched in the same orbit, apart from the
inclination which should be switched by 180 deg in order to minimize the competing classical precessions due to the centrifugal oblateness of the
Earth. \citet{IorioPL} showed that such an orbital configuration would allow, in principle, to use the difference of the perigees as well.
\textrm{Test calculations were performed by} \citet{tanti,Ciuetal97a} with  the
LAGEOS and LAGEOS II satellites,
according to a strategy by \citet{Ciu96} involving the use of a suitable linear combination of the nodes $\Omega$ of both
satellites and the perigee $\omega$ of LAGEOS II in order to remove the impact of the first two multipoles of the non-spherical gravitational potential
of the Earth. Latest tests have been reported by \citet{Ciu04}, \citet{Ciu06}, \citet{Ciu07}, \citet{Ciu10}, \citet{Luc07a}, and  \citet{Ries08,Riesetal}
with only the nodes of \textrm{both satellites} according to a combination of them explicitly proposed by \citet{IorMor}. The total uncertainty reached is
still \textrm{a} matter of debate \citep{crit1,Ciu05,Luc05,IorJoG,Ior07,Iordrag,Ior10b,Ciu10} because of the lingering uncertainties in the Earth's multipoles and in how
to evaluate their biasing impact; it may be as large as $\approx 20-30\%$ according to conservative evaluations \citep{crit1,IorJoG,Ior07,Iordrag,Ior10b},
while more optimistic views \citep{Ciu04,Ciu05,Ciu06,Ries08,Riesetal,Ciu10} point towards $\approx 10-15\%$.

\subsection{The LAGEOS-LAGEOS II tests}\lb{OBLA}
LAGEOS \citep{LAGO} was put into orbit in March 1976, followed by its twin LAGEOS II \citep{Zerbini} in October 1992; they are passive, spherical spacecraft entirely
covered by retroreflectors which allow for their accurate tracking through laser pulses sent from Earth-based ground stations.
They orbit at altitudes of about 6000 km  in nearly circular paths markedly inclined
to the Earth's equator; see Table \ref{OSIGNUR} for their orbital geometries and Lense-Thirring node precessions\footnote{\citet{Cug78} quoted
40 mas yr$^{-1}$ for the Lense-Thirring node precession of LAGEOS by modeling the Earth as a spinning
homogeneous sphere. The correct value for the Lense-Thirring node precession of LAGEOS was obtained by \citet{Ciu86}. }. The corresponding linear shifts amount to
about 1.7 m yr$^{-1}$ in the cross-track direction\footnote{A perturbing acceleration like $\bds A_{\rm GM}$ is customarily projected onto the
radial ${\hat r}$, transverse ${\hat \tau}$ and cross-track ${\hat \nu}$ directions of an orthogonal  frame comoving with the
satellite \citep{Sof}; it turns out that the Lense-Thirring node precession affects the cross-track component
of the orbit according to $\Delta\nu_{\rm LT} \approx a\sin I\Delta\Omega_{\rm LT}$ (eq. (A65), p. 6233 in \citep{Cri}).} at the LAGEOS altitudes.

\begin{table}[t]
\caption{Orbital parameters and Lense-Thirring node precessions of LAGEOS, LAGEOS II, LARES and GRACE for
$S_{\oplus} = 5.86\times 10^{33}$ kg m$^2$ s$^{-1}$ \protect\citep{IERS}.  The semimajor axis $a$ is in km, the inclination $I$ is in deg, and the Lense-Thirring rate $\dot\Omega_{\rm LT}$ is in mas yr$^{-1}$.}\label{OSIGNUR}
\begin{tabular}{@{}lllll}
\hline
Satellite & $a$  & $e$ & $I$  & $\dot\Omega_{\rm LT}$  \\
\hline
LAGEOS & 12270 &  0.0045 & 109.9 & 30.7\\
LAGEOS II & 12163 &  0.014 & 52.65 & 31.5 \\
LARES & 7828 &  0.0 & 71.5 & 118.1 \\
GRACE & 6835 & 0.001 & 89.02 & 177.4\\
\hline
\end{tabular}
\end{table}
Since earlier studies \citep{Bogo59,Cug78}, researchers were aware \textrm{that a major source of systematic errors is represented by
the even ($\ell=2,4,6,\dots$) zonal ($m=0$)} harmonic coefficients\footnote{The relation among the  even zonals $J_{\ell}$ and
the normalized Stokes gravity coefficients $\overline{C}_{\ell 0}$,
which are customarily determined in the global Earth's gravity solutions, is $J_{\ell}\doteq -\sqrt{2\ell + 1}\ \overline{C}_{\ell 0}$.} $J_{\ell}, \ell=2,4,6$ of the
multipolar expansion of the classical part of the terrestrial gravitational potential, accounting for its departures from
spherical symmetry  due to the Earth's diurnal rotation, induce competing secular
precessions\footnote{Also the mean anomaly $\mathcal{M}$ experiences secular precession due to the even zonals, but it is
not involved in the measurement of the Lense-Thirring effect with the LAGEOS satellites.} of the node and the perigee of
satellites \citep{Kau} whose nominal sizes are several orders of magnitude larger than the Lense-Thirring ones. They cannot be removed from the
time series of data without affecting the Lense-Thirring pattern itself as well. The only thing that can be done is to model such a
corrupting effect as most accurately as possible and assessing the impact of the residual mismodelling on the measurement of the gravitomagnetic effect.
In the case of the node,
the secular precessions induced by the even zonals of the geopotential can be written as
\begin{equation}
\dot\Omega^{\rm geopot}=\sum_{\ell  =2}\dot\Omega_{.\ell}J_{\ell},
\end{equation}
where the coefficients $\dot\Omega_{.\ell}, \ell=2,4,6,\textrm{\dots}$ depend on the parameters of the Earth ($GM$ and the equatorial radius $R$) and
on the semimajor axis $a$, the inclination $I$ and the eccentricity $e$ of the satellite. For example, for $\ell=2$
\textrm{the largest precession is} due to the first even zonal harmonic $J_2$ which,
in the small eccentricity approximation valid for  geodetic satellites, is \textrm{given by}
\eqi \dot\Omega_{J_2}=-\rp{3}{2}n\left(\rp{R_{\oplus}}{a}\right)^2\rp{\cos I J_2}{(1-e^2)^2}.\lb{trazi}\eqf
The mismodelling in the \textrm{geopotential}-induced precessions can be written as
\begin{equation}\delta\dot\Omega^{\rm geopot}\leq \sum_{\ell  =2}\left|\dot\Omega_{.\ell}\right|\delta J_{\ell},\lb{mimo}\end{equation}
where $\delta J_{\ell}$ represents our uncertainty in the knowledge of the even zonals $J_{\ell}$. The coefficients $\dot\Omega_{.\ell}$ of the aliasing
classical node precessions \citep{Kau} $\dot\Omega^{\rm geopot}$ induced by the even zonals  have been analytically
worked out up to $\ell=20$ in the small eccentricity approximation by, e.g. \citet{Ior03}.

The three-elements combination used by \citet{tanti} allowed for removing the uncertainties in $J_2$ and $J_4$.
In \citet{Ciu98} a $\approx 20\%$ test was reported
by using the\footnote{Contrary to the subsequent models based on the dedicated satellites CHAMP (\url{http://www-app2.gfz-potsdam.de/pb1/op/champ/})
and GRACE (\url{http://www-app2.gfz-potsdam.de/pb1/op/grace/index_GRACE.html}), EGM96 relies upon multidecadal tracking of SLR data of a constellation of
geodetic satellites including LAGEOS and LAGEOS II as well; thus the possibility of a sort of $a-priori$ `imprinting' of the Lense-Thirring effect itself,
not solved-for in EGM96, cannot be neglected.} EGM96 Earth gravity model \citep{Lem98}; subsequent detailed analyses showed that such an evaluation of
the total error budget was overly optimistic in view of the likely unreliable computation of the total bias due to the even
zonals \citep{Ior03,Ries03a,Ries03b}.  An analogous, huge underestimation turned out to hold also for the effect of the
non-gravitational perturbations \citep{Mil87} like the direct solar radiation pressure, the Earth's albedo, various subtle thermal
effects depending on  the physical properties of the satellites' surfaces and their rotational state
\citep{Inv94,Ves99,Luc01,Luc02,Luc03,Luc04,Lucetal04,Ries03a}, which the perigees  of LAGEOS-like satellites are particularly sensitive to.
As a result, the realistic total error budget in the test reported in \citep{Ciu98} might be as large as $60-90\%$ or (by considering EGM96 only) even more.

The observable used by \citet{Ciu04} with the EIGEN-GRACE02S model \citep{eigengrace02s} and by \citet{Ries08} with other
more recent Earth gravity models was a linear combination\footnote{See also \citep{Pav02,Ries03a,Ries03b}.} of
the nodes of LAGEOS and LAGEOS II
\begin{equation}
f\doteq\dot\Omega^{\rm LAGEOS}+
c_1\dot\Omega^{\rm LAGEOS\ II }, \lb{combi}
\end{equation}
which was explicitly computed by \citet{IorMor} following the approach put forth by \citet{Ciu96}.
\textrm{Here,
\begin{equation}
c_1  \doteq -\rp{\dot\Omega^{\rm LAGEOS}_{.2}}{\dot\Omega^{\rm
LAGEOS\ II }_{.2}},
\end{equation}
and by means of \rfr{trazi} we find
{\small{\begin{equation}
c_1 =-\rp{\cos I_{\rm LAGEOS}}{\cos I_{\rm LAGEOS\
II}}\left(\rp{1-e^2_{\rm LAGEOS\ II}}{1-e^2_{\rm
LAGEOS}}\right)^2\left(\rp{a_{\rm LAGEOS\ II}}{a_{\rm LAGEOS}}\right)^{\rp{7}{2}}\lb{coff},
\end{equation}}}
where the values of Table \ref{OSIGNUR} approximately yield $c_1=0.544$.}
%
 The Lense-Thirring signature of \rfr{combi} amounts to 47.8 mas yr$^{-1}$. The combination  of \rfr{combi} allows, by construction, to remove the aliasing
 effects due to the static and time-varying parts of the first even zonal $J_2$. The nominal (i.e. computed with the estimated values of
 $J_{\ell}$, $\ell=4,6\textrm{,\dots}$) bias due to the remaining higher degree even zonals would amount to  about $10^5$ mas yr$^{-1}$; the need
 \textrm{for} a
 careful and reliable modeling of such an important source of systematic bias \textrm{is thus quite} apparent. Conversely, the nodes of the LAGEOS-type
 spacecraft are directly affected by the non-gravitational accelerations at a $\approx 1\%$ level of the Lense-Thirring effect
 \citep{Luc01,Luc02,Luc03,Luc04,Lucetal04}. For a comprehensive, up-to-date overview of the numerous and subtle issues concerning the
 measurement of the Lense-Thirring effect see, e.g., \citep{IorNOVA}.

 \subsubsection{Conservative evaluation of the impact of the mismodelling in the even zonal harmonics}\lb{gravAZZA}
 A common feature of all the competing evaluations  so far published is that the systematic bias due to the static component of  the
  geopotential was always calculated  by using the released (more or less accurately calibrated) sigmas $\sigma_{J_{\ell}}$  of one Earth gravity model
  solution at a time for the uncertainties $\delta J_{\ell}$. Thus, it was said that the model X yields a $x\%$ error, the model Y
  yields a $y\%$ error, and so on.

Since a trustable calibration of the formal, statistical uncertainties in the estimated zonals of the covariance matrix of a global solution is
always a difficult task to be implemented in a reliable way, a much more realistic and conservative approach consists, instead,
of taking the difference\footnote{See Fig.5 of \citet{Luc07b} for a comparison
of the estimated $\overline{C}_{40}$ in different models.}
\eqi
\Delta J_{\ell}=\left|J_{\ell}(\rm X) - J_{\ell}(\rm Y)\right|,\ \ell=2,4,6,\textrm{\dots}
\eqf
of the
estimated even zonals for different pairs of Earth gravity field solutions as representative of the real uncertainty $\delta J_{\ell}$ in the zonals \citep{Lerch}. In \textrm{Tables \ref{tavola1}--\ref{tavolanuova2}} we present our results for the most recent GRACE-based models released so far by different institutions and retrievable \textrm{in} the Internet at\footnote{L. I. thanks M Watkins (JPL) for having provided me with the even zonals and their sigmas of the  JEM01-RL03B model.}
\url{http://icgem.gfz-potsdam.de/ICGEM/}.
The models used are EIGEN-GRACE02S \citep{eigengrace02s}  from GFZ (Potsdam, Germany), GGM02S \citep{ggm02} and GGM03S \citep{ggm03} from CSR (Austin, Texas),
ITG-Grace02s \citep{ITG}, ITG-Grace03s
\citep{itggrace03s} and ITG-Grace2010s \citep{ITG010} from IGG (Bonn, Germany), JEM01-RL03B from JPL (NASA, USA),  AIUB-GRACE01S \citep{aiub} and AIUB-GRACE02S \citep{aiub2} from AIUB (Switzerland).
This approach was explicitly followed also, e.g., by \citet{Mil87} with the GEM-L2 and GEM 9 models, and by \citet{Ciu96} with the JGM3 and GEMT-2 models. Note that we do not consider
models including data from CHAMP, LAGEOS itself\footnote{It is just one of the devices with which the Lense-Thirring effect is measured:
using Earth's gravity models
including its data would yield  a-priori \virg{imprint} of GTR itself. Note that the latest models by GFZ  are unsuitable for our purposes because
they make use of LAGEOS data.}
and Earth-based data.
In \textrm{Tables \ref{tavola1}--\ref{tavolanuova2}} we quote both the sum $\sum_{\ell=4}^{20}f_{\ell}$ of the absolute
values of the individual mismodelled terms \textrm{(denoted by SAV)}
\eqi
f_{\ell}\doteq\left|\dot\Omega_{.\ell}^{\rm LAGEOS} + c_1\dot\Omega_{.\ell}^{\rm LAGEOS\ II}\right|\Delta J_{\ell}
\eqf
\textrm{and the square root} of the sum of their squares $\sqrt{\sum_{\ell=4}^{20}f^2_{\ell}}$ (RSS); in both cases we normalized them
to the combined Lense-Thirring total precession of 47.8 mas yr$^{-1}$.
\begin{table*}[t]
   \small\caption{Impact of the mismodelling in the even zonal harmonics on $f_{\ell}=\left|\dot\Omega^{\rm LAGEOS}_{\ell} + c_1\dot\Omega^{\rm LAGEOS\ II}_{.\ell}\right|\Delta J_{\ell},\ \ell=4,\dots,20$, in mas yr$^{-1}$. Recall that $J_{\ell}=-\sqrt{2\ell + 1}\ \overline{C}_{\ell 0}$; for the uncertainty in the even zonals we have taken here the difference $\Delta\overline{C}_{\ell 0}=\left|\overline{C}_{\ell 0}^{\rm (X)}-\overline{C}_{\ell 0}^{\rm (Y)}\right|$ between the model X=GGM02S \protect\citep{ggm02} and the model Y=ITG-Grace02s \protect\citep{ITG}.
   GGM02S is based on 363 days of GRACE-only data   (GPS and intersatellite tracking, neither constraints nor regularization applied)
spread between April 4, 2002 and Dec 31, 2003. The $\sigma$ are formal for both models. $\Delta \overline{C}_{\ell 0}$ are always larger than the linearly added sigmas, apart from   $\ell=12$ and $\ell=18$. Values of $f_{\ell}$ smaller than 0.1 mas yr$^{-1}$ have not been quoted. The Lense-Thirring precession of the combination of \rfr{combi} amounts to 47.8 mas yr$^{-1}$. The percent bias $\delta\mu$ have been computed by normalizing the linear sum of $f_{\ell}, \ell=4,\dots,20$ (SAV) and the square root of the sum of $f_\ell^2, \ell=4,\dots,20$ to the Lense-Thirring combined precessions.
}\label{tavola1}
\begin{tabular}{@{}llll@{}}
\tableline
$\ell$ & $\Delta\overline{C}_{\ell 0}$ (GGM02S-ITG-Grace02s) & $\sigma_{\rm  X}+\sigma_{\rm Y}$ & $f_{\ell}$  (mas yr$^{-1}$)\\
\tableline
4 & $1.9\times 10^{-11}$ &  $8.7\times 10^{-12}$ & 7.2\\
6 & $2.1\times 10^{-11}$ &  $4.6\times 10^{-12}$ & 4.6\\
8 & $5.7\times 10^{-12}$ &  $2.8\times 10^{-12}$ & 0.2\\
10 & $4.5\times 10^{-12}$ &  $2.0\times 10^{-12}$ & -\\
12 & $1.5\times 10^{-12}$ &  $1.8\times 10^{-12}$ & -\\
14 & $6.6\times 10^{-12}$ &  $1.6\times 10^{-12}$ & -\\
16 & $2.9\times 10^{-12}$ &  $1.6\times 10^{-12}$ & -\\
18 & $1.4\times 10^{-12}$ &  $1.6\times 10^{-12}$ & -\\
20 & $2.0\times 10^{-12}$ &  $1.6\times 10^{-12}$ & -\\

\tableline
 &    $\delta\mu = 25\%$ (SAV) & $\delta\mu = 18\%$ (RSS) &   \\  %
\tableline %
\end{tabular}
\end{table*}
%
%
%
%
%
%
%
%
%
%
%
%
%
%
%
%
%
%
%

\begin{table*}[t]
   \small\caption{Bias due to the mismodelling in the even zonals of the models X=ITG-Grace03s \protect\citep{itggrace03s}, based on GRACE-only accumulated normal equations from data out of September 2002-April 2007 (neither apriori information nor regularization used), and Y=GGM02S \protect\citep{ggm02}.  The $\sigma$ for both models are formal. $\Delta \overline{C}_{\ell 0}$ are always larger than the linearly added sigmas, apart from  $\ell=12$ and $\ell=18$.}\label{tavola11}
\begin{tabular}{@{}llll@{}}
\tableline
$\ell$ & $\Delta\overline{C}_{\ell 0}$ (ITG-Grace03s-GGM02S) & $\sigma_{\rm  X}+\sigma_{\rm Y}$ & $f_{\ell}$  (mas yr$^{-1}$)\\
\tableline
4 & $2.58\times 10^{-11}$ &  $8.6\times 10^{-12}$ & 9.6\\
6 & $1.39\times 10^{-11}$ &  $4.7\times 10^{-12}$ & 3.1\\
8 & $5.6\times 10^{-12}$ &  $2.9\times 10^{-12}$ & 0.2\\
10 & $1.03\times 10^{-11}$ &  $2\times 10^{-12}$ & -\\
12 & $7\times 10^{-13}$ &  $1.8\times 10^{-12}$ & -\\
14 & $7.3\times 10^{-12}$ &  $1.6\times 10^{-12}$ & -\\
16 & $2.6\times 10^{-12}$ &  $1.6\times 10^{-12}$ & -\\
18 & $8\times 10^{-13}$ &  $1.6\times 10^{-12}$ & -\\
20 & $2.4\times 10^{-12}$ &  $1.6\times 10^{-12}$ & -\\

\tableline
&    $\delta\mu = 27\%$ (SAV) & $\delta\mu = 21\%$ (RSS) &   \\  %
\tableline %
\end{tabular}
\end{table*}
\begin{table*}[t]
   \small\caption{Bias due to the mismodelling in the even zonals of the models  X = GGM02S \protect\citep{ggm02} and Y = GGM03S \protect\citep{ggm03} retrieved from data spanning January 2003 to December 2006.
    The $\sigma$ for GGM03S are calibrated. $\Delta \overline{C}_{\ell 0}$ are larger than the linearly added sigmas for $\ell = 4,6$. (The other zonals are of no concern)}\label{tavola03S}
\begin{tabular}{@{}llll@{}}
\tableline
$\ell$ & $\Delta\overline{C}_{\ell 0}$ (GGM02S-GGM03S) & $\sigma_{\rm  X}+\sigma_{\rm Y}$ & $f_{\ell}$  (mas yr$^{-1}$)\\
\tableline
4 & $1.87\times 10^{-11}$ &  $1.25\times 10^{-11}$ & 6.9\\
6 & $1.96\times 10^{-11}$ &  $6.7\times 10^{-12}$ & 4.2\\
8 & $3.8\times 10^{-12}$ &  $4.3\times 10^{-12}$ & 0.1\\
10 & $8.9\times 10^{-12}$ &  $2.8\times 10^{-12}$ & 0.1\\
12 & $6\times 10^{-13}$ &  $2.4\times 10^{-12}$ & -\\
14 & $6.6\times 10^{-12}$ &  $2.1\times 10^{-12}$ & -\\
16 & $2.1\times 10^{-12}$ &  $2.0\times 10^{-12}$ & -\\
18 & $1.8\times 10^{-12}$ &  $2.0\times 10^{-12}$ & -\\
20 & $2.2\times 10^{-12}$ &  $1.9\times 10^{-12}$ & -\\

\tableline
&    $\delta\mu = 24\%$ (SAV) & $\delta\mu = 17\%$ (RSS) &   \\  %
\tableline %
\end{tabular}
\end{table*}
\begin{table*}[t]
   \small\caption{Bias due to the mismodelling in the even zonals of the models  X = EIGEN-GRACE02S \protect\citep{eigengrace02s} and Y = GGM03S \protect\citep{ggm03}.
    The $\sigma$ for both models are calibrated. $\Delta \overline{C}_{\ell 0}$ are always larger than the linearly added sigmas apart from $\ell = 14,18$.}\label{tavola033S}
 \begin{tabular}{@{}llll@{}}
\tableline
$\ell$ & $\Delta\overline{C}_{\ell 0}$ (EIGEN-GRACE02S-GGM03S) & $\sigma_{\rm  X}+\sigma_{\rm Y}$ & $f_{\ell}$  (mas yr$^{-1}$)\\
\tableline
4 & $2.00\times 10^{-11}$ &  $8.1\times 10^{-12}$ & 7.4\\
6 & $2.92\times 10^{-11}$ &  $4.3\times 10^{-12}$ & 6.3\\
8 & $1.05\times 10^{-11}$ &  $3.0\times 10^{-12}$ & 0.4\\
10 & $7.8\times 10^{-12}$ &  $2.9\times 10^{-12}$ & 0.1\\
12 & $3.9\times 10^{-12}$ &  $1.8\times 10^{-12}$ & -\\
14 & $5\times 10^{-13}$ &  $1.7\times 10^{-12}$ & -\\
16 & $1.7\times 10^{-12}$ &  $1.4\times 10^{-12}$ & -\\
18 & $2\times 10^{-13}$ &  $1.4\times 10^{-12}$ & -\\
20 & $2.5\times 10^{-12}$ &  $1.4\times 10^{-12}$ & -\\

\tableline
&    $\delta\mu = 30\%$ (SAV) & $\delta\mu = 20\%$ (RSS) &   \\  %
\tableline
\end{tabular}
\end{table*}
\begin{table*}[t]
   \small\caption{Bias due to the mismodelling in the even zonals of the models  X = JEM01-RL03B, based on 49 months of GRACE-only data, and Y = GGM03S \protect\citep{ggm03}.
    The $\sigma$ for GGM03S are calibrated. $\Delta \overline{C}_{\ell 0}$ are always larger than the linearly added sigmas apart from $\ell = 16$.}\label{tavolaJEM1}
\begin{tabular}{@{}llll@{}}
\tableline
$\ell$ & $\Delta\overline{C}_{\ell 0}$ (JEM01-RL03B-GGM03S) & $\sigma_{\rm  X}+\sigma_{\rm Y}$ & $f_{\ell}$  (mas yr$^{-1}$)\\
\tableline
4 & $1.97\times 10^{-11}$ &  $4.3\times 10^{-12}$ & 7.3\\
6 & $2.7\times 10^{-12}$ &  $2.3\times 10^{-12}$ & 0.6\\
8 & $1.7\times 10^{-12}$ &  $1.6\times 10^{-12}$ & -\\
10 & $2.3\times 10^{-12}$ &  $8\times 10^{-13}$ & -\\
12 & $7\times 10^{-13}$ &  $7\times 10^{-13}$ & -\\
14 & $1.0\times 10^{-12}$ &  $6\times 10^{-13}$ & -\\
16 & $2\times 10^{-13}$ &  $5\times 10^{-13}$ & -\\
18 & $7\times 10^{-13}$ &  $5\times 10^{-13}$ & -\\
20 & $5\times 10^{-13}$ &  $4\times 10^{-13}$ & -\\

\tableline
&    $\delta\mu = 17\%$ (SAV) & $\delta\mu = 15\%$ (RSS) &   \\  %
\tableline %
\end{tabular}
\end{table*}
\begin{table*}[t]
   \small\caption{Bias due to the mismodelling in the even zonals of the models  X = JEM01-RL03B and Y = ITG-Grace03s \protect\citep{itggrace03s}.
    The $\sigma$ for ITG-Grace03s are formal. $\Delta \overline{C}_{\ell 0}$ are always larger than the linearly added sigmas.}\label{tavolaJEM2}
\begin{tabular}{@{}llll@{}}
\tableline
$\ell$ & $\Delta\overline{C}_{\ell 0}$ (JEM01-RL03B-ITG-Grace03s) & $\sigma_{\rm  X}+\sigma_{\rm Y}$ & $f_{\ell}$  (mas yr$^{-1}$)\\
\tableline
4 & $2.68\times 10^{-11}$ &  $4\times 10^{-13}$ & 9.9\\
6 & $3.0\times 10^{-12}$ &  $2\times 10^{-13}$ & 0.6\\
8 & $3.4\times 10^{-12}$ &  $1\times 10^{-13}$ & 0.1\\
10 & $3.6\times 10^{-12}$ &  $1\times 10^{-13}$ & -\\
12 & $6\times 10^{-13}$ &  $9\times 10^{-14}$ & -\\
14 & $1.7\times 10^{-12}$ &  $9\times 10^{-14}$ & -\\
16 & $4\times 10^{-13}$ &  $8\times 10^{-14}$ & -\\
18 & $4\times 10^{-13}$ &  $8\times 10^{-14}$ & -\\
20 & $7\times 10^{-13}$ &  $8\times 10^{-14}$ & -\\

\tableline
&    $\delta\mu = 22\%$ (SAV) & $\delta\mu = 10\%$ (RSS) &   \\  %
\tableline
\end{tabular}
\end{table*}
\begin{table*}[t]
   \small\caption{Aliasing effect of the mismodelling in the even zonal harmonics estimated in the X=ITG-Grace03s \protect\citep{itggrace03s} and the Y=EIGEN-GRACE02S \protect\citep{eigengrace02s} models.  The covariance matrix $\sigma$ for ITG-Grace03s are formal, while the ones of EIGEN-GRACE02S are calibrated. $\Delta \overline{C}_{\ell 0}$ are larger than the linearly added sigmas for $\ell =4,...,20$, apart from $\ell=18$. }\label{tavola7}
\begin{tabular}{@{}llll@{}}
\tableline
$\ell$ & $\Delta\overline{C}_{\ell 0}$ (ITG-Grace03s-EIGEN-GRACE02S) & $\sigma_{\rm  X}+\sigma_{\rm Y}$ & $f_{\ell}$  (mas yr$^{-1}$)\\
\tableline
4 & $2.72\times 10^{-11}$ &  $3.9\times 10^{-12}$ & 10.1\\
6 & $2.35\times 10^{-11}$ &  $2.0\times 10^{-12}$ & 5.1\\
8 & $1.23\times 10^{-11}$ &  $1.5\times 10^{-12}$ & 0.4\\
10 & $9.2\times 10^{-12}$ &  $2.1\times 10^{-12}$ & 0.1\\
12 & $4.1\times 10^{-12}$ &  $1.2\times 10^{-12}$ & -\\
14 & $5.8\times 10^{-12}$ &  $1.2\times 10^{-12}$ & -\\
16 & $3.4\times 10^{-12}$ &  $9\times 10^{-13}$ & -\\
18 & $5\times 10^{-13}$ &  $1.0\times 10^{-12}$ & -\\
20 & $1.8\times 10^{-12}$ &  $1.1\times 10^{-12}$ & -\\

\tableline
&    $\delta\mu = 37\%$ (SAV) & $\delta\mu = 24\%$ (RSS) &   \\  %
\tableline %
\end{tabular}

\end{table*}
%
%
%
%
           %
 %
%
%
%
%
%
%
%
\begin{table*}[t]
   \small\caption{Bias due to the mismodelling in the even zonals of the models  X = JEM01-RL03B, based on 49 months of GRACE-only data, and Y = AIUB-GRACE01S \protect\citep{aiub}. The latter one was obtained from GPS satellite-to-satellite tracking data and K-band range-rate data out of the
period January 2003 to December 2003 using the Celestial Mechanics Approach.
No accelerometer data, no de-aliasing products, and no regularisation was
applied.
    The $\sigma$ for AIUB-GRACE01S are formal.
    $\Delta \overline{C}_{\ell 0}$ are always larger than the linearly added sigmas.
    }\label{tavolaAIUB1}
\begin{tabular}{@{}llll@{}}
\tableline
$\ell$ & $\Delta\overline{C}_{\ell 0}$ (JEM01-RL03B$-$AIUB-GRACE01S) & $\sigma_{\rm  X}+\sigma_{\rm Y}$ & $f_{\ell}$  (mas yr$^{-1}$)\\
\tableline
4 & $2.95\times 10^{-11}$ &  $2.1\times 10^{-12}$ & 11\\
6 & $3.5\times 10^{-12}$  &  $1.3\times 10^{-12}$ & 0.8\\
8 & $2.14\times 10^{-11}$ &  $5\times 10^{-13}$ & 0.7\\
10 & $4.8\times 10^{-12}$ &  $5\times 10^{-13}$ & -\\
12 & $4.2\times 10^{-12}$ &  $5\times 10^{-13}$ & -\\
14 & $3.6\times 10^{-12}$ &  $5\times 10^{-13}$ & -\\
16 & $8\times 10^{-13}$ &    $5\times 10^{-13}$ & -\\
18 & $7\times 10^{-13}$  &    $5\times 10^{-13}$ & -\\
20 & $1.0\times 10^{-12}$ &    $5\times 10^{-13}$ & -\\

\tableline
&   $\delta\mu = 26\%$ (SAV)&    $\delta\mu = 23\%$ (RSS)& \\  %
\tableline %
\end{tabular}
\end{table*}
\begin{table*}[t]
   \small\caption{Bias due to the mismodelling in the even zonals of the models  X = EIGEN-GRACE02S \protect\citep{eigengrace02s} and Y = AIUB-GRACE01S \protect\citep{aiub}. The $\sigma$ for AIUB-GRACE01S are formal, while those of EIGEN-GRACE02S are calibrated.
    $\Delta \overline{C}_{\ell 0}$ are  larger than the linearly added sigmas for $\ell=4,6,8,16$.
    }\label{tavolaAIUB2}
\begin{tabular}{@{}llll@{}}
\tableline
$\ell$ & $\Delta\overline{C}_{\ell 0}$ (EIGEN-GRACE02S$-$AIUB-GRACE01S) & $\sigma_{\rm  X}+\sigma_{\rm Y}$ & $f_{\ell}$  (mas yr$^{-1}$)\\
\tableline
4 & $2.98\times 10^{-11}$ &  $6.0\times 10^{-12}$ & 11.1\\
6 & $2.29\times 10^{-11}$  &  $3.3\times 10^{-12}$ & 5.0\\
8 & $1.26\times 10^{-11}$ &  $1.9\times 10^{-12}$ & 0.4\\
10 & $6\times 10^{-13}$ &  $2.5\times 10^{-12}$ & -\\
12 & $5\times 10^{-13}$ &  $1.6\times 10^{-12}$ & -\\
14 & $5\times 10^{-13}$ &  $1.6\times 10^{-12}$ & -\\
16 & $2.9\times 10^{-12}$ &    $1.4\times 10^{-12}$ & -\\
18 & $6\times 10^{-13}$  &    $1.4\times 10^{-12}$ & -\\
20 & $2\times 10^{-13}$ &    $1.5\times 10^{-12}$ & -\\

\tableline
&   $\delta\mu = 34\%$ (SAV)&    $\delta\mu = 25\%$ (RSS)& \\  %
\tableline %
\end{tabular}
\end{table*}
\begin{table*}[t]
   \small\caption{Bias due to the mismodelling in the even zonals of the models  X = JEM01-RL03B  and Y = AIUB-GRACE02S \protect\citep{aiub2}.
   The $\sigma$ for both AIUB-GRACE02S and JEM01-RL03B are formal.
    $\Delta \overline{C}_{\ell 0}$ are  larger than the linearly added sigmas for $\ell=4-20$.
    }\label{tavolaAIUB2jem}
\begin{tabular}{@{}llll@{}}
\tableline
$\ell$ & $\Delta\overline{C}_{\ell 0}$ (JEM01-RL03B$-$AIUB-GRACE02S) & $\sigma_{\rm  X}+\sigma_{\rm Y}$ & $f_{\ell}$  (mas yr$^{-1}$)\\
\tableline
4 & $1.58\times 10^{-11}$ &  $2\times 10^{-13}$ & 5.9\\
6 & $5.9\times 10^{-12}$  &  $1\times 10^{-13}$ & 1.3\\
8 & $5.8\times 10^{-12}$ &  $1\times 10^{-13}$ & 0.2\\
10 & $1.27\times 10^{-11}$ &  $1\times 10^{-13}$ & 0.1\\
12 & $4.3\times 10^{-12}$ &  $1\times 10^{-13}$ & -\\
14 & $2.7\times 10^{-12}$ &  $1\times 10^{-13}$ & -\\
16 & $1.6\times 10^{-12}$ &    $1\times 10^{-13}$ & -\\
18 & $1.8\times 10^{-12}$  &    $1\times 10^{-13}$ & -\\
20 & $1.9\times 10^{-12}$ &    $1\times 10^{-13}$ & -\\

\tableline
&   $\delta\mu = 16\%$ (SAV)&    $\delta\mu = 13\%$ (RSS)& \\  %
\tableline %
\end{tabular}
\end{table*}
\begin{table*}[t]
   \small\caption{Bias due to the mismodelling in the even zonals of the models  X = ITG-Grace2010s \protect\citep{ITG010}  and Y = AIUB-GRACE02S \protect\citep{aiub2}. The ITG-Grace2010s model has been obtained by processing  7 yr (2002-2009) of GRACE data.
   The $\sigma$ for both models are formal.
    $\Delta \overline{C}_{\ell 0}$ are  larger than the linearly added sigmas for all the degrees considered.
    }\label{tavolanuova1}
\begin{tabular}{@{}llll@{}}
\tableline
$\ell$ & $\Delta\overline{C}_{\ell 0}$ (ITG-Grace2010s$-$AIUB-GRACE02S) & $\sigma_{\rm  X}+\sigma_{\rm Y}$ & $f_{\ell}$  (mas yr$^{-1}$)\\
\tableline
4 & $2.665\times 10^{-11}$ &  $1.9\times 10^{-13}$ & 9.9\\
6 & $4.66\times 10^{-12}$  &  $1.1\times 10^{-13}$ & 1.0\\
8 & $3.86\times 10^{-12}$ &  $9\times 10^{-14}$ & 0.1\\
10 & $9.50\times 10^{-12}$ &  $8\times 10^{-14}$ & 0.1\\
12 & $1.67\times 10^{-12}$ &  $7\times 10^{-14}$ & -\\
14 & $2.36\times 10^{-12}$ &  $7\times 10^{-14}$ & -\\
16 & $8.7\times 10^{-13}$ &    $6\times 10^{-14}$ & -\\
18 & $8.0\times 10^{-13}$  &    $6\times 10^{-14}$ & -\\
20 & $1.09\times 10^{-12}$ &    $7\times 10^{-14}$ & -\\

\tableline
&   $\delta\mu = 23\%$ (SAV)&    $\delta\mu = 21\%$ (RSS)& \\  %
\tableline %
\end{tabular}
\end{table*}
\begin{table*}[t]
   \small\caption{Bias due to the mismodelling in the even zonals of the models  X = ITG-Grace2010s \protect\citep{ITG010}  and Y = GGM03S \protect\citep{ggm03}.
   The $\sigma$ of ITG-Grace2010s are formal, while those for GGM03S are calibrated.
    $\Delta \overline{C}_{\ell 0}$ are  larger than the linearly added sigmas for $\ell=4,10,12,16,18$.
    }\label{tavolanuova2}
\begin{tabular}{@{}llll@{}}
\tableline
$\ell$ & $\Delta\overline{C}_{\ell 0}$ (ITG-Grace2010s$-$GGM03S) & $\sigma_{\rm  X}+\sigma_{\rm Y}$ & $f_{\ell}$  (mas yr$^{-1}$)\\
\tableline
4 & $3.05\times 10^{-11}$ &  $4.3\times 10^{-12}$ & 11.3\\
6 & $1.4\times 10^{-12}$  &  $2.3\times 10^{-12}$ & 0.3\\
8 & $3\times 10^{-13}$ &  $1.6\times 10^{-12}$ & -\\
10 & $9\times 10^{-13}$ &  $8\times 10^{-13}$ & -\\
12 & $1.9\times 10^{-12}$ &  $6\times 10^{-13}$ & -\\
14 & $6\times 10^{-13}$ &  $6\times 10^{-13}$ & -\\
16 & $9\times 10^{-13}$ &    $4\times 10^{-13}$ & -\\
18 & $1.7\times 10^{-12}$  &    $4\times 10^{-13}$ & -\\
20 & $2\times 10^{-13}$ &    $4\times 10^{-13}$ & -\\

\tableline
&   $\delta\mu = 25\%$ (SAV)&    $\delta\mu = 24\%$ (RSS)& \\  %
\tableline %
\end{tabular}
\end{table*}
The systematic bias evaluated with a more realistic approach is about 3 to 4 times larger than \textrm{the one obtained by using only
a single particular model}.
The scatter is still quite large and far from the $5-10\%$ claimed in \textrm{\citet{Ciu04}}. In particular, it appears that $J_4$, $J_6$, and to a
lesser extent $J_8$, which are just the most relevant zonals for us because of their impact on the combination of \rfr{combi}, are the most
uncertain ones, with discrepancies $\Delta J_{\ell}$ between different models, in general, larger than the sum of their sigmas $\sigma_{J_{\ell}}$,
\textrm{whether} calibrated or not.  

Such an approach has been criticized by \citet{Ciu10} by writing that one should not compare models with different intrinsic accuracies.
Moreover, \citet{Ciu10} claim that our method would be exactly equivalent to compare a modern value of the Newtonian gravitational
 constant $G$, accurate to
$10^{-5}$, to the earlier results \textrm{obtained in the 18$^{\rm th}$ century}, accurate to $10^{-2}$,
and conclude that the present-day accuracy would be wrong by a factor 1000.
Such criticisms are incorrect for the following reasons. According to CODATA\footnote{See \url{http://physics.nist.gov/cuu/Constants/} on the WEB.},
the present-day relative accuracy
in $G$ is $1.0\times 10^{-4}=0.01\%$, not $0.0015\%$, corresponding to $1.5\times 10^{-5}$, as claimed by \citet{Ciu10}; thus,
according to their reasoning corrected
for this error, our approach would be exactly equivalent to state that the present-day accuracy in $G$ would be wrong by a
factor \textrm{of} 100.
Moreover, the relative uncertainties in the Earth's gravity models considered are, in fact, all of the same order of magnitude;
for example, the relative uncertainty in $\overline{C}_{40}$ is $7.2\times 10^{-6}$ from
EIGEN-GRACE02S \citep{eigengrace02s} and $7.8\times 10^{-6}$ in the more recent GGM03S model\footnote{The relative uncertainty in $\overline{C}_{40}$ is $2\times 10^{-7}$ for ITG-Grace2010s \citep{ITG010}, but it must be recalled that for such a model the available errors are the formal, statistical ones.} \citep{ggm03}. Thus, the comparison drawn by \citet{Ciu10}
between $G$ and the even zonals
is misleading. \textrm{Even more important}, it is well known that the rejection  of a \virg{suspect} experimental result
from a sample of data is always a very delicate
matter \citep{Taylor}, and quantitative criteria are needed to reject one or more outliers \citep{Peirce,Chauvenet}.
Instead, \citet{Ciu10} do not apply any of them to support their claims against
the comparison of various geopotential models. The most famous rejection criterion \textrm{is perhaps the} one devised by \citet{Chauvenet}. Anyway, it relies
upon an arbitrary assumption that a measurement may be rejected if the probability of
obtaining the deviation from the mean for that value is less than the inverse of twice the
number of measurements; moreover, it makes no
distinction between the case of one or several suspicious data values. The criterion by \citet{Peirce}, instead, is
a rigorous theory that can be easily applied in the case of several suspicious data values
using the table in \citet{Ross}. Let us apply it to the case of $G$.
If we consider the set of modern measurements in Table 13 of \citet{codata}, accurate to $10^{-3}-10^{-4}$, and the result by \citet{Cavendish},
accurate to $10^{-2}$, reported by \citet{deBoer} and \citet{Ruff}, it turns out that, according to the \citet{Peirce} criterion,
the oldest value must be rejected.
On the contrary, if we apply the \citet{Peirce} criterion
to all the values of, e.g., $\overline{C}_{40}$ from the models considered, no one of them has to be rejected.
Interestingly, if we also considered models including CHAMP, LAGEOS and Earth-based data, and even the latest SLR-based solutions, the result would not change:
indeed, concerning  $\overline{C}_{40}$, only the CHAMP-based TUM-2S
model \citep{TUM} would not pass the  \citet{Peirce} criterion.
Thus, we do not see
any founded, quantitative reasons
to \textrm{decline} the comparison among different Earth's gravity models followed here.

Another way to evaluate the uncertainty in the LAGEOS-LAGEOS II node test may consist of computing the nominal values of the total
combined precessions for different models  and comparing them, i.e. by taking
\eqi
 \left|\sum_{\ell = 4}\left(\dot\Omega_{.\ell}^{\rm LAGEOS} + c_1\dot\Omega_{.\ell}^{\rm LAGEOS\ II}\right) [J_{\ell}({\rm  X})-J_{\ell}(\rm Y)]\right|.
 \eqf
The results for each pair of models are shown in
Table \ref{tavolaa}.
 \begin{table*}[t]\small\caption{ Systematic uncertainty $\delta\mu$ in the LAGEOS-LAGEOS II test evaluated by taking the absolute value of the difference between the nominal values of the total combined node precessions due to the even zonals for different models X and Y, i.e. $\left|
 \dot\Omega^{\rm geopot}({\rm X})-\dot\Omega^{\rm geopot}({\rm Y})\right|$. The average is $\approx 17\%$.}\label{tavolaa}

\begin{tabular}{@{}ll@{}}
\tableline
Models compared & $\delta\mu$  \\
\tableline

ITG-Grace2010s$-$JEM01-RL03B  & $7\%$\\
ITG-Grace2010s$-$GGM02S  & $0.3\%$\\
ITG-Grace2010s$-$GGM03S  & $24\%$\\
ITG-Grace2010s$-$ITG-Grace02  & $25\%$\\
ITG-Grace2010s$-$ITG-Grace03  & $27\%$\\
ITG-Grace2010s$-$AIUB-GRACE02S  & $23\%$\\
ITG-Grace2010s$-$AIUB-GRACE01S  & $27\%$\\
ITG-Grace2010s$-$EIGEN-GRACE02S  & $5\%$\\
AIUB-GRACE02S$-$JEM01-RL03B & $16\%$\\
%
%
AIUB-GRACE02S$-$GGM02S & $23\%$\\
AIUB-GRACE02S$-$GGM03S & $17\%$\\
AIUB-GRACE02S$-$ITG-Grace02 & $2\%$\\
AIUB-GRACE02S$-$ITG-Grace03 & $13\%$\\
%
%
AIUB-GRACE02S$-$EIGEN-GRACE02S & $28\%$\\
AIUB-GRACE01S$-$JEM01-RL03B & $20\%$\\
%
%
AIUB-GRACE01S$-$GGM02S & $27\%$\\
AIUB-GRACE01S$-$GGM03S & $3\%$\\
AIUB-GRACE01S$-$ITG-Grace02 & $2\%$\\
AIUB-GRACE01S$-$ITG-Grace03 & $0.1\%$\\
%
%
AIUB-GRACE01S$-$EIGEN-GRACE02S & $33\%$\\
%
%
%
%
%
%
%
%
%
JEM01-RL03B$-$GGM02S & $7\%$\\
JEM01-RL03B$-$GGM03S & $17\%$\\
JEM01-RL03B$-$ITG-Grace02 & $18\%$\\
JEM01-RL03B$-$ITG-Grace03s & $20\%$\\
%
%
JEM01-RL03B$-$EIGEN-GRACE02S & $13\%$\\
%
%
GGM02S$-$GGM03S & $24\%$\\
GGM02S$-$ITG-Grace02& $25\%$\\
GGM02S$-$ITG-Grace03s& $27\%$\\
%
%
GGM02S$-$EIGEN-GRACE02S & $6\%$\\
%
%
GGM03S$-$ITG-Grace02 & $1\%$\\
GGM03S$-$ITG-Grace03s & $3\%$\\
%
%
GGM03S$-$EIGEN-GRACE02S & $30\%$\\
%
%
ITG-Grace02$-$ITG-Grace03s & $2\%$\\
%
%
ITG-Grace02$-$EIGEN-GRACE02S & $31\%$\\
%
%
%
ITG-Grace03s$-$EIGEN-GRACE02S & $33\%$\\
%
%
%
%
%
\tableline

\end{tabular}

\end{table*}
Their average is about $17\%$.

A further, different approach that could be followed to take into account the scatter among the various solutions consists
in computing mean and standard deviation of the entire set of values of the even zonals for the models considered so far, degree by degree, and
taking the standard deviations as representative of the uncertainties $\delta J_{\ell}, \ell = 4,6,8,\textrm{\dots}$.
It yields $\delta\mu = 15\%$, a figure slightly larger that that by \citet{Ries08}. Anyway,
in evaluating mean and standard deviation for each even zonals, \citet{Ries08} also used global gravity
solutions like EIGEN-GL04C \citep{gl04c} and EIGEN-GL05C \citep{gl05c} which
include data from the LAGEOS satellite itself; this may
likely have introduced a sort of favorable a priori \virg{imprint}
of the Lense-Thirring effect itself. Moreover, \citet{Ries08} gave only a RSS evaluation of the total bias.

It should be recalled that also the further bias due to the cross-coupling between $J_2$ and the orbit inclination, evaluated to be about $9\%$ in \citet{Ior07},
must be added.

\subsubsection{An a-priori, \virg{imprinting} effect?}\lb{impr}
GRACE recovers the spherical harmonic coefficients of the geopotential  from the tracking of both satellites
by GPS  and from the observed intersatellite
distance variations \citep{eigengrace02s}. A potential critical issue is, thus, a possible \virg{memory} effect of the gravitomagnetic force. Its
impact in the satellite-to-satellite tracking was preliminarily addressed
in \citet{crit1}; here we will focus on the  \virg{imprint} coming from the GRACE orbits which is more important for us because it mainly resides
in the low degree even zonals \citep{conet}.
%
%

Concerning that issue, \citet{Ciu05} write that such a kind of leakage of the Lense-Thirring signal itself into the even zonals retrieved by GRACE is
completely negligible because the GRACE satellites move along (almost)  polar orbits. Indeed, for perfectly polar ($I=90$ deg) trajectories, the
gravitomagnetic force is entirely out-of-plane, while the perturbing action of the even zonals is confined to the orbital plane itself.
According to \citet{Ciu05}, the deviations of the orbit of GRACE from the ideal  polar  orbital configuration would have negligible
consequences on the \virg{imprint} issue. In particular, they write: \virg{the values of the even zonal harmonics
determined by the GRACE orbital perturbations are substantially independent
on the a priori value of the Lense–Thirring effect. [...] The small deviation from a polar orbit of the GRACE satellite, that
is $1.7\times 10^{-2}$ rad, gives only rise, $at\ most$, to a very small correlation with
a factor $1.7\times 10^{-2}$}.
The meaning of such a statement is unclear; anyway, we will show below that such a conclusion is incorrect.

The relevant orbital parameters of GRACE are quoted in Table \ref{OSIGNUR}; variations of the orders of about 10 km in the semimajor axis $a$ and 0.001 deg in
the inclination $I$ may occur, but it turns out that they are irrelevant in our \textrm{discussion}
(\url{http://www.csr.utexas.edu/grace/ground/globe.html}).
The orbital plane  of GRACE is, in fact, shifted by 0.98 deg from the ideal polar configuration, and, contrary to \textrm{what is
claimed} in \citet{Ciu05}, this does matter because its classical secular node precessions are far from being negligible with respect to our issue.
The impact of the Earth's gravitomagnetic force on the even zonals retrieved by GRACE can be quantitatively evaluated by computing the \virg{effective}
value
 $\overline{C}^{\rm LT}_{\ell 0}$ of the normalized even zonal gravity coefficients which would induce classical secular node precessions for GRACE as large as those due to its Lense-Thirring effect, which is independent of the inclination $I$.
To be more precise, $\overline{C}^{\rm LT}_{\ell 0}$ \textrm{comes} from solving the following equation which connects the classical even zonal precession
of degree $\ell$ $\dot\Omega_{J_{\ell}}\doteq \dot\Omega_{.\ell}J_{\ell}$ to the Lense-Thirring node precession $\dot\Omega_{\rm LT}$
\eqi \dot\Omega_{.\ell}J_{\ell}=\dot\Omega_{\rm LT}.\eqf
Table \ref{tibulo2}
\begin{table*}[t]
\caption{$\overline{C}_{\ell 0}^{\rm LT}$: effective \virg{gravitomagnetic}  normalized gravity coefficients for GRACE ($\ell=4,6;\ m=0$).
They have been obtained by comparing the GRACE classical node precessions to the Lense-Thirring rate. Thus,
they may be viewed as a quantitative measure of the leakage of the Lense-Thirring effect itself into
the second and third  even zonal harmonics of the global gravity solutions from GRACE.
Compare them with the much smaller calibrated errors $\sigma_{\overline{C}_{\ell 0}}$ in $\overline{C}_{40}$ and $\overline{C}_{60}$ of
the GGM03S model \citep{ggm03}. \label{tibulo2}
}
\begin{tabular}{@{}llll}
\hline
$\overline{C}_{40}^{\rm LT}$ & $\overline{C}_{60}^{\rm LT}$ & $\sigma_{\overline{C}_{40}}$ & $\sigma_{\overline{C}_{60}}$  \\
 \hline
$2.23\times 10^{-10}$ & $-2.3\times 10^{-11}$ & $4\times 10^{-12}$ & $2\times 10^{-12}$ \\
 \hline
\end{tabular}
\end{table*}
lists $\overline{C}^{\rm LT}_{\ell 0}$ for  degrees $\ell=4,6$, which are the most effective in affecting the combination of \rfr{combi}.
Thus, the gravitomagnetic field of the Earth contributes to the value of the second even zonal of the geopotential retrieved from the orbital motions of
GRACE by an amount of the order of $2\times 10^{-10}$, while for $\ell=6$ the imprint is one order of magnitude smaller. Given the present-day level
of accuracy of the latest GRACE-based solutions, which is of the order of $10^{-12}$, effects as large as those of
Table \ref{tibulo2} cannot be neglected. Thus, we conclude that the influence of the Earth's gravitomagnetic field on the low-degree even zonal
harmonics of the global gravity solutions  from GRACE may exist, falling well within the present-day level of measurability.

A further, crucial step consists of evaluating the impact of such an a-priori \virg{imprint} on the test conducted with the LAGEOS satellites and the combination of \rfr{combi}: if the LAGEOS-LAGEOS II uncancelled combined classical geopotential precession computed with the GRACE-based a-priori \virg{imprinted} even zonals of Table \ref{tibulo2}  is a relevant part of \textrm{-- or even larger than --} the combined Lense-Thirring precession, it will be demonstrated that the doubts concerning the a-priori gravitomagnetic \virg{memory} effect  are founded. It turns out that this is just the case because  \rfr{combi} and Table \ref{tibulo2} yield a
combined geopotential precession whose magnitude is 77.8 mas yr$^{-1}$ ($-82.9$ mas yr$^{-1}$ for $\ell=4$ and $5.1$ mas yr$^{-1}$ for $\ell=6$), i.e. just 1.6 times the Lense-Thirring signal itself. This means that the part of the LAGEOS-LAGEOS II uncancelled classical combined node precessions  which is affected by the \virg{imprinting} \textrm{of} the Lense-Thirring force through the GRACE-based geopotential's spherical harmonics  is as large as the LAGEOS-LAGEOS II combined gravitomagnetic signal itself.

We, now, comment on how \citet{Ciu05}  reach a different conclusion. They write: \virg{However, the Lense-Thirring effect depends on the third
power of the inverse of the distance from the central body, i.e., $(1/r)^3$, and the
$J_2, J_4, J_6 \textrm{,\dots}$ effects depend on the powers $(1/r)^{3.5}$, $(1/r)^{5.5}$, $(1/r)^{7.5}$ \textrm{,\dots} of the
distance; then, since the ratio of the semimajor axes of the GRACE satellites
to the LAGEOS' satellites is $\sim\rp{6780}{12270}\cong
1.8$, any conceivable \virg{Lense-Thirring
imprint} on the spherical harmonics at the GRACE altitude becomes quickly,
with increasing distance, a negligible effect, especially for higher harmonics of
degree $l>4$. Therefore, any conceivable \virg{Lense-Thirring imprint} is negligible
at the LAGEOS' satellites altitude.} From such statements it seems that they compare the classical GRACE precessions \textrm{with}
 the gravitomagnetic LAGEOS' ones. This is meaningless since, as we have shown, one has, first, to compare the classical and relativistic precessions of GRACE itself, with which the Earth's gravity field is solved for \textrm{and only after to} compute the impact of the relativistically \virg{imprinted} part of the GRACE-based even zonals on the combined LAGEOS nodes. These two stages have to be kept separate, with the first one which is fundamental; if different satellite(s) Y were to be used to measure the gravitomagnetic field of the Earth, the impact of the Lense-Thirring effect itself on them should be evaluated by using the \virg{imprinted} even zonals evaluated in the first stage.
Finally, in their latest statement \citet{Ciu05} \textrm{write}: \virg{In addition, in \citep{Ciuetal97}, it was proved with several
simulations that by far the largest part of this \virg{imprint} effect is absorbed in
the by far largest coefficient $J_2$.} Also such a statement, in the present context, has no validity since the cited work refers to a pre-GRACE era. Moreover,
no quantitative details at all were explicitly
released concerning the quoted simulations, so that it is  \textrm{impossible to form
an opinion.}

\subsubsection{A new approach to extract the Lense-Thirring signature from the data}\lb{approach}
The technique adopted so far by \citet{Ciu04} and \citet{Ries08} to extract the gravitomagnetic signal from the
LAGEOS and LAGEOS II data  is described in detail in
\citep{LucBal06,Luc07b}. In \textrm{both approaches} the Lense-Thirring force is not included in the dynamical force models
used to fit the satellites' data. In the data reduction process no dedicated gravitomagnetic parameter is estimated, contrary to, e.g., station
coordinates, state vector,
\textrm{satellite} drag and radiation coefficients $C_D$ and $C_R$, respectively, etc.; its effect is retrieved with a sort of post-post-fit analysis in
which the time series of the computed\footnote{The expression ``residuals of the nodes'' is used, strictly speaking, in an improper sense because the
Keplerian orbital elements are not directly measured.} ``residuals'' of the nodes with the difference between the orbital elements of consecutive arcs,
combined with \rfr{combi}, is fitted with a straight line.

In regard to possible other approaches which could be followed, it would be useful to, e.g., estimate
(in the least square sense), among other solve-for parameters, purely phenomenological corrections $\Delta\dot\Omega$ to
the LAGEOS/LAGEOS II node precessions as well, without modelling the Lense-Thirring \textrm{effect} itself, so that it will be, in principle,
contained in $\Delta\dot\Omega$, and combine them according to \rfr{combi}.
Something similar has been done \textrm{-- although for different scopes --} for the perihelia of the inner planets of the solar system
\citep{Pit05} (see Section \ref{sole}) and the periastron of the pulsars \citep{Kra06}.
To be more definite, various solutions with a complete suite of dynamical models, apart from the gravitomagnetic force
itself, should be produced in which one inserts a further solve-for parameter, i.e. a correction  $\Delta\dot\Omega$ to the standard
Newtonian modelled precessions.  One could see how the outcome varies by changing the data sets and/or the parameters to be solved for.
Maybe it could be done for each arc, so to have a collection of such node extra-rates. Such a strategy would be much more model-independent.

As previously suggested by \citet{Nor01}, another way to tackle the problem consists of
looking at a Lense-Thirring-dedicated parameter to be estimated along with all the
zonals in a new global solution for the gravity field  incorporating the gravitomagnetic
component as well; instead, in all the so far produced global gravity solutions no relativistic
parameter(s) have been included in the set of the estimated ones. As shown in Section \ref{impr}, this would also cure the impact
of possible forms of a-priori \virg{imprinting} effects.

A first, tentative step  towards the implementation of the
strategy of the first point mentioned above with the LAGEOS satellites in terms  of the PPN parameter $\gamma$ has been
recently taken by \citet{Poz}.

%

\subsection{The LARES mission}
\citet{vpe76a}  proposed to measure the Lense-Thirring precession of the nodes $\Omega$ of a pair of counter-orbiting spacecraft
to be launched in terrestrial polar orbits and endowed with \textrm{a} drag-free apparatus. A somewhat equivalent, cheaper version of such an idea was
put forth ten years later by \citet{Ciu86,Ciu89} who proposed to launch a passive, geodetic
satellite in an orbit identical to that of LAGEOS  apart from the orbital planes which should have been displaced by 180 deg apart. The measurable
quantity was, in the case of the proposal by \citet{Ciu86}, the sum of the nodes of LAGEOS and of the new
spacecraft, later named LAGEOS III \citep{Ciu94}, LARES \citep{LARES}, \textrm{and} WEBER-SAT\footnote{In memory of Dr J. Weber, US Naval Academy (USNA) Class
of 1940.}, in order to cancel to a high level of accuracy the corrupting
effect of the multipoles of the Newtonian part of the terrestrial gravitational potential which represent the
major source of systematic \textrm{errors}. Although extensively studied by various groups \citep{CSR,LARES}, such an idea was not implemented for many years.
\citet{Ioretal02}  proposed to include also the data from LAGEOS II by using a
different observable.  Such an approach was proven in \citet{IorNA} to be, in principle, potentially useful in making the constraints
on the orbital configuration of the new SLR satellite less stringent than it was originally required in view of the recent improvements in our knowledge of
the classical part of the terrestrial gravitational potential due to the dedicated CHAMP and, especially, GRACE  missions.

Since reaching high altitudes and minimizing the unavoidable orbital injection errors is expensive, \textrm{the possibility of discarding
LAGEOS and LAGEOS II using a low-altitude, nearly polar orbit for LARES \citep{LucPao01,Ciu06b} was explored. However, in}
\citet{Ior02,Ior07c} it was proven that such alternative approaches are not feasible. It was also suggested that LARES
would be able to probe alternative theories of gravity \citep{Ciu04b}, but also in this case it turned out to be impossible \citep{IorJCAP,Ior07d}.

The stalemate came to an end when ASI recently approved the LARES mission,
although with a different orbital geometry with respect to the original configuration: now the orbital
altitude is 1450 km corresponding to a semimajor axis $a=7828$ km
\citep{Ciu10}. See Table \ref{OSIGNUR} for the new orbital parameters and the related Lense-Thirring node precession.
LARES should be launched in late 2010/early 2011 with the first qualification flight of the VEGA rocket (\url{http://www.spacenews.com/civil/100115-asi-expects-budget-remain-flat-2010.html}).

The combination that should be used for measuring the Lense-Thirring effect with LAGEOS, LAGEOS II and LARES is \citep{IorNA}
\eqi \dot\Omega^{\rm LAGEOS}+k_1\dot\Omega^{\rm LAGEOS\ II}+ k_2\dot\Omega^{\rm LARES},
\lb{combaz}
\eqf
\textrm{where the} coefficients $k_1$ and $k_2$ entering \rfr{combaz} are defined as
\begin{equation}
\begin{array}{lll@{}}
k_1 = \rp{\cf 2{LARES}\cf4{LAGEOS}-\cf 2{LAGEOS}\cf 4{LARES}}{\cf 2{LAGEOS\ II}\cf 4{LARES}-\cf 2{LARES}\cf 4{LAGEOS\ II}}= 0.3586,\\\\
k_2 =  \rp{\cf 2{LAGEOS}\cf4{LAGEOS\ II}-\cf 2{LAGEOS\ II}\cf 4{LAGEOS}}{\cf 2{LAGEOS\ II}\cf 4{LARES}-\cf 2{LARES}\cf 4{LAGEOS\ II}}= 0.0751.
\end{array}\lb{cofis}
 \end{equation}
\textrm{By construction, the combination \rfr{combaz} cancels the impact of the first two even zonals; we have used $a_{\rm LR}=7828$ km
 and $I_{\rm LR}=71.5$ deg.}
The total Lense-Thirring effect, according to \rfr{combaz} and \rfr{cofis}, amounts to 50.8 mas yr$^{-1}$.
\subsubsection{A conservative evaluation of the impact of the geopotential on the LARES mission}
The systematic error due to the uncancelled even zonals $J_6, J_8,\textrm{\dots}$ can be conservatively evaluated as
\eqi\delta\mu\leq \sum_{\ell = 6}\left|\dot\Omega^{\rm LAGEOS}_{.\ell}+k_1\dot\Omega^{\rm LAGEOS\ II}_{.\ell}+
k_2\dot\Omega^{\rm LARES}_{.\ell}\right|\delta J_{\ell}\lb{biass}\eqf

Of crucial importance is how to assess $\delta J_{\ell}$.  By proceeding as in Section \ref{gravAZZA} and by using the
same models up to degree $\ell = 60$  because of the lower altitude of LARES with respect to LAGEOS and LAGEOS II which brings into
play more even zonals, \textrm{we come up with} the results presented in Table \ref{tavolq}. They have been obtained with the standard and widely used
Kaula approach \citep{Kau} in the following way. We, first, calibrated our numerical calculation with the analytical ones performed
with the explicit expressions for $\dot\Omega_{.\ell}$ worked out up to $\ell=20$ in \citet{Ior03}; then, after having obtained
identical results, we confidently extended our numerical calculation to higher degrees by means of two different softwares (Matlab and MATHEMATICA).
 \begin{table*}[t]
 \small\caption{ Systematic percent uncertainty $\delta\mu$ in the combined Lense-Thirring effect with LAGEOS, LAGEOS II and
 LARES according to \rfr{biass} and $\delta J_{\ell}= \Delta J_{\ell}$ up to degree $\ell = 60$ for the global Earth's gravity solutions
 considered here; the approach by \protect\citep{Kau} has been followed. For LARES we adopted $a_{\rm LR}=7828$ km, $I_{\rm LR}=71.5$ deg,
 $e_{\rm LR} = 0.0$.}\label{tavolq}
\begin{tabular}{@{}lll@{}}
\tableline
Models compared ($\delta J_{\ell}=\Delta J_{\ell}$) & $\delta\mu$ (SAV) & $\delta\mu$ (RSS)\\
\tableline
ITG-Grace2010s$-$JEM01-RL03B  & $4\%$ & $2\%$\\
ITG-Grace2010s$-$GGM02S  & $14\%$ & $8\%$\\
ITG-Grace2010s$-$GGM03S  & $2\%$ & $1\%$\\
ITG-Grace2010s$-$ITG-Grace02  & $2\%$ & $1\%$\\
ITG-Grace2010s$-$ITG-Grace03  & $2\%$  & $1\%$\\
ITG-Grace2010s$-$AIUB-GRACE02S  & $5\%$  & $2\%$\\
ITG-Grace2010s$-$AIUB-GRACE01S  & $22\%$  & $14\%$\\
ITG-Grace2010s$-$EIGEN-GRACE02S  & $24\%$  & $13\%$\\
AIUB-GRACE02S$-$JEM01-RL03B & $9\%$ & $4\%$\\
%
%
AIUB-GRACE02S$-$GGM02S & $16\%$ & $7\%$\\
AIUB-GRACE02S$-$GGM03S & $4\%$ & $2\%$\\
AIUB-GRACE02S$-$ITG-Grace02 & $5\%$ & $2\%$\\
AIUB-GRACE02S$-$ITG-Grace03 & $6\%$ & $3\%$\\
%
%
AIUB-GRACE02S$-$EIGEN-GRACE02S & $24\%$ & $13\%$\\
AIUB-GRACE02S$-$AIUB-GRACE01S & $23\%$ & $13\%$\\
AIUB-GRACE01S$-$JEM01-RL03B & $23\%$ & $16\%$\\
%
%
AIUB-GRACE01S$-$GGM02S & $16\%$ & $8\%$\\
AIUB-GRACE01S$-$GGM03S & $22\%$ & $13\%$\\
AIUB-GRACE01S$-$ITG-Grace02 & $24\%$ & $15\%$\\
AIUB-GRACE01S$-$ITG-Grace03 & $22\%$ & $14\%$\\
%
%
AIUB-GRACE01S$-$EIGEN-GRACE02S & $14\%$ & $7\%$\\
%
%
%
%
%
%
%
%
%
JEM01-RL03B$-$GGM02S & $14\%$ & $9\%$  \\
JEM01-RL03B$-$GGM03S & $5\%$ & $3\%$  \\
JEM01-RL03B$-$ITG-Grace02 & $4\%$ & $2\%$  \\
JEM01-RL03B$-$ITG-Grace03s & $5\%$ & $2\%$  \\
%
%
JEM01-RL03B$-$EIGEN-GRACE02S & $26\%$ & $15\%$  \\
%
%
GGM02S$-$GGM03S & $13\%$ & $7\%$  \\
GGM02S$-$ITG-Grace02& $16\%$ & $8\%$  \\
GGM02S$-$ITG-Grace03s& $14\%$ & $7\%$  \\
%
%
GGM02S$-$EIGEN-GRACE02S & $14\%$ & $7\%$  \\
%
%
GGM03S$-$ITG-Grace02 & $3\%$ & $2\%$  \\
GGM03S$-$ITG-Grace03s & $2\%$ & $0.5\%$  \\
%
%
GGM03S$-$EIGEN-GRACE02S & $24\%$ & $13\%$  \\
%
%
ITG-Grace02$-$ITG-Grace03s & $3\%$ & $2\%$  \\
%
%
ITG-Grace02$-$EIGEN-GRACE02S & $25\%$ & $14\%$  \\
%
%
%
ITG-Grace03s$-$EIGEN-GRACE02S & $24\%$ & $13\%$  \\
%
%
%
%
%
\tableline
\end{tabular}
\end{table*}

It must be stressed that our results may be still optimistic: indeed, computations for $\ell > 60$ become unreliable because of numerical instability of the results.

In Table \ref{tavolx} we repeat the calculation by using for $\delta J_{\ell}$ the covariance matrix sigmas $\sigma_{J_{\ell}}$; also in this case we use the approach by \citet{Kau} up to degree $\ell = 60$.
 \begin{table*}[t]
 \small\caption{ Systematic percent uncertainty $\delta\mu$ in the combined Lense-Thirring effect with LAGEOS, LAGEOS II and
 LARES according to \rfr{biass} and $\delta J_{\ell}= \sigma_{J_{\ell}}$ up to degree $\ell = 60$ for the global Earth's gravity solutions
 considered here; the approach by
 \protect\citep{Kau} has been followed. For LARES we adopted $a_{\rm LR}=7828$ km, $I_{\rm LR}=71.5$ deg, $e_{\rm LR} = 0.0$.}\label{tavolx}
\begin{tabular}{@{}lll@{}}
\tableline
Model ($\delta J_{\ell}=\sigma_{\ell}$) & $\delta\mu$ (SAV) & $\delta\mu$ (RSS)\\
\tableline
ITG-Grace2010s (formal) & $0.2\%$ & $0.1\%$\\
AIUB-GRACE02S (formal) & $1\%$ &  $0.9\%$\\
AIUB-GRACE01S (formal) & $11\%$ & $9\%$\\
%
 %
JEM01-RL03B (formal) & $1\%$ & $0.9\%$\\
GGM03S (calibrated) & $5\%$ & $4\%$\\
GGM02S (formal) & $20\%$ & $15\%$\\
ITG-Grace03s (formal) & $0.3\%$ & $0.2\%$\\
ITG-Grace02s (formal) & $0.4\%$ & $0.2\%$\\
%
%
EIGEN-GRACE02S (calibrated) & $21\%$ & $17\%$ \\
 \tableline
\end{tabular}
\end{table*}

If, instead, one assumes $\delta J_{\ell}=\sigma_{\ell},\ \ell=2,4,6,\textrm{\dots}$ i.e., the standard deviations of
the sets of all the best estimates of $J_{\ell}$ for the models considered here the systematic bias, up to $\ell=60$,
amounts to $12\%$ (SAV) and $6\%$ (RSS). Again, also this result may turn out to be optimistic for the same reasons as before.


It must be pointed out that the evaluations presented here rely upon calculations of the coefficients $\dot\Omega_{.\ell}$ performed with the well
known standard approach by \citet{Kau}; it would be important to try to follow also different computational strategies in order to test them.
In regard to this point, \citet{Ciu10} state that, in reality, the bias due to the even zonals is of the order of $1\%$ or less,
and repeatedly write that the results of one of us \textrm{(L. I.) are based on
\virg{miscalculations}.}
In fact, \citet{Ciu10}, do not demonstrate
their allegations by explicitly disclosing the
alleged error(s).
\subsubsection{The impact of some non-gravitational perturbations}
It is worthwhile noting that also the impact of the subtle non-gravitational perturbations will be different with respect to the
original proposal because LARES will fly in a  lower orbit and its thermal behavior will  probably be different with respect to
LAGEOS and LAGEOS II.  The reduction of the impact of the thermal accelerations, like the Yarkovsky-Schach effects, should have  been reached
with two concentric spheres. However, as explained by \citet{Andres}, this solution will increase the floating potential of LARES because of the much higher electrical resistivity and, thus, the perturbative effects produced by the charged particle drag. Moreover, the atmospheric drag will increase also because of the lower orbit of the satellite, both in its neutral and charged components.
Indeed, although it does not affect directly the node $\Omega$, it induces a secular decrease of the inclination $I$ of a LAGEOS-like satellite \citep{Mil87} which translates into a further bias for the node itself according to
\eqi\delta\dot\Omega_{\rm drag}=\rp{3}{2}n\left(\rp{R}{a}\right)^2 \rp{\sin I\ J_2}{(1-e^2)^2}\delta I,\eqf in which $\delta I$
accounts not only for the measurement errors in the inclination, but also for any unmodelled/mismodelled dynamical effect on it. According to
\citep{acppb}, the secular decrease for LARES would amount to \eqi \left\langle\dert I t\right\rangle_{\rm LR}\approx -0.6\ {\rm mas}\ {\rm yr}^{-1}\eqf yielding a systematic uncertainty in the Lense-Thirring signal of \rfr{combaz} of about $3-9\%$ yr$^{-1}$. An analogous indirect node effect via the inclination could be induced by the thermal Yarkovski-Rubincam force as well \citep{acppb}.
Also the Earth's albedo, with its anisotropic components, may have a non-negligible  effect.

\citet{Ciu10} objected that, in fact, the disturbing effect examined would not appear in the real data
analysis procedure because the inclination along with all the other Keplerian orbital elements would be measured arc by arc, so that
one should only have to correct the signal for the measured value of the inclination; after all, the same problems, if not even larger, would
occur with the semimajor axes of the LAGEOS satellites, which are known to undergo still unexplained secular decrease of 1.1 mm d$^{-1}$ \citep{Ruby}
and their consequent mappings onto the node rates. The problem is that while a perturbation $\Delta a$ pertains the in-plane, radial
component \citep{Cri} of the \textrm{LAGEOS} orbits, both the Lense-Thirring node precession and the shifts in the inclination affect the out-of-plane,
normal component \textrm{of the orbit \citep{Cri}}; thus, even if repeated corrections to the semimajor axis could be applied without affecting
the gravitomagnetic signal of interest, the same would not hold for the inclination. This is particularly true in
view of the fact that, for still unexplained reasons, the Lense-Thirring effect itself has never been estimated, either as a short-arc or as a global parameter.
Moreover, it has been claimed that the recent improvements in atmospheric refraction modelling would allow to measure the inclination of the
\textrm{LAGEOS} satellites at
a level of accuracy, on average, of $30$ $\mu$as for LAGEOS  and $10$ $\mu$as for LAGEOS II \citep{Ciu10}. Firstly,  the
tracking of a relatively low satellite is always more difficult than for higher targets, so that
caution would be needed in straightforwardly extrapolating results valid for LAGEOS to the still non-existing LARES. Second,
it is difficult to understand the exact
sense of such claims because they would imply an accuracy $\delta r\approx a\delta I$ in reconstructing the orbits of LAGEOS and LAGEOS II, on average, of
$0.2$ cm and $0.06$ cm, respectively.

Let us point out the following issue as well. At present, it is not yet clear how the data of LAGEOS, LAGEOS II and LARES will be finally used by the proponent team \textrm{in order to detect} the Lense-Thirring effect. This could turn out to be a non-trivial matter because of the non-gravitational perturbations. Indeed, if, for instance, a combination\footnote{The impact of the geopotential is, by construction, unaffected with respect to the combination of \rfr{combaz}.}
\eqi\dot\Omega^{\rm LARES}+h_1\dot\Omega^{\rm LAGEOS}+ h_2\dot\Omega^{\rm LAGEOS\ II}\lb{altr}\eqf was adopted instead of that of \rfr{combaz}, the coefficients of the nodes of LAGEOS and LAGEOS II, in view of the lower altitude of LARES, would be
 \begin{equation}
\begin{array}{lll}
h_1 = \rp{\cf 2{LAGEOS\ II}\cf4{LARES}-\cf 2{LARES}\cf 4{LAGEOS\ II}}{\cf 2{LARES}\cf 4{LAGEOS\ II}-\cf 2{LAGEOS\ II}\cf 4{LAGEOS}}= 13.3215,\\\\
h_2 =  \rp{\cf 2{LARES}\cf4{LAGEOS}-\cf 2{LAGEOS}\cf 4{LARES}}{\cf 2{LAGEOS}\cf 4{LAGEOS\ II}-\cf 2{LAGEOS\ II}\cf 4{LAGEOS}}= 4.7744.
\end{array}\lb{cofcaz}
 \end{equation}
and the combined Lense-Thirring signal would amount to 676.8 mas yr$^{-1}$.
As a consequence, the direct and indirect effects of the non-gravitational\footnote{The same may hold also for time-dependent gravitational perturbations affecting the nodes of LAGEOS and LAGEOS II, like the tides.} perturbations on the nodes of LAGEOS and LAGEOS II would be enhanced by such larger coefficients and this may yield a lower total obtainable accuracy.
\section{In search of the Sun's gravitomagnetic field}\lb{sole}
\subsection{General considerations}
Recent determinations of the Sun's proper angular momentum
\eqi S_{\odot}=(190.0\pm 1.5)\times 10^{39}\ {\rm kg\ m}^2\ {\rm s}^{-1}
\eqf
from helioseismology \citep{Pij1,Pij2}, accurate to $0.8\%$, yield a value about one order of magnitude smaller than that  obtained by assuming
a homogeneous and uniformly rotating  Sun, as done in the \textrm{pioneering} work by \citet{DeS}, and also by \citet{Sof} and \citet{Cug78} who concluded
that, at their time, it was not possible to measure the solar Lense-Thirring effect.
\textrm{Despite the reduced magnitude of the solar gravitomagnetic field with respect to the earlier predictions, the present and
near future situation seems more promising.}
\textrm{The characteristic length with which the accuracy of the determination of the orbits of the particles should be compared}
%
is \eqi l^{\odot}_g \doteq \rp{S_{\odot}}{M_{\odot}c} =319\ {\rm m};\eqf
in the case of gravitoelectric effects, $l_g$ is usually replaced by the Schwarzschild radius $R_g\doteq 2GM/c^2=3$ km for the Sun.
The present-day accuracy in knowing, e.g., the inner planets' mean \textrm{orbital} radius  \eqi\left\langle r\right\rangle=a\left(1+\rp{e^2}{2}\right),\eqf is shown in
Table \ref{chebol}.
\begin{table*}[t]
\small
\caption{First line: uncertainties \textrm{(in m)} in the average heliocentric distances of the inner planets obtained
by propagating the formal errors in $a$ and $e$ according to Table 3 of \protect\citet{Pit08}; the EPM2006 ephemerides
were used by \protect\citet{Pit08}. Second line: maximum differences \textrm{(in m)} between the EPM2006 and the DE414
\protect\citep{DE414} ephemerides for the inner planets in the time interval 1960-2020 according to Table 5 of
\protect\citet{PROO}.
Third line: maximum differences \textrm{(in m)} between the EPM2008 \protect\citep{PROO} and the DE421
\protect\citep{DE421} ephemerides for the inner planets in the time interval 1950-2050 according to Table 5 of
\protect\citet{PROO}.
They have to be compared \textrm{with} the characteristic gravitomagnetic length of the Sun $l_g^{\odot}=319$ m.\label{chebol}
}
\begin{tabular}{@{}lllll@{}}
\tableline
Type of orbit uncertainty & {Mercury} & {Venus} & {Earth} & {Mars}  \\
 \tableline
$\delta\left\langle r \right\rangle$ (EPM2006)  & 38 & 3 & 1 & 2 \\
EPM2006$-$DE414 & 256 & 131 & 17.2 & 78.7 \\
EPM2008$-$DE421 & 185 & 4.6 & 11.9 & 233\\
 \tableline
\end{tabular}
\end{table*}
Such values have been obtained by linearly propagating the formal, statistical errors in $a$ and $e$ according to Table 3 of \citet{Pit08}; even by re-scaling them by a a factor of, say, $2-5$, the gravitomagnetic effects due to the Sun's rotation fall, in principle, within the measurability domain.
Another possible way to evaluate the present-day uncertainty in the planetary orbital motions consists of looking at different ephemerides of comparable accuracy.
In Table \ref{chebol} we do that for the EPM2006/EPM2008 \citep{Pit08,PROO}, and the DE414/DE421 \citep{DE414,DE421} ephemerides; although, larger than $\delta\left\langle r\right\rangle$,
the maximum differences between such ephemerides are smaller than the solar gravitomagnetic length $l_g^{\odot}$.

\textrm{With regard to the currently collected ranging data to the Venus Express spacecraft\footnote{See on the WEB \url{http://www.esa.int/esaMI/Venus_Express/}.}}
and in view of the ongoing Messenger \citep{Balo} and the future BepiColombo\footnote{It is an ESA mission, including
two spacecraft, one of which provided by Japan, to be put into orbit around Mercury. The launch is scheduled for 2014. The construction of the
instruments is currently ongoing.}  \citep{Milani0,Balo} missions to Mercury and of the developments
in the Planetary Laser Ranging (PLR) technique,
\textrm{the planetary orbit accuracy is likely to be further improved.}
More specifically,
recent years have seen increasing efforts towards the implementation of PLR accurate to cm-level \citep{ILR1,ILR2,ILR,Degn,ILR3,PLR,MOLA,recent}. It would allow to reach major improvements in three related fields: solar system dynamics, tests of general relativity and alternative theories of gravity, and physical properties of the target planet itself. In principle, any solar system body endowed  with a solid surface and a transparent atmosphere would be a suitable platform for a PLR system, but some targets are more accessible than others. Major efforts have been practically devoted so far to Mercury \citep{ILR1} and Mars \citep{ILR2,ILR3}, although simulations reaching 93 \textrm{AU} or more have been undertaken as well \citep{Degn,MOLA}.
In 2005 two interplanetary laser transponder experiments were successfully
demonstrated by the Goddard Geophysical Astronomical Observatory (GGAO). The first utilized the non-optimized Mercury Laser Altimeter
(MLA) on the Messenger spacecraft \citep{ILR1,ILR}, obtaining a formal error in the laser range solution of 0.2 m, or one part in $10^{11}$. The second utilized the Mars Orbiting Laser
Altimeter (MOLA) on the Mars Global Surveyor spacecraft \citep{Abshire,ILR}. A precise \textrm{value} of the Earth-Mars distance, measured between their centers of mass and taken over an extended period (five years or more), would support, among other things, a better determination of several parameters of the solar system. Sensitivity analyses point towards measurement uncertainties between 1 mm and 100 mm \citep{ILR2}.
\textrm{A recent analysis of the future BepiColombo mission to Mercury,}
aimed to accurately determining, among other things, several key parameters of post-Newtonian gravity and the solar quadrupole moment from Earth-Mercury distance data collected with  a multi-frequency radio link \citep{Milani0,Milani}, points toward a maximum uncertainty of $4.5-10$ cm in determining the Earth-Mercury range over a multi-year time span (1-8 yr) \citep{Milani0,Ashby,Milani}. A proposed spacecraft-based mission aimed to accurately measure also  the gravitomagnetic field of the Sun and its  $J_2$  along with other PPN parameters like $\gamma$ and $\beta$ by means of interplanetary ranging is the Astrodynamical Space Test of
Relativity using Optical Devices\footnote{Its cheaper version ASTROD I makes use of one
spacecraft in a Venus-gravity-assisted solar orbit, ranging optically with ground
stations \textrm{\citep{astrod1}}.} (ASTROD) \citep{astrod}.
Another space-based mission proposed to accurately test several aspects of  the gravitational interaction via interplanetary laser ranging is the Laser Astrometric Test of Relativity (LATOR) \citep{lator}.

It is remarkable to note that the currently available estimate of
$S_{\odot}$	from helioseismology is accurate enough to allow, in principle, a genuine Lense-Thirring test. Moreover, it was determined in a relativity-free
fashion from astrophysical techniques which do not rely on
the dynamics of planets in the gravitational field of the
Sun. Thus, there is no any a priori \virg{memory} effect of general relativity itself  in the adopted value of $S_{\odot}$.

\subsection{The Lense-Thirring perihelion precessions of the planets and their measurability}
The action of the solar gravitomagnetic field \textrm{on  Mercury's} longitude of perihelion\footnote{Since $\Omega$ and $\omega$ do not lie, in general, in the same plane, $\varpi$ is a \virg{dogleg} angle.}
$\varpi$ was calculated for the first
time by \citet{DeS} who, by assuming a homogenous and
uniformly rotating Sun, found a secular rate of $-0.01$
arcseconds per century (arcsec cy$^{-1}$ in the following). This value is
also quoted by \citet{Sof}; \citet{Cug78}
yield $-0.02$ arcsec cy$^{-1}$  for the argument of perihelion $\omega$ of Mercury.

The recent estimate of the Sun's angular momentum \citep{Pij1,Pij2} from helioseismology \textrm{implies} a
precessional effect \textrm{for Mercury which is one order of magnitude smaller}; see Table \ref{chebolas} for the predicted
Lense-Thirring precessions $\dot\varpi_{\rm LT}$ of the longitudes of the perihelia of the inner
planets \textrm{ which are} of the order of $10^{-3}-10^{-5}$ arcsec cy$^{-1}$.
\textrm{They are obtained by taking} into account the fact that the
inclinations $I$ of the planets usually quoted in \textrm{the} literature refer to the mean ecliptic
at a given epoch\footnote{It is J2000 (JD 2451545.0).}  \citep{Roy}, while the Sun's equator is tilted by $\epsilon_{\odot} = 7.15$ deg to
the mean ecliptic at J2000 \citep{Carri}.
\begin{table*}[t]
\small
\caption{First line: corrections ${\Delta\dot\varpi}$, in ${ 10^{-4} }$ {arcsec cy}${^{-1}}$ (1 arcsec cy$^{-1}=10$ mas yr$^{-1}=1.5\times 10^{-15}$ s$^{-1}$),
to the standard Newton/Einstein secular precessions of the perihelia
of the {inner} planets estimated by \protect\citet{Pit05} with the {EPM2004} ephemerides. The result for Venus (E.V. Pitjeva, private communication, 2008) has
been obtained by recently processing radiometric
data from Magellan spacecraft with the EPM2006 ephemerides \protect\citep{Pit08}; it has also been  cited in Table 4 of \citet{IAUFIENGA} and, as far as the uncertainty is concerned, in Table 1 of \citet{indiani}. The errors in {square brackets} are
the $1-\sigma$ { formal}, { statistical} ones. Second line:
predicted Lense-Thirring perihelion precessions ${\dot\varpi_{\rm LT}}$, in ${ 10^{-4} }$ {arcsec cy}${^{-1}}$
\protect\citep{IorS1}.
Third line: nominal values of the perihelion precessions due to the Sun's oblateness for $J_2^{\odot}=2\times 10^{-7}$
\protect\citep{obloK,Pit05}; the current level of uncertainty in it is about $10\%$ \protect\citep{oblo2}.
Fourth line: nominal values of the perihelion precessions due to the ring of the minor asteroids for $m_{\rm ring}=5\times 10^{-10}\ M_{\odot}$
\protect\citep{ring}; the uncertainty in it amounts to $\delta m_{\rm ring}=1\times 10^{-10}\ M_{\odot}$ \protect\citep{ring}. Anyway, more recent estimates yield a smaller uncertainty: it is of the order of $\delta m_{\rm ring}=0.1-0.3\times 10^{-10}\ M_{\odot}$ \protect\citep{INPO}.
Fifth line: nominal values of the perihelion precessions due to a massive ring modelling the action of the Classical Kuiper Belt Objects
(CKBOs) for $m_{\rm CKBOs}=0.052\ m_{\oplus}=1.562\times 10^{-7}M_{\odot}$ \protect\citep{ckbo}. Actually, they may be smaller since direct estimates of the mass of the TNOs, modelled as a ring in the EPM2008 ephemerides, place an upper limit of $5.26\times 10^{-8}M_{\odot}$ \protect\citep{PROO}.\label{chebolas}
}
\begin{tabular}{@{}lllll@{}}
\tableline
& {Mercury} & {Venus} & {Earth} & {Mars}  \\
 \tableline
$\Delta\dot\varpi$ & $-36\pm 50\ [42]$ & $-4\pm 5\ [1]$ & $-2\pm 4\ [1]$ & $1\pm 5\ [1]$\\
\tableline
$\dot\varpi_{\rm LT}$& $-20$ & $-3$  & $-1$ & $-0.3$\\
$\dot\varpi_{J_2^{\odot}}$ & $+254$ & $+26$ & $+8$ & $+2$\\
$\dot\varpi_{\rm  ring}$ & $+3$ & $+7$ & $+11$ & $+24$\\
$\dot\varpi_{\rm CKBOs}$ & $+0.2$ & $+0.6$ & $+1$ & $+2$\\
\tableline
\end{tabular}
\end{table*}
So far, the solar Lense-Thirring effect on the orbits of the inner planets
was believed to be too small to be detected \citep{Sof}.
However, the situation \textrm{seems} now favorably changing. \citet{IorAA} preliminarily investigated the possibility of
measuring such tiny effects in view of recent important
developments in the planetary ephemerides generation.

First attempts to measure the Sun's Lense-Thirring effect have recently been
implemented by \citet{IorS1,LTsun} with the Ephemerides of Planets and the
Moon \textrm{(EPM2004)} \textrm{produced by \textrm{\citet{PitSS}}}. They are based on a data set
of more than 317,000 observations (1913$-$2003) including
radiometric measurements of planets and spacecraft,
astrometric CCD observations of the outer planets and
their satellites, and meridian and photographic observations.
Such ephemerides were constructed by the simultaneous
numerical integration of the equations of motion for
all planets, the Sun, the Moon,
\textrm{and the 301 largest asteroids. The
rotations of the Earth and the Moon, the perturbations from the solar quadrupolar mass moment
$J_2^{\odot}$ and that of the asteroid ring that lies in the ecliptic plane and which
consists of the remaining smaller asteroids are included. With}
regard to the
post-Newtonian dynamics, only the gravitoelectric, Schwarzschild-like terms of order  $\mathcal{O}(c^{-2})$,
in the harmonic gauge \citep{NewH}, were included; the gravitomagnetic force of the Sun was not modelled.

The EPM2004 ephemerides were used by \citet{Pit05} to phenomenologically
estimate corrections $\Delta\dot\varpi$ to the known standard Newtonian/Einsteinian secular
precessions of the longitudes of perihelia of the inner planets
as fitted parameters of a particular solution.
\textrm{Table \ref{chebolas} displays these corrections, which
are obtained by comparing computer simulations based on the constructed ephemerides
with actual observations}. In determining such
extra-precessions the PPN parameters  $\gamma$ and $\beta$
and the solar even zonal harmonic coefficient $J_2^{\odot}$ were not
fitted; they were held fixed to their general relativistic and Newtonian values, i.e.
$\gamma=\beta=1$, $J_2^{\odot} = 2\times 10^{-7}$. In the EPM2006 ephemerides \citep{Pit08}, used to accurately estimate the perihelion
precession of Venus as well, were constructed by also modelling the actions of Eris and of the other 20 largest Trans-Neptunian Objects (TNOs);
the database was enlarged by including, among other things, ranging data to \textrm{the} Magellan and Cassini spacecraft.

Although the original purpose\footnote{The goal of \citet{Pit05} was to make a test of the quality of the
previously obtained general solution in which certain values of $\beta,\gamma,J_2^{\odot}$
were obtained. If the construction of the ephemerides was satisfactory,
very small `residual' effects due to such parameters should have been
found. She writes: \virg{At present, as a test, we can determine [...] the
corrections to the motions of the planetary perihelia, which allows us to
judge whether the values of $\beta,\gamma$ and $J_2^{\odot}$ used to construct the ephemerides
are valid.}. The smallness of the extra-perihelion precessions found in her
particular test-solution is interpreted by Pitjeva as follows: \virg{Table 3 [of \citet{Pit05}] shows
that the parameters $\gamma=\beta=1,J_2^{\odot}=2\times 10^{-7}$ used to construct the
EPM2004 ephemerides are in excellent agreement with the observations.}} of the estimation of the
corrections $\Delta\dot\varpi$ was not the measurement of the Lense-Thirring effect, the
results of Table \ref{chebolas} can be used to take first
steps towards an observational corroboration of the existence
of the solar gravitomagnetic force.

From Table \ref{chebolas} it turns out that the  magnitude of the Lense-Thirring perihelion precessions of the inner planets
just lies at the edge of the accuracy in determining  $\Delta\dot\varpi$. All the predicted Lense-Thirring precessions are compatible
with the estimated corrections $\Delta\dot\varpi$.
It \textrm{should} be noted that for Venus the $1-\sigma$ {formal}, statistical error in $\Delta\dot\varpi$ is smaller than the gravitomagnetic effect.

It must be considered that the inclusion of new data sets, modeling of further dynamical effects and new data processing techniques may have an impact on the estimation of the corrections $\Delta\dot\varpi$.
\begin{table*}[t]
\small
\caption{ Corrections ${\Delta\dot\varpi}$, in ${ 10^{-4} }$ {arcsec cy}${^{-1}}$ (1 arcsec cy$^{-1}=10$ mas yr$^{-1}=1.5\times 10^{-15}$ s$^{-1}$),
to the standard Newton/Einstein secular precessions of the perihelia
of the {inner} planets estimated by \protect\citet{PROO} with the {EPM2008} ephemerides including also radiometric data from Venus Express, Cassini and Magellan for Venus. The uncertainties are realistic. Source: Table 8 of \protect\citet{PROO}. Cfr. with Table \ref{chebolas}. \label{chebolas2}
}
\begin{tabular}{@{}lllll@{}}
\tableline
& {Mercury} & {Venus} & {Earth} & {Mars}  \\
 \tableline
$\Delta\dot\varpi$ & $-40\pm 50$ & $240\pm 330$ & $60\pm 70$ & $-70\pm 70$\\
\tableline
\end{tabular}
\end{table*}
Thus, we are still in a preliminary, ongoing phase. As an example, in Table \ref{chebolas2}, retrieved from Table 8 of \citet{PROO}, we show new results for $\Delta\dot\varpi$ of the inner planets.

Concerning the systematic alias due to the various competing dynamical effects listed in Table \ref{chebolas}, the present-day level of
\textrm{their mismodelling} would make them not particularly insidious (at $\approx 10\%$ level of accuracy or even less, taking into account the recent improvements in constraining the masses of the rings of the minor asteroids and of the TNOs), at least for Mercury and, especially, Venus,
with the exception of the impact of the Sun's oblateness on Mercury. \textrm{The latter could be removed anyway} by suitably designing a linear combination
of the perihelia of Mercury and Venus, as done for the LAGEOS satellites.
Moreover,
\textrm{an improvement of the determination of $J_2^{\odot}$ by one order of magnitude with respect to the present-day level
of uncertainty is one} of the goals of the BepiColombo mission \citep{Milani0,Ashby,Milani}. At present, it is known with an uncertainty of about $10\%$ \citep{INPO}.

In principle, it would be possible to use also the nodes \citep{IorAA}, if only the
corrections $\Delta\dot\Omega$ to their standard precessions were available.
\textrm{Once other teams of astronomers have independently estimated their own corrections to the standard perihelion
precessions (and, hopefully, to the nodal precessions as well)}
with different ephemerides, it will be possible to fruitfully repeat the present test.
\subsection{The interplanetary ranges}
An alternative approach to the perihelion precessions consists of looking at the Earth-planet ranges $|\bds \rho|$, which are, among other things, directly observable quantities\footnote{Recall that the perihelia, like all the other Keplerian orbital elements,  cannot be directly measured.}. We numerically \textrm{investigated the} solar Lense-Thirring effect on such an observable for the inner planets \citep{Iorioranges} by taking the difference between the perturbed range $(|\vec{\rho}_{\rm P}|)$ and the unperturbed, reference range $(|\vec{\rho}_{\rm R}|)$.
The temporal interval of the numerical integrations has been taken equal to $\Delta t = 2$ yr because most of the present-day
available time series of range residuals from several spacecraft approximately cover similar temporal
extensions; moreover, also the typical operational time spans forecasted for
future PLR technique are similar.
\subsubsection{Earth-Mercury}
At present, the 1-way range residuals of Mercury from radar-ranging span 30 yr (1967-1997) and are at a few km-level (Figure B-2 of \citet{DE421}); the same holds for the 1-way Mercury radar closure residuals covering  8 yr (1989-1997, Figure B-3 a) of \cite{DE421}). There are also a pair of Mariner 10 range residuals in the 70s at Mercury at 0.2 km level (Figure B-3 b) of \citet{DE421}). Ranging to BepiColombo should be accurate to $4.5-10$ cm \citep{Ashby,Milani} over a few years.

Figure \ref{EMB_Mercury_LT} depicts the range perturbation due to the Sun's Lense-Thirring effect, neither considered so far  in the dynamical force models of the planetary ephemerides nor in the BepiColombo analyses so far performed.
\begin{figure}[t]
\includegraphics[width=\columnwidth]{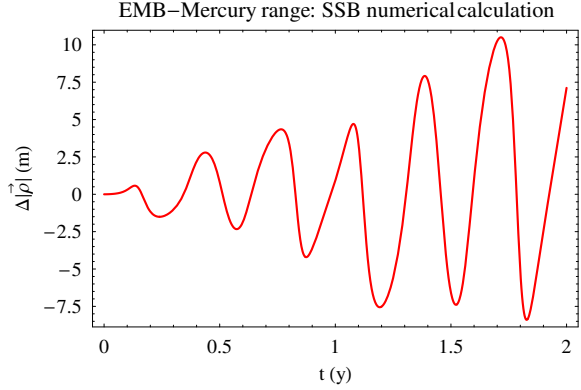}
 \caption{Difference $\Delta |\vec{\rho}|\doteq |\vec{\rho}_{\rm P}|-|\vec{\rho}_{\rm R}|$ in the numerically integrated EMB-Mercury ranges with $(\vec{\rho}_{\rm P})$ and without $(\vec{\rho}_{\rm R})$ the perturbation due to the Sun's Lense-Thirring field over $\Delta t=2$ yr. The same initial conditions (J2000) have been used for \textrm{both integrations}. The state vectors at the reference epoch have been retrieved from the NASA JPL Horizons system. The integrations have been performed in the  ICRF/J2000.0 reference frame, with the mean equinox of the reference epoch and the reference $\{xy\}$ plane rotated from the mean ecliptic of the epoch to the Sun's equator, centered at the Solar System Barycenter (SSB). }\lb{EMB_Mercury_LT}
\end{figure}
The peak-to-peak amplitude of the Lense-Thirring signal  is up to $17.5$ m over 2 yr, which, if on the one hand is unmeasurable from currently available radar-ranging to Mercury, on the other hand corresponds to a potential relative accuracy in measuring it with BepiColombo of $2-5\times 10^{-3}$; this clearly
shows that the solar gravitomagnetic  field should be taken into account in future analyses and data processing. Otherwise, it would alias the recovery of other effects.

Figure \ref{EMB_Mercury_J2} shows the nominal signature of the Sun's quadrupolar mass moment  on the Mercury range for $J_2^{\odot}=2\times 10^{-7}$.
Its action has been modeled as \citep{VRBIK}
\eqi
\bds A_{J_2^{\odot}}=-\rp{3J_2^{\odot} R_{\odot}^2 GM_{\odot}}{2r^4}\left[{\hat r}-5\left({\hat r}{\boldmath\cdot} {\hat k}\right)^2+2\left({\hat r}{\boldmath\cdot}{\hat k}\right){\hat k}\right], \lb{U2}
\eqf
where $R_{\odot}$ is the Sun's mean equatorial radius and ${\hat k}$ is the unit vector of the $z$ axis directed along the body's rotation axis.
\begin{figure}[t]
\includegraphics[width=\columnwidth]{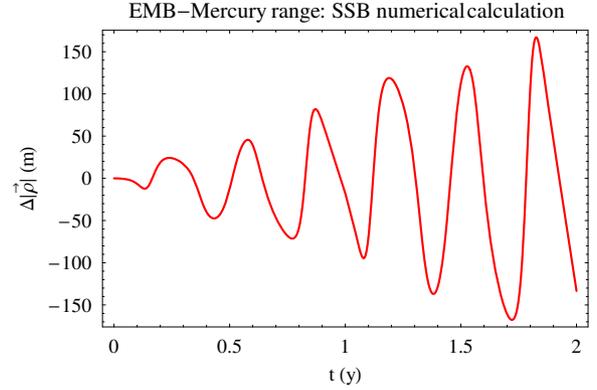}
 \caption{Difference $\Delta |\vec{\rho}|\doteq |\vec{\rho}_{\rm P}|-|\vec{\rho}_{\rm R}|$ in the numerically integrated EMB-Mercury ranges with $(\vec{\rho}_{\rm P})$ and without $(\vec{\rho}_{\rm R})$ the nominal perturbation due to the Sun's quadrupole mass moment $J_2^{\odot}=2.0\times 10^{-7}$ over $\Delta t=2$ yr. The same initial conditions (J2000) have been used for both  integrations. The state vectors at the reference epoch have been retrieved from the NASA JPL Horizons system. The integrations have been performed in the  ICRF/J2000.0 reference frame, with the mean equinox of the reference epoch and the reference $\{xy\}$ plane rotated from the mean ecliptic of the epoch to the Sun's equator, centered at the Solar System Barycenter (SSB). }\lb{EMB_Mercury_J2}
\end{figure}
 The signal \textrm{in} Figure \ref{EMB_Mercury_J2} has a maximum span of 300 m, corresponding to an accuracy measurement of $3\times 10^{-4}$. A measure of the solar $J_2^{\odot}$ accurate to  $10^{-2}$  is one of the goals of BepiColombo \citep{Milani0}; knowing precisely $J_2^{\odot}$ would yield important insights on the internal rotation of the Sun.

Concerning the impact of neglecting the gravitomagnetic field of the Sun in future data analyses, it may affect the determination of $J_2^{\odot}$ at \textrm{the} $12\%$ level. On the other hand, in order to allow for a determination of the Lense-Thirring effect,
the Sun's quadrupole mass moment should be known with an accuracy better than \textrm{at present} by at least one order of magnitude; this is just one of the goals of BepiColombo. Anyway, also the time signatures of the two signals would play a role.
\subsubsection{Earth-Venus}
\textrm{Although at present, contrary to Mercury and Mars, no tests of interplanetary ranging to Venus have been
practically performed, we include this case} not only for completeness but also because simulations of interplanetary transponder and laser communications experiments via dual station ranging to SLR satellites covering also Venus have been implemented \citep{Degn,MOLA}.
Currently available radar-ranging normal points to Venus cover about 33 yr, from 1962 to 1995. The range residuals are depicted in Figure B-6 of \cite{DE421}; after having been as large as 15 km in the first 10 yr, they \textrm{have dropped} below 5 km since. Figure B-4 \textrm{of \citet{DE421}} shows the range residuals to Venus Express at Venus from 2006 to 2008; the are below the 10 m level.

Figure \ref{EMB_Venus_LT} shows the Lense-Thirring perturbation of the Venus range, integrated in a frame aligned with the Sun's  equator.
\begin{figure}[t]
\includegraphics[width=\columnwidth]{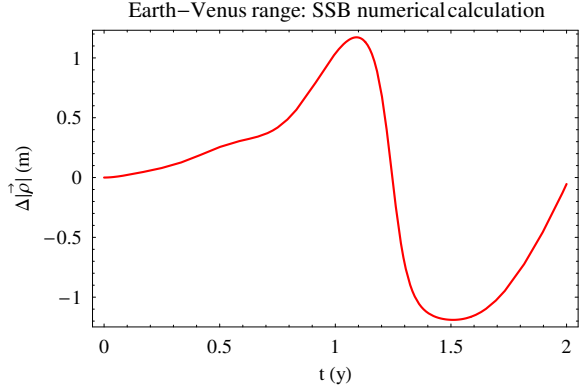}
 \caption{Difference $\Delta |\vec{\rho}|\doteq |\vec{\rho}_{\rm P}|-|\vec{\rho}_{\rm R}|$ in the numerically integrated EMB-Venus ranges with $(\vec{\rho}_{\rm P})$ and without $(\vec{\rho}_{\rm R})$ the perturbation due to the Sun's Lense-Thirring field over $\Delta t=2$ yr. The same initial conditions (J2000) have been used for both the integrations. The state vectors at the reference epoch have been retrieved from the NASA JPL Horizons system. The integrations have been performed in the  ICRF/J2000.0 reference frame, with the mean equinox of the reference epoch and the reference $\{xy\}$ plane rotated from the mean ecliptic of the epoch to the Sun's equator, centered at the Solar System Barycenter (SSB). }\lb{EMB_Venus_LT}
\end{figure}
The peak-to-peak amplitude  is 2 m, which would  be measurable with a future accurate cm-level ranging device with a relative accuracy of $2-5\times 10^{-2}$.  The Lense-Thirring signature is still too small to be detected nowadays with the current spacecraft ranging.

The range perturbation due to the Sun's oblateness is depicted in Figure \ref{EMB_Venus_J2} for the nominal value $J_2^{\odot}=2\times 10^{-7}$. Also in this case a barycentric frame rotated to the Sun's equator has been adopted.
\begin{figure}[ht!]
\includegraphics[width=\columnwidth]{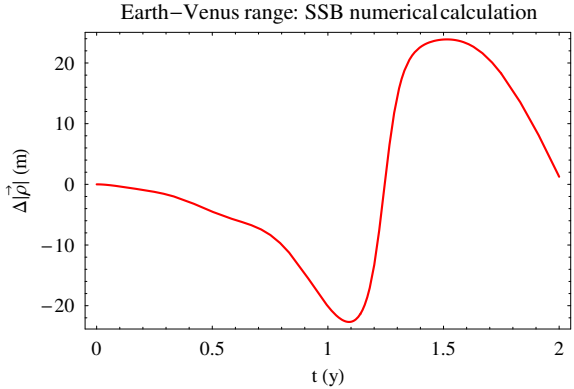}
 \caption{Difference $\Delta |\vec{\rho}|\doteq |\vec{\rho}_{\rm P}|-|\vec{\rho}_{\rm R}|$ in the numerically integrated EMB-Venus ranges with $(\vec{\rho}_{\rm P})$ and without $(\vec{\rho}_{\rm R})$ the nominal perturbation due to the Sun's quadrupole mass moment $J_2^{\odot}=2.0\times 10^{-7}$ over $\Delta t=2$ yr. The same initial conditions (J2000) have been used for both the integrations. The state vectors at the reference epoch have been retrieved from the NASA JPL Horizons system. The integrations have been performed in the  ICRF/J2000.0 reference frame, with the mean equinox of the reference epoch and the reference $\{xy\}$ plane rotated from the mean ecliptic of the epoch to the Sun's equator, centered at the Solar System Barycenter (SSB). }\lb{EMB_Venus_J2}
\end{figure}
The nominal maximum shift is 40 m, so that a measure accurate to $2.5\times 10^{-3}$ would be possible with a future 10 cm-level ranging technique.

If not modeled, the Sun's gravitomagnetic field would impact a determination of $J_2^{\odot}$ at \textrm{the} $5\%$ level. Conversely, the present-day $10\%$ uncertainty in the Sun's oblateness  would yield  a mismodeled signal two times larger than the gravitomagnetic one. \textrm{However},
their temporal signatures are different \textrm{and would therefore allow to seperate these two effects.}
\subsubsection{Earth-Mars}
For Mars we have at our disposal long time series of range residuals accurate to about $1-10$ m-level thanks to several spacecraft (Viking, Mars Pathfinder, Mars Global Surveyor, Mars Odyssey, Mars Reconnaissance Orbiter, Mars Express) which have orbited or are still orbiting the red planet. Figure B-10 of \citep{DE421} depicts the 1-way range residuals of the Viking Lander at Mars spanning from 1976 to 1982; they are
\textrm{approximately at the 20 m level}. Figure B-11 of  \citep{DE421} shows the 1-way range residuals of several post-Viking spacecraft; they generally cover a few years and are accurate to $5-10$ m.

The Lense-Thirring range perturbation, computed in a frame aligned with the Sun's equator, is shown in Figure \ref{EMB_Mars_LT}. Its peak-to-peak amplitude is about 1.5 m, not too far from the present-day range accuracy; thus, its existence as predicted by general relativity is not in contrast with the range residuals currently available. It could be measured with a future cm-level ranging system at a $3-6\%$ level.
\begin{figure}[t]
\includegraphics[width=\columnwidth]{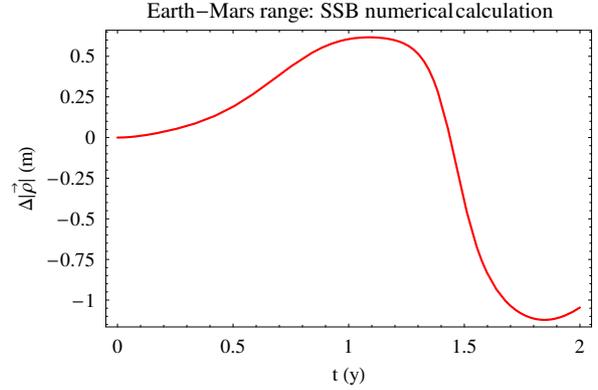}
 \caption{Difference $\Delta |\vec{\rho}|\doteq |\vec{\rho}_{\rm P}|-|\vec{\rho}_{\rm R}|$ in the numerically integrated EMB-Mars ranges with $(\vec{\rho}_{\rm P})$ and without $(\vec{\rho}_{\rm R})$ the perturbation due to the Sun's Lense-Thirring field over $\Delta t=2$ yr. The same initial conditions (J2000) have been used for both integrations. The state vectors at the reference epoch have been retrieved from the NASA JPL Horizons system. The integrations have been performed in the  ICRF/J2000.0 reference frame, with the mean equinox of the reference epoch and the reference $\{xy\}$ plane rotated from the mean ecliptic of the epoch to the Sun's equator, centered at the Solar System Barycenter (SSB). }\lb{EMB_Mars_LT}
\end{figure}

Figure \ref{EMB_Mars_J2} illustrates the nominal signal due to the Sun's quadrupole mass moment for $J_2^{\odot}=2\times 10^{-7}$ computed in a frame aligned with the Sun's equator. Its peak-to-peak amplitude amounts to 25 m; thus, its effect would be well measurable at a $2-4\times 10^{-3}$ level by means of a new ranging facility with an accuracy of the order of cm.  Since $J_2^{\odot}$ is nowadays accurate to $10^{-1}$, the corresponding mismodeled signature would be as large as about $2.5$ m.
\begin{figure}[t]
\includegraphics[width=\columnwidth]{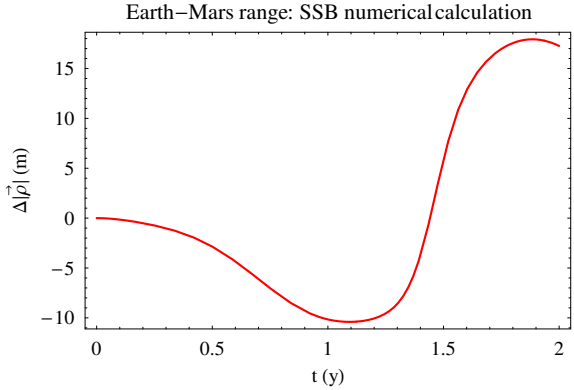}
 \caption{Difference $\Delta |\vec{\rho}|\doteq |\vec{\rho}_{\rm P}|-|\vec{\rho}_{\rm R}|$ in the numerically integrated EMB-Mars ranges with $(\vec{\rho}_{\rm P})$ and without $(\vec{\rho}_{\rm R})$ the nominal perturbation due to the Sun's quadrupole mass moment $J_2^{\odot}=2.0\times 10^{-7}$ over $\Delta t=2$ yr. The same initial conditions (J2000) have been used for both  integrations. The state vectors at the reference epoch have been retrieved from the NASA JPL Horizons system. The integrations have been performed in the  ICRF/J2000.0 reference frame, with the mean equinox of the reference epoch and the reference $\{xy\}$ plane rotated from the mean ecliptic of the epoch to the Sun's equator, centered at the Solar System Barycenter (SSB). }\lb{EMB_Mars_J2}
\end{figure}
If not properly modeled, the Lense-Thirring effect would bias the $J_2^{\odot}$ signal at a $6\%$ level. Conversely, \textrm{considering} $J_2^{\odot}$ as a potential source of systematic bias for the recovery of the gravitomagnetic effect, the mismodeled signature of the Sun's quadrupolar mass moment would be $1.6$ times larger than it. An improvement in its knowledge by one order of magnitude, as expected from, e.g., BepiColombo, would push its bias on the Lense-Thirring signal at $16\%$. \textrm{However}, it must be noted that their temporal evolutions are different.

\section{The gravitomagnetic field of Mars: perspectives for its detection}\lb{marte}
\section{General overview}
Since the angular momentum of Mars can be evaluated \textrm{to be}
 \eqi
 S_{\rm M}=(1.92\pm 0.01)\times 10^{32}\ {\rm kg\ m}^2\ {\rm s}^{-1}
 \eqf
from the latest spacecraft-based determinations of the areophysical parameters \citep{Kon},
 \textrm{the corresponding gravitomagnetic length reads}
 \eqi
 l^{\rm M}_g \doteq \rp{S_{\rm M}}{M_{\rm M}c} =1.0\ {\rm m}.
 \eqf
\textrm{This} has to be compared with the present-day accuracy in determining the orbit of a spacecraft like, e.g., {Mars Global Surveyor}
(MGS) \textrm{which is about $0.15$ m in the radial direction \citep{Kon} and}
not affected by the gravitomagnetic force itself. Thus, it makes sense, in principle,
to investigate the possibility of measuring the Lense-Thirring effect in the gravitational field of Mars as well.

In fact,
\citet{LTmars,IorMGS2} proposed an interpretation of the time
series of the RMS orbit overlap differences \citep{Kon} of the out-of-plane part $\nu$
of the orbit of
MGS over a time span $\Delta P$ of about 5 years (14 November
1999-14 January 2005 in \citet{IorMGS2})
in terms of the Lense-Thirring effect. It turned out that the average of such a time series over
$\Delta P$, normalized to the predicted Lense-Thirring
out-of-plane mean shift over the same time span, is $\mu=
1.0018\pm 0.0053$.
The interpretation by \citet{LTmars,IorMGS2} has recently been questioned by \citet{LTkrogh}; a reply has been set in \citet{LTreplytokrogh}.
Basically,
various linear fits to {\rm  different data sets} including, among others,
the {\rm  full} time series of the {\rm  entire MGS data}  (4 February 1999--14 January 2005) \textrm{were made as well};
the predictions of general relativity turn out to be {\rm  always confirmed}. The analytical calculation of the competing aliasing effects due to both
the gravitational and non-gravitational perturbations, which affect the {\rm  in-plane} orbital components of MGS, do {\rm  not} show up in the real data.
Moreover, the non-conservative forces, whose steadily refined modeling mainly improved the {\rm  in-plane} orbital components of MGS, {\rm  not} the
{\rm  normal} one, exhibit high-frequency, non-cumulative in time variations \citep{For06}.

In view of the never fading interest for planetological missions to the red planet, \textrm{the} preliminary design of a spacecraft-based dedicated mission to Mars has been investigated by \citet{Ior09}. Here we will deal with such an issue in some detail.
In particular, we will concentrate  on the  the multipolar expansion of the martian gravitational
potential  in order
to see what are the critical issues in view of the present-day knowledge of
the Martian space environment. We will not discuss here the perturbations
of non-gravitational origin which depend on the shape, the instrumentation
and the orbital maneuvers of such probes. They would be strongly related to
possible other tasks, more consistent with planetology, which could be fruitfully
assigned to such a mission in order to enhance the possibility that it
may become something more than a mere, although-hopefully-interesting,
speculation; in this respect the medium-long term ambitious programs of
NASA to Mars may turn out to be useful also for the purpose discussed
here.

\subsection{The use of one nearly polar spacecraft}
\begin{figure}[t]
   \includegraphics[width=\columnwidth]{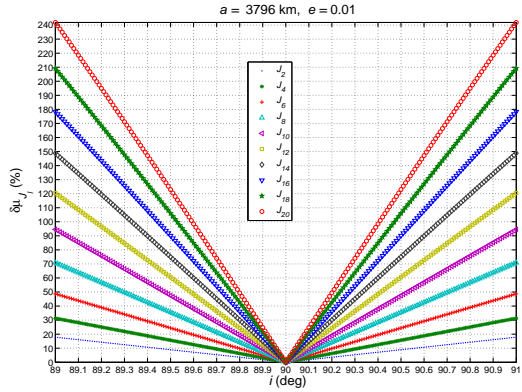}
   \caption{Systematic percent errors $\delta\mu_{J_{\ell}}$ per degree $\ell$ due to the mismodelling in the even zonal harmonics $\delta J_{\ell},\ell=2,4,6...$ for $a=3796$ km, $e=0.01$, $89$ deg $\leq I\leq 91$ deg according to the calibrated covariance sigmas of the MGS95J global gravity solution up to $\ell=20$. The Lense-Thirring effect amounts to 34 mas yr$^{-1}$ corresponding to a shift in the out-of-plane direction of 0.62 m yr$^{-1}$.}
   \label{figura1}
   \end{figure}

Let us start with a polar orbital geometry and examine  the systematic error $\delta\mu$ induced by the uncertainty in the even zonals $\delta J_{\ell}$ on the node $\Omega$ by assuming for them the calibrated covariance sigmas of the latest Mars gravity model MGS95J \citep{Kon}.  We will evaluate it as
 \eqi
 \delta\mu_{J_{\ell}}\leq |\dot\Omega_{.\ell}|\delta J_{\ell},\ \ell=2,4,\textrm{\dots}
 \eqf
 Because  of the presence of $\cos I$ in all the coefficients $\dot\Omega_{.\ell}$ of the oblateness-induced node precessions, we will concentrate on nearly polar ($I\approx 90$ deg) orbital configurations to minimize such a corrupting effect.
It turns out that  altitudes of a few hundreds of km typical of the majority of the currently ongoing martian missions are definitely not suited for our scope, as shown by Figure \ref{figura1}.
It must also be pointed out that likely more even zonals of degree higher than $\ell=20$ would come into play for \textrm{such} low orbital altitudes, as in the case of the forthcoming terrestrial LARES satellite.
Note  that inserting a probe into an areocentric orbit is not an easy task, so that we decided to allow for a departure of up to 1 deg from the ideal polar orbital configuration to account for unavoidable orbital injection errors. Also typical mission requirements pull the inclination some degrees apart from 90 deg: for Mars Global Surveyor (MGS) $I=92.86$ deg. The systematic bias per degree increases for higher degrees $\ell=2,4,6,\textrm{\dots}$ and it turns out that only a very narrow range for $I$, i.e. $\Delta I\approx 10^{-3}$ deg, unlikely to obtain, might push the systematic errors below the $1\%$ level.

\begin{figure}[t]
   \includegraphics[width=\columnwidth]{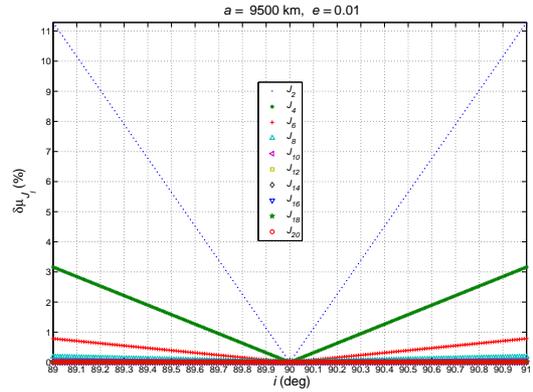}
   \caption{Systematic percent errors $\delta\mu_{J_{\ell}}$ per degree $\ell$ due to the mismodelling in the even zonal harmonics $\delta J_{\ell},\ell=2,4,6...$ for $a=9500$ km, $e=0.01$, $89$ deg $\leq I\leq 91$ deg according to the calibrated covariance sigmas of the MGS95J gravity model up to $\ell=20$. The Lense-Thirring effect amounts to 2 mas yr$^{-1}$ corresponding to a shift in the out-of-plane direction of 0.10 m yr$^{-1}$.}
   \label{figura2}
   \end{figure}

 By keeping a near polar geometry, much larger values of the semimajor axis, comparable with that of Phobos ($a=9380$ km), one of the two natural satellites of Mars, yield reasonable results, as shown \textrm{in} Figure \ref{figura2}.
 Indeed, now the noise decreases with high degree terms and the most serious contribution is due to the first even zonal $\ell=2$ yielding a bias up to about 11$\%$ for $\Delta I=1$ deg; $J_4$ affects the the Lense-Thirring precessions at most at \textrm{the} $3\%$ level for $\Delta I=1$ deg. Note also \textrm{that a} much larger departure from 90 deg ($\Delta I\approx 0.1$ deg) would allow a further reduction of the noise \textrm{below
  1\%}. It must be noted that for such a high-altitude orbital configuration the gravitomagnetic shift in the out-of-plane direction would amount to 0.10 m yr$^{-1}$; it is certainly a small figure, but, perhaps, not too small if one considers that the average shift in the out-of-plane direction of the much lower Mars Global Surveyor is 1.6 m after about 5 yr. It poses undoubtedly challenging requirements in terms of sensitivity and overall orbit determination accuracy, \textrm{but future} improvements may allow to detect such a small displacement.

 Finally, let us remark that, in principle, also the temporal variations of $J_2$ should be accounted for; however, since at present no secular trends have been detected, such changes, mainly seasonal, annual and semi-annual \citep{Kon} would not seriously impact our measurement.

\begin{figure}[t]
   \includegraphics[width=\columnwidth]{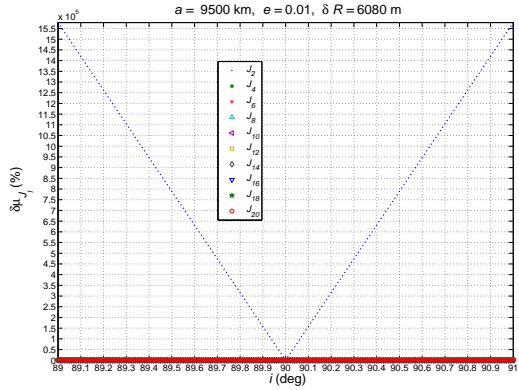}
   \caption{Systematic percent errors $\delta\mu_{J_{\ell}}$ per degree $\ell$ due to the uncertainty in Mars' radius $R$, assumed to be $\delta R = 6080$ m, for $a=9500$ km, $e=0.01$, $89$ deg $\leq i\leq 91$ deg. The errors for $\ell=4,6$ are as large as $6000\%$ and $500\%$, respectively.}
   \label{figura3}
   \end{figure}

\begin{figure}[t]
 \includegraphics[width=\columnwidth]{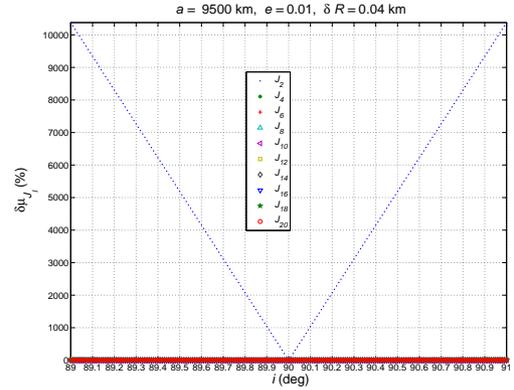}
   \caption{Systematic percent errors $\delta\mu_{J_{\ell}}$ per degree $\ell$ due to the uncertainty in Mars' radius $R$, assumed to be $\delta R = 0.04$ km, for $a=9500$ km, $e=0.01$, $89$ deg $\leq i\leq 91$ deg. The errors for $\ell=4,6$ are as large as $40\%$ and $5\%$, respectively.}
   \label{figura4}
   \end{figure}

 Another martian parameter which must be taken into account is the equatorial radius $R$ along with its uncertainty.
Since
\eqi\dot\Omega_{.\ell}\propto R^{\ell},\eqf the systematic error per degree induced by $\delta R$
can be written as
\eqi
\delta\mu_{J_{\ell}}\leq \ell\left(\rp{\delta R}{R}\right)\left|\dot\Omega_{.\ell}J_{\ell}\right|,\ \ell=2,4,6\textrm{\dots}
\eqf
If we assume conservatively for it $\delta R = 6080$ m, i.e. the difference between the reference value of MGS95J \citep{Kon} and  the one in \citet{Yod95}, the result is depicted in Figure \ref{figura3} for $a=9500$ km.
If, instead, one takes $\delta R=0.04$ km the errors per degree are as in Figure \ref{figura4} for $a=9500$ km.
\textrm{Obviously}, $R$ is the most serious source of systematic error, even for high altitudes.  Moreover,
 \textrm{it might be} unrealistic to expect improvements in the determination of $R$ by future areocentric missions able to push $\delta R$ below terrestrial values, i.e. $1-0.1$ m.  Thus,
$R$ will likely remain an insurmountable obstacle if only one probe is to be used.

Concerning $GM$, it turns out that the impact of its mismodelling   is of no concern being $\lesssim 1\%$.
 %
 %
 %
 %
 %
 %

In summary, in this Section we investigated a nearly polar, high-altitude Phobos-like orbit and noted that the Lense-Thirring effect amounts to a shift of 0.10 m yr$^{-1}$ in the out-of-plane direction. Concerning the systematic errors, the
most crucial source of aliasing is the radius $R$ ($\delta\mu\approx 10000 \%$ for $\delta R = 0.04$ km) and, to a lesser extent, the first even zonal harmonic $J_{2}$ of the areopotential ($\delta\mu\leq 11\%$). To  reduce them the probe should be inserted into an orbit with an inclination close to 90 deg within $10^{-3} $ deg or less. In conjunction with such a very tight constraint, it must also be hoped that future missions to Mars will improve our knowledge of the fundamental parameters of the red planet to a sufficient extent to allow for larger departures from the ideal polar geometry: after all, the gravity solution MGS95J represents an improvement of one order of magnitude with respect to the previous MGS75D model \citep{Yuan01}. However, this may be valid for the spherical harmonic coefficients of the areopotential, not for the radius of Mars for which an uncertainty of the order of \textrm{one} meter would not be enough. We must conclude that the use of only one probe for measuring its nodal Lense-Thirring precession is unfeasible.
\subsection{Including Phobos and Deimos}
A possible solution to reduce the systematic errors would be to consider a linear combination of the nodes of the proposed probe and of the two natural satellites of Mars, i.e. Phobos ($a = 9380$ km, $I = 1.075$ deg, $e = 0.0151$) and Deimos ($a = 23460$ km, $I = 1.793$ deg, $e = 0.0002$), suitably designed to cancel out the impact of just $\delta J_2$ and $\delta R/R$, according to an approach \textrm{suggested} for the first time in the context of the terrestrial LAGEOS-LAGEOS II test \citep{Ciu96}.
A possible combination  is
\eqi \delta\dot\Omega^{\rm Deimos} + \alpha_1\delta\dot\Omega^{\rm Probe}+\alpha_2\delta\dot\Omega^{\rm Phobos},\lb{Kombi}\eqf
with $\delta\dot\Omega$ denoting an  Observed-minus-Calculated ($O-C$) quantity  accounting for  unmodelled/mismodelled features of motion and extracted from the data processing with full dynamical force models. By purposely leaving the gravitomagnetic force unmodelled, it can be
\textrm{written as}
\eqi\delta\dot\Omega = \mu X_{\rm LT} + \dot\Omega_{.2}\delta J_{2} + \dot\Omega_{.R}\left(\rp{\delta R}{R}\right) + \Delta\textrm{;}
\eqf
here $\mu=1$ in general relativity \textrm{and the} coefficient $\dot\Omega_{.R}$ is defined as
\eqi
\dot\Omega_{.R} = R\derp{\dot\Omega^{\rm (obl)}}{R}=\sum_{\ell=2}\ell\dot\Omega_{.\ell}J_{\ell},
\eqf
and $\Delta$ represents all other remaining \textrm{mismodelled/un\-modelled} effects acting on the node
(e.g. $\delta J_4, \delta J_6,\textrm{\dots;}$ $\delta GM$). By solving for $\mu$ one gets
\begin{equation}
\left\{
\begin{array}{lll}
\alpha_1 = \rp{\cf 2{\rm Phobos}\dot\Omega_{.R}^{\rm Deimos}-\cf 2{\rm Deimos}\dot\Omega_{.R}^{\rm Phobos}}{\cf 2{\rm Probe}\dot\Omega_{.R}^{\rm Phobos}-\cf 2{Phobos}\dot\Omega_{.R}^{\rm Probe}},\\\\
\alpha_2 =  \rp{\cf 2{\rm Deimos}\dot\Omega_{.R}^{\rm Probe}-\cf 2{\rm Probe}\dot\Omega_{.R}^{\rm Deimos}}{\cf 2{\rm Probe}\dot\Omega_{.R}^{\rm Phobos}-\cf 2{\rm Phobos}\dot\Omega_{.R}^{\rm Probe}}.
\end{array}\lb{cofi}
\right.
\end{equation}
Note that $\alpha_1$, contrary to $\alpha_2$, is not defined for $I=90$ deg because $\dot\Omega_{.\ell}^{\rm Probe}$, which \textrm{is}
proportional to $\cos I$, \textrm{vanishes} for all $\ell$ at $I=90$ \textrm{deg.
 %
 %
 %
 %
 %
 %
Now,} the systematic bias of $J_4, J_6, J_8\textrm{,\dots}$ on the combination of \rfr{Kombi} turns out to be $\delta\mu\lesssim 6\%$ \textrm{for $89$ deg $\leq I\leq 89.9$ deg} \textrm{where} the major contribution is due to $J_4$. It is an acceptable result, \textrm{especially in} view of the fact that the admissible range for the inclination \textrm{amounts now to} about 1 deg and that \rfr{Kombi} is, by construction, immune to the uncertainty in the martian radius; indeed, although it is likely that many physical properties and parameters like, e.g., the even zonals, of the red planets will be determined with increasing accuracy by the
many ongoing and planned missions, it is unlikely that the radius will be known to a sufficient accuracy to change at an acceptable level the bias induced by it (see Figure \ref{figura3} and Figure \ref{figura4}).

The combination of \rfr{Kombi} is also able to remove \textrm{a} large part of the effect of the mismodelling in $GM$
\eqi
\delta\mu_{GM}\leq\rp{1}{2}\left(\rp{\delta GM}{GM}\right)\sum_{\ell=2}^{20}\left|\dot\Omega_{.\ell}J_{\ell}\right|.\lb{ergm}
\eqf
Indeed, the main contribution to \rfr{ergm}  is due to $J_2$, which is canceled out by \rfr{Kombi} with the coefficients of \rfr{cofi}.  The other terms in \rfr{ergm}, not canceled by the combination of \rfr{Kombi}, yield negligible \textrm{errors well} below $1\%$.

However, it must be stressed that the node of Phobos undergoes secular precessions due to  other perturbations of gravitational origin \citep{Lai07} (\textrm{e.g., the }non-sphericity of Phobos itself, \textrm{the martian} tidal bulge and nutation) which have not been considered here and that would affect the combination of \rfr{Kombi}. It turns out that the most relevant one is that due to the spherical harmonic coefficients $c_{20}$ and $c_{22}$ of Phobos itself \citep{Bor90} whose induced secular precession amounts nominally to 200 km over 3 yr \citep{Lai07}. The nutation perturbation nominally amounts to 0.3 km over 3 yr \citep{Lai07}. The tidally induced precession, parameterized in terms of the martian Love number $k_2$ would amount nominally \textrm{to} 0.06 km after 3 yr \citep{Lai07}, but being $k_2=0.152\pm 0.009$   \citep{Kon}, its modelling would left a $\approx 1$ m yr$^{-1}$ mismodelled trend.
The Phobos gravitomagnetic shift is 0.1 m yr$^{-1}$.

In terms of the detectability of the Lense-Thirring signal with the combination of \rfr{Kombi}, \textrm{undoubtedly a} major drawback of the strategy of including Phobos and Deimos is the poor accuracy with which their orbits can be reconstructed with respect to their Lense-Thirring signal. Indeed, while their gravitomagnetic shifts are of the order of $1-10$ cm yr$^{-1}$, the latest NASA martian ephemerides\footnote{See on the WEB \url{http://ssd.jpl.nasa.gov/?sat_ephem}.} for them yield an accuracy of $1-10$ km in the radial, transverse and out-of-plane orbital components \citep{MAR}. Thus, according to the present-day level of \textrm{accuracy, including} the natural satellites of Mars in the combination of \rfr{Kombi} would introduce a noise which would overwhelm the relativistic trend of interest. This situation may become more favorable in the near future in view of the \textrm{planned Phobos-Grunt space mission}\footnote{See on the WEB \url{http://www.esa.int/esaMI/ESA_Permanent_Mission_in_Russia/SEMIJFW4QWD_0.html}.} \citep{Grunt} which should allow, among other things, to improve the orbit determination of Phobos as well.
 \subsection{Two dedicated probes}
In principle, the linear combination approach could be followed by using the nodes of two \textrm{other spacecraft}\footnote{For a polar geometry the Lense-Thirring precession of the pericentre $\omega$ of a spacecraft vanishes being proportional to $\cos I$.}, although sending to Mars three new probes would increase the costs and the difficulties   of such a demanding mission.

It is interesting to consider a scenario involving only two  probes, named P1 and P2, in a nearly polar counter-orbiting configuration, proposed for the first time by \citet{vpe76a} in the framework of the attempts to design a suitable terrestrial mission, and later generalized by \citet{Ciu86} for other inclinations.
By assuming for, say, P1 $a_1=9500$ km, $I_1=89$ deg, $e_1=0.01$, it turns out that the sum of their nodes would be an observable relatively insensitive to departures from the ideal configuration for P2, i.e.  $a_2=9500$ km, $I_2=91$ deg, $e_2=0.01$, at least in regard to the mismodelling in the even zonals. \textrm{For} $a_2=9600$ km, $90.05$ deg $\leq I_2\leq 92$ deg, $e_2=0.03$ the bias due to $J_2$ is up to $7\%$,
while the other even zonals would have an impact of the order of $1\%$ or less.
%
 %
 %
 %
 %
 %
Unfortunately, the uncertainty in the martian radius makes such constraints much more stringent. Indeed, $\delta\mu_R\approx 10^5\%$ for $\delta R=6080$ m and $\delta\mu_R\approx 4000\%$  for $\delta R=0.04$ km. Also in this case, the \textrm{improvement in $R$ expected} from future missions to the red \textrm{planet} may be not sufficient.

\begin{figure}[t]
   \includegraphics[width=\columnwidth]{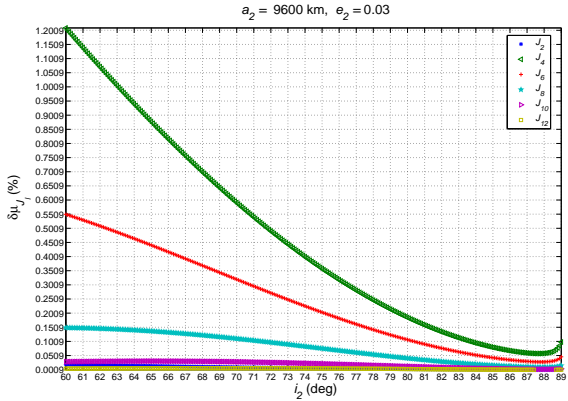}
   \caption{Systematic errors $\delta\mu_{J_{\ell}}$ per degree due to the mismodelling $\delta J_2, \delta J_4, \delta J_6,...$ (MGS95J model) in the uncancelled even zonal harmonics in the combination of  the nodes of probe P1 ($a_1=9500$ km, $I_1=89$ deg, $e_1=0.01$) and probe P2 ($a_2=9600$ km, $60$ deg $\leq I_2\leq 89$ deg, $e_2=0.03$) with \rfr{koff} which cancels out entirely the bias due to $\delta R$.}
   \label{figura14}
   \end{figure}

 An approach that could be \textrm{successfully} followed within the framework of the linear combination strategy with two probes P1 and P2, especially in view of likely future improvements
 in the even zonal harmonics of the areopotential, consists in designing a combination which, by construction, entirely cancels out  the bias due to the uncertainty in $R$ \eqi\delta\mu_R\leq \left|\dot\Omega_{.R}^{\rm P2} + k_1\dot\Omega_{.R}^{\rm P1}\right|\left(\rp{\delta R}{R}\right),\eqf being, instead, affected by $\delta J_2, \delta J_4, \delta J_6$. It is analogous to the three-nodes combination of \rfr{combi}, with
 \eqi
 k_1 = -\rp{\dot\Omega_{.R}^{\rm P2}}{\dot\Omega_{.R}^{\rm P1}}.\lb{koff}
 \eqf
 Indeed, as can be \textrm{seen from} Figure \ref{figura14}, $\delta\mu_{J_{\ell}}$ would be \textrm{rather small} ($\lesssim 1\%$), according to the present-day MGS95J model, provided that inclinations $I\ll 90$ deg are adopted for the second probe.
 It turns out that the most relevant even zonals would be $J_4$ and $J_6$ with a bias of the order of $\approx 1-0.5\%$.
With \rfr{koff} an inclination range as large as ten degrees would be admissible: \textrm{this} is really a great advantage with respect to the other cases examined so far.

\subsection{Summary}
The main source of systematic bias is the mismodelling in some of the parameters entering the multipolar expansion of the
Newtonian part of the areopotential, especially the Mars' equatorial radius $R$ and, to a lesser but non-negligible extent,
the even zonal harmonics $J_{\ell},\ell=2,4,6,\textrm{\dots}$. Since the resulting aliasing node precessions are proportional to $R^{\ell}a^{-(3/2 + \ell)}\cos I$, a high-altitude ($a\approx 9500$ km), polar ($I=90$ deg) probe \textrm{would be an} optimal solution, but the
\textrm{estimation} of the unavoidable orbit injection errors showed that such an option is unfeasible because it would require too stringent constraints on the departures from the ideal case $I=90$ deg.
\textrm{In principle it would be possible} to suitably combine the nodes of such a probe with those of the natural satellites of Mars, Phobos and Deimos. However, the Lense-Thirring shifts of \textrm{these} bodies ($1-10$ cm yr$^{-1}$) are orders of magnitude too small with respect to the present accuracy in reconstructing their orbits ($1-10$ km).

 A viable option consists in using two probes P1 and P2 at high-altitudes ($a_1=9500$ km, $a_2=9600$ km) and different inclinations
 ($I_1\approx 90$ deg, $60$ deg $\leq I_2 \leq 89$ deg), and combining their nodes so as to entirely cancel out the bias due to $R$: the resulting bias due to $\delta J_2, \delta J_4, \delta J_6\textrm{,\dots}$  would be $\lesssim 1\%$, according to the present-day MGS95J gravity model, over a  range of values for the inclination as large as ten degrees.

 A major challenge would certainly be \textrm{to reach a satisfying} accuracy in reconstructing the orbits of such probes whose Lense-Thirring out-of-plane shifts would amount to about $10$ cm yr$^{-1}$; for example, the spacecraft should remain operative for many years, without any failure in the communication with the Earth, and it is likely that one or more landers would be required as well.

 Finally, we \textrm{emphasize} that a different type of analysis should be done to complete the studies presented here,
 \textrm{based on} direct observable quantities like range and range-rates.

\section{The gravitomagnetic field of Jupiter: perspectives in measuring the jovian angular momentum via the Lense-Thirring effect}
\subsection{General overview}
After the original proposal by \citet{LT} of looking at the Galilean satellites\footnote{Their semimajor axes range from 421800 km (Io) to 1882700 km (Callisto), the eccentricities being of the order of $10^{-3}$. See on the WEB \url{http://ssd.jpl.nasa.gov/?sat_elem}. } of Jupiter as natural probes for measuring the gravitomagnetic precessions of their longitudes of perijove $\varpi$, \citet{Bogo59}
considered the  centennial  shift of the argument of perijove $\omega$ of the inner moon Amalthea ($a=181400$ km, $e=0.0032$) amounting to 112 arcsec, while \citet{Sof} looked at centennial node shift of Io of about 9 arcsec.
No evaluations of the systematic bias posed by several competing dynamical effects of the Jovian space environment have been examined in such earlier studies.
In more recent times, \citet{IorLan} revisited the Jovian system of the Galilean moons in view of the recent advancements in their orbit
determination. It turned out \textrm{that most of their} gravitomagnetic signals would be absorbed in the estimation procedure of the satellites' initial state vectors; the remaining signatures would amount to some tens \textrm{of meters}, while the current best observations
have an accuracy of a few tens of kilometers\footnote{See on the WEB \url{http://ssd.jpl.nasa.gov/?sat_ephem#ref3}.}. \citet{IorLan} concluded that a spacecraft orbiting Jupiter would be required to reveal its gravitomagnetic field; as we will see below, such an opportunity \textrm{may be
 offered} in the near future \citep{IorioNA010}.

\textrm{In an attempt to measure gravitomagnetism by means of the deflection of electromagnetic waves by Jupiter due to
its orbital motion\footnote{In this case, the mass currents inducing a gravitomagnetic action are
not those related to the Jupiter's proper rotation (intrinsic gravitomagnetism), but are due to its translational orbital motion
(extrinsic gravitomagnetism).}, a dedicated VLBI-based radio-interferometric experiment has been performed \citep{KopFom,Fom08}.}
With regard to other suggested non-gravitomagnetic  tests of general
relativity in the jovian gravitational field, \citet{His} proposed to measure the much larger
gravitoelectric Einstein pericenter precessions  of the natural satellites of Jupiter and Saturn.
There exist also plans for performing a test of the light bending due \textrm{to Jupiter's} monopole and quadrupole mass moments with the
forthcoming astrometric mission GAIA \citep{Cro}. Finally, \textrm{also the proposal} by \citet{Haas} \textrm{to measure Schiff's}
gyroscope precession in the gravitational field of Jupiter \textrm{should be mentioned}.

\textrm{
Another} opportunity to fruitfully  use the Jovian system  may open after
\textrm{the approval of
the Juno mission\footnote{See on the WEB \url{http://juno.wisc.edu/index.html}} by NASA \citep{Mat05} to Jupiter.}
Juno is a spinning, solar powered spacecraft to be placed in a highly eccentric polar orbit around Jupiter (see Table \ref{tavolajun} for its
relevant orbital parameters)
specifically designed to avoid its highest radiation regions. Understanding the formation, evolution and
structure of Jupiter is the primary science goal of
Juno.
\begin{table}[t]
\caption{Planetocentric nominal orbital parameters of Juno. $a,e,I$ are the semi-major axis (in jovian radii $R=71492$ km), the
eccentricity and the inclination (in deg) \textrm{with respect to Jupiter's} equator, respectively. $\dot\Omega_{\rm LT} $ and $\dot\omega_{\rm LT} $
\textrm{are Juno's} Lense-Thirring node and perijove precessions (in mas yr$^{-1}$) for
$S = 6.9\times 10^{38}$ kg m$^2$ s$^{-1}$ \protect\citep{Sof03}. $P$ is the orbital period \textrm{in days}. $T$ is
the mission duration \textrm{in years}. }\label{tavolajun}
%
\begin{tabular}{@{}lllllll}
\hline
$a$  & $e$ & $I$  & $\dot\Omega_{\rm LT}$  & $\dot\omega_{\rm LT}$  & $P$  & $T$ \\
\hline
20.03 & 0.947 & 90  & $68.5$ & 0 & 11 & 1\\
\hline
\end{tabular}
\end{table}
It will carry onboard a dual frequency gravity/radio
science system, a six wavelength microwave radiometer for atmospheric sounding and composition, a
dual-technique magnetometer, plasma detectors, energetic particle detectors, a radio/plasma wave
experiment, and an ultraviolet imager/spectrometer. The nominal mission's lifetime is 1 year. Juno is aimed, among other things, at accurately mapping
the  gravitational field of Jupiter \citep{And76} with unprecedented accuracy \citep{And04} by exploiting the slow apsidal precession of its 11-day orbit.

Since the Lense-Thirring \textrm{precession is} due to the proper angular momentum $\bds S$ of the
orbited central body, one may also \textrm{take its existence as granted} and use it as a direct, dynamical measurement
\textrm{of Jupiter's} angular momentum through the Lense-Thirring effect;
this would yield further, important information concerning the interior of Jupiter.  Indeed, the moment of inertia ratio $C/MR^2$ entering $S$ is
a measure of the concentration of mass towards the center of the planet \citep{Irw03}. Such a figure, together with the measured values
of the zonal\footnote{They preserve the axial symmetry.} coefficients of the gravity field accounting for its deviations from spherical symmetry
may be fitted with internal models \textrm{of the variation of the density, pressure, temperature and composition with depth}  \citep{Irw03,Gui05,Hor08}.
Moreover, a dynamical, model-independent determination of $S$ would be important also for a better knowledge of the  history and
formation of Jupiter \citep{Mac08}.

\subsection{Sensitivity analysis}
\textrm{Jupiter's} proper angular momentum amounts to  \citep{Sof03}
\eqi S \approx 6.9\times 10^{38}\ {\rm kg\ m^2\ s^{-1}}\lb{LTJUP}
\eqf
\textrm{and} the corresponding Lense-Thirring precessions \textrm{of Juno's} node and perijove \textrm{are listed in Table \ref{tavolajun}} . It turns out that only the node
experiences a GM precession
%
%
  %
  %
%
%
%
corresponding to  a shift $\Delta\nu$ of the  cross-track component  of the planetocentric position  \citep{Cri}   %

\eqi
  \Delta \nu_{\rm LT} = a\sqrt{1+\rp{e^2}{2}}\sin I\Delta\Omega_{\rm LT} = 572\ {\rm m}\ (T=1\ {\rm yr})
  %
  %
\eqf
over the entire duration of the mission.
A total accuracy of the order of 1-10 m with respect to the km-level of the past Jupiter missions in \textrm{reconstructing Juno's}
orbit in a planetocentric frame does not seem
an unrealistic target, although much work is clearly required in order to have a more firm answer. Note that a 1-10 m accuracy implies a $0.2-2\%$ error
in measuring the gravitomagnetic shift

\textrm{To consider the possible detection of the Lense-Thirring effect by means of the Juno mission is also suggested
by a different approach with respect to the cumulative measurement over the full mission duration previously outlined}.
Indeed, a gravity-science pass for Juno is defined by a continuous, coherent Doppler range-rate measurement plus and minus three hours of closest
approach; in practice, most of the Lense-Thirring precession takes place just during such a six-hours pass, a near optimum condition. Another
crucial factor is the orientation of the Earth \textrm{with respect to Juno's} orbit: our planet will be aligned 67 deg from the probe's orbital plane at approximately
two degrees south latitude on the jovian equator. Preliminary numerical simulations \textrm{of Juno's} Lense-Thirring Doppler range-rate signal show that
such an orbital geometry \textrm{represents} a perfect compromise for measuring \textrm{both Jupiter's} even zonal harmonics and the gravitomagnetic  signal itself.
Indeed, it turns out that the maximum Lense-Thirring Doppler signal over a single six-hour gravity pass is of the order of hundred $\mu$m s$^{-1}$, while
the limit of accuracy for Juno's Ka-band Doppler system is about one  $\mu$m s$^{-1}$ over such a pass. Thus, even by taking 25 repeated passes
\textrm{out of a total
of approximately 33}, it would be possible to reach a measurement precision below the percent level.

\textrm{It should be noted that there are estimates for $S$ in the literature}
\textrm{in favor of smaller values than given in}
\rfr{LTJUP} by a factor $1.5-1.6$; for example, \citet{Mac08} \textrm{report}
\eqi
S = 4.14 \times 10^{38}\ {\rm kg\ m^2\ s^{-1}};\lb{SJap}
\eqf
the ratio of \textrm{of the values of \rfr{LTJUP} to \rfr{SJap} is 1.6, i.e. close} to 1.5 coming from the ratio  of $C/MR^2=2/5=0.4$, valid for a homogenous sphere,
to $C/MR^2=0.264$ by \citet{Irw03} who assumes a concentration of mass \textrm{towards Jupiter's} center.
Here we \textrm{consider} only the systematic uncertainty induced by the imperfect knowledge of the Newtonian part of \textrm{Jupiter's gravitational field and use the value of} \rfr{LTJUP} for $S$.

Also in this case, a major source of systematic uncertainty \textrm{is due to the deviation of Jupiter's}
 gravitational field from spherical symmetry \citep{And76}.

\subsubsection{Analytical calculations}
As seen in Section \ref{OBLA}, the  zonal ($m=0$) harmonic coefficients $J_{\ell}$ of the multipolar expansion of the Newtonian part of the planet's gravitational
potential give rise to long-period, \textrm{(i.e. averaged over one orbital revolution)}, orbital effects  on the longitude of the ascending
node $\Omega$, the argument of pericentre $\omega$ and the mean anomaly $\mathcal{M}$  of the form
\begin{equation}
\left\langle\dot\Psi\right\rangle = \sum_{\ell=2}\dot\Psi_{.\ell} J_{\ell},\ \Psi=\Omega,\omega,\mathcal{M},
\end{equation}
where $\dot\Psi_{.\ell}$ are coefficients depending on the planet's $GM$ and equatorial
radius $R$, and on the spacecraft's inclination $I$ and eccentricity $e$ through the inclination $F_{\ell m p}(I)$ and eccentricity $G_{\ell p q}(e)$
functions, respectively \citep{Kau}.
Note that one of the major scientific goals of the Juno mission is  a greatly improved determination of just the harmonic coefficients of the jovian
gravity potential; for the present-day values of the zonals\footnote{The Jupiter gravity field is essentially determined by the Pioneer 11
flyby at 1.6$R_{\rm Jup}$ \citep{And76}; Voyager added little, and Galileo, which never got close to Jupiter, added nothing.} for $\ell=2,3,4,6$ see
Table \ref{zonalsjup}.
\begin{table*}[t]
\small
\caption{Zonal harmonics of the Jupiter's gravity field according to the JUP230 orbit solution \protect\citep{JUP230}. They are in unit of $10^{6}$.}\label{zonalsjup}
\begin{tabular}{@{}llll@{}}
\tableline
$J_2$ & $J_3$ & $J_4$  & $J_6$\\
\tableline
$14696.43\pm 0.21$ & $-0.64\pm 0.90$ & $-587.14\pm 1.68$  & $34.25\pm 5.22$ \\
\tableline
\end{tabular}
\end{table*}
According to \citet{And04}, it might be possible to determine the first three even zonals with an accuracy of $10^{-9}$ and the other ones up
to $\ell=30$  at a $10^{-8}$ level. Concerning $J_2$, this would be an improvement of two orders of magnitude with respect to Table \ref{zonalsjup},
while the improvements in $J_4$ and $J_6$ would be \textrm{of about} three orders of magnitude.
By using the results we are going to present below for the long-period node and pericenter precessions, it can be shown that determining the
low degree zonals at \textrm{the} $10^{-9}$ level of accuracy translates into an accuracy of the order of $0.5-1$ mas yr$^{-1}$ in $\dot\Omega$ and $\dot\omega$,
thus confirming the expectations of the previous Section.

\textrm{Long-period terms fulfill the condition
\eqi
\ell-2p+q=0
\eqf
which results in\footnote{Note that $2(\ell+1)$ in \rfr{corrz} corrects a wrong $6$ in (16), pag. 556  of \citet{IorioNA010}.} \citep{Kau}}
{{\begin{eqnarray}
  \dot\Omega_{.\ell} &=& -\rp{n}{\sqrt{1-e^2}\sin I}\left(\rp{R}{a}\right)^{\ell}\times
  \nonumber\\
  & &
  \times\sum_{p=0}^{\ell}\left[F^{'}_{\ell 0 p}G_{\ell p (2p-\ell)}
  W_{\ell 0 p(2p-\ell)}\right], \\
  \nonumber\\
  \dot\omega_{.\ell} &=& -\rp{n}{\sqrt{1-e^2}}\left(\rp{R}{a}\right)^{\ell}\sum_{p=0}^{\ell}\left[-\cot I F^{'}_{\ell 0 p}G_{\ell p (2p-\ell)}+\right.
  \nonumber\\
  & & \left. +
  \rp{(1-e^2)}{e}F_{\ell 0 p}G^{'}_{\ell p(2p-\ell)}\right]W_{\ell 0 p(2p-\ell)}, \\
  \nonumber\\
  \dot{\mathcal{M}}_{.\ell}  &=& n \left\{1 - \left(\rp{R}{a}\right)^{\ell}\sum_{p=0}^{\ell}F_{\ell 0 p}\left[2(\ell+1)G_{\ell p(2p-\ell)}-\right.\right.
  \nonumber\\
  & & \left.\left.
  -\rp{(1-e^2)}{e}G^{'}_{\ell p(2p-\ell)}\right]W_{\ell 0 p(2p-\ell)}\right\}\lb{corrz},
\end{eqnarray}
}
}
where $W_{\ell 0p(2p-\ell)}$ are trigonometric functions having the pericentre as their argument.
Contrary to \textrm{the} small eccentricity satellites like the LAGEOS ones previously examined, in this case
we will be forced to keep all the terms of order $\mathcal{O}(e^k)$ with $k>2$ in computing the eccentricity functions
for given pairs of $\ell$ and $p$. Moreover, \textrm{since all the non-zero eccentricity and inclination functions for a given degree $\ell$
are needed, we have to consider} all the
non-vanishing terms with $0\leq p\leq \ell$. First, we will extend our calculations to the  even zonals so far determined, i.e.  $\ell=2,4,6$.
In this case $W$ reads
\eqi W_{\ell 0 p(2p-\ell)}=\cos[(\ell -2p)\dot\omega t]=\cos(q\dot\omega t).
\eqf
It must be noted that the terms with
\eqi
p = \rp{\ell}{2},\ q=0,
\eqf
\textrm{i.e. $W_{\ell 0 \rp{\ell}{2} 0}=1$,} yield secular precessions, while those with
\eqi
q=2p-\ell\neq 0
\eqf
\textrm{induce} harmonic signals with circular frequencies $-q\dot\omega$.

\textrm{For degree $\ell=2$ only secular precessions occur and} the non-vanishing inclination and eccentricity functions and their derivatives are\footnote{Here and in the following $cI\doteq \cos I$, $sI\doteq\sin I$. Note that \rfr{correz} corrects a typo in (23), pag. 556 of \citet{IorioNA010}. }
\eqi F_{201} = \rp{3}{4}sI^2  - \rp{1}{2}.\eqf

\eqi F^{'}_{201} = \rp{3}{2}sIcI.\eqf

\eqi G_{210}=\rp{1}{(1-e^2)^{3/2}}.\eqf

\eqi G^{'}_{210} = \rp{3 e}{(1-e^2)^{5/2}}.\lb{correz}\eqf

\textrm{For $\ell=4$ we find}\footnote{Note that \rfr{corrza} and \rfr{corrza2} correct typos in (30) and (31), \textrm{p.} 557 of \citet{IorioNA010}.}
\begin{eqnarray}
  F_{401} &=&  -\rp{35}{32}sI^4  + \rp{15}{16}sI^2    \ = \ F_{403},\\
  \nonumber\\
  F_{402} &=& \rp{105}{64}sI^4  - \rp{15}{8}sI^2  + \rp{3}{8}.
\end{eqnarray}

\begin{eqnarray}
  F^{'}_{401} &=& \left(-\rp{35}{8}sI^3  + \rp{15}{8}sI\right)cI=  F^{'}_{403},\\
  \nonumber\\
  F^{'}_{402} &=& \left(\rp{105}{16}sI^3  - \rp{15}{4}sI\right)cI.
\end{eqnarray}

\begin{eqnarray}
  G_{41-2} &=& \rp{3}{4}\rp{e^2}{(1-e^2)^{7/2}} \ = \ G_{432},\\
  \nonumber\\
  G_{420} &=& \rp{1+\rp{3}{2}e^2}{(1-e^2)^{7/2}}.
\end{eqnarray}

\begin{eqnarray}
  G^{'}_{41-2} &=&  \rp{3}{2}\rp{e\left(1+\rp{5}{2}e^2\right)}{(1-e^2)^{9/2}}\ = \ G^{'}_{432},\lb{corrza}\\
  \nonumber\\
   G^{'}_{420} &=& \rp{10 e\left(1+\rp{3}{4}e^2\right)}{(1-e^2)^{9/2}}\lb{corrza2},
\end{eqnarray}
\textrm{where in} addition to secular terms, also harmonic signals with the frequencies $\pm 2\dot\omega$ are present.

\textrm{For} $\ell = 6$ the inclination and eccentricity functions, along with their \textrm{derivatives read}\footnote{Note that \rfr{Korrza}, \rfr{Korrza2} and \rfr{Korrza3} correct typos in (41), (42), (43), \textrm{p.} 557 of \citet{IorioNA010}.}
\begin{eqnarray}
  F_{601} &=& \rp{693}{512}sI^6  - \rp{315}{256}sI^4   =  F_{605}, \\
  \nonumber\\
  F_{602} &=& -\rp{3465}{1024}sI^6  + \rp{315}{64}sI^4  - \rp{105}{64}sI^2  =  F_{604}, \\
  \nonumber\\
  F_{603} &=& \rp{1155}{256}sI^6  - \rp{945}{64}sI^4  + \rp{105}{32}sI^2  -\rp{5}{16}.
\end{eqnarray}

{\small{
\begin{eqnarray}
  F^{'}_{601} &=& \left(\rp{2079}{256}sI^5  - \rp{315}{64}sI^3 \right)cI=  F^{'}_{605},\\
  \nonumber\\
  F^{'}_{602}  &=& \left(-\rp{10395}{512}sI^5  + \rp{315}{16}sI^3  - \rp{105}{32}sI\right)cI=  F^{'}_{604}, \\
  \nonumber\\
  F^{'}_{603}  &=& \left(\rp{3465}{128}sI^5  - \rp{945}{16}sI^3  + \rp{105}{16}sI\right)cI.
\end{eqnarray}
}
}

\begin{eqnarray}
  G_{61-4} &=& \rp{5}{16}\rp{e^4}{(1-e^2)^{11/2}}   =  G_{654},\\
  \nonumber\\
  G_{62-2} &=& \rp{5}{2}\rp{e^2}{(1-e^2)^{11/2}}\left(1 + \rp{e^2}{8}\right)   =  G_{642},\\
  \nonumber\\
  G_{630} &=& \rp{1}{(1-e^2)^{11/2}}\left( 1 + 5e^2 + \rp{15}{8}e^4 \right) .
\end{eqnarray}

\begin{eqnarray}
G^{'}_{61-4} & = &\rp{5}{4}\rp{e^3\left(1+\rp{7}{4}e^2\right)}{(1-e^2)^{13/2}} =  G^{'}_{654},\lb{Korrza}\\
 \nonumber\\
G^{'}_{62-2} & = &  \rp{5 e\left(1+\rp{19}{4}e^2 + \rp{7}{16}e^4\right)}{(1-e^2)^{13/2}}  =  G^{'}_{642},\lb{Korrza2}\\
\nonumber\\
G^{'}_{630} &=& \rp{21 e\left(1 + \rp{5}{2}e^2 + \rp{5}{8} e^4\right)}{(1-e^2)^{13/2}}\lb{Korrza3}.
\end{eqnarray}
 In addition to the secular rates, also harmonic signals with frequencies $\pm 4\dot\omega,\pm 2\dot\omega$ \textrm{do occur}.

\textrm{Let us now focus} on the action of the odd ($\ell=3,5,7,\textrm{\dots}$) zonal ($m=0$) harmonics.
\textrm{In this case}
\eqi W_{\ell 0 p(2p-\ell)}=s[(\ell-2p)\dot\omega t]=-s(q\dot\omega t),\eqf
 so that only harmonic terms \textrm{exist} for $q\neq 0$.

\textrm{Finally, for $\ell=3$ we find}\footnote{Note that \rfr{mammamia} corrects a typo in (48), \textrm{p.} 557 of \citet{IorioNA010}.}

\begin{eqnarray}
F_{301} &=&  \rp{15}{16}sI^3  -\rp{3}{4}sI= -F_{302}, \\
\nonumber\\
F^{'}_{301} & = & \left( \rp{45}{16}sI^2  -\rp{3}{4}\right)cI= -F^{'}_{302}.
\end{eqnarray}

\begin{eqnarray}
  G_{31-1} &=& \rp{e}{(1-e^2)^{5/2}} = G_{321}, \\
  \nonumber\\
  G^{'}_{31-1} &=&  \rp{1+4e^2}{(1-e^2)^{7/2}} = G^{'}_{321}\lb{mammamia}.
\end{eqnarray}

Thus, we have long-period effects varying \textrm{with frequencies} $\pm\dot\omega$.

\textrm{We note that the long-period even and odd zonal harmonic terms can be} approximated by secular precessions with a
high level of accuracy over the expected 1-yr lifetime $T$ of \textrm{the Juno mission} because the period of \textrm{Junos's}
perijove is of the order of $\approx 500$ yr, i.e.
\begin{eqnarray}
\cos (q\omega) & = & \cos(q\dot\omega t + q\omega_0)\approx\cos(q\omega_0),\ 0\leq t\leq T,\\
 \nonumber\\
 \sin\omega   & = & \sin(\dot\omega t + \omega_0)\approx\sin\omega_0,\ 0\leq t\leq T.
 \end{eqnarray}
 Thus, the choice of the initial condition $\omega_0$ will be crucial in determining \textrm{the impact of the long-period terms}.

Now,  it would be possible, in principle, to use the node of Juno to measure the gravitomagnetic effect. Indeed, the Lense-Thirring node
precession is independent of $I$, while all the zonal precessions  of $\Omega$ vanish for $I=90$ deg. \textrm{This is not the case} for the perijove
and the mean anomaly, but they are not affected by the gravitomagnetic force for $I=90$ deg. In reality, the situation will be \textrm{less
favorable}
because of the unavoidable orbit injection errors which will induce some departures $\delta I$ \textrm{of Juno's} orbital plane from the ideal polar
configuration. Thus, unwanted, corrupting node zonal secular precessions will appear; their mis-modeling due to the uncertainties $\delta J_{\ell}$
may swamp the recovery of the Lense-Thirring effect if their determination by Juno will not be accurate enough.
Note that there is no risk of some sort of a-priori \virg{imprint} effect of the Lense-Thirring effect itself on the values of the zonals retrieved from the
Juno's perijove motion because the gravitomagnetic pericenter precession vanishes for polar orbits.

\begin{figure}[t]
\includegraphics[width=\columnwidth]{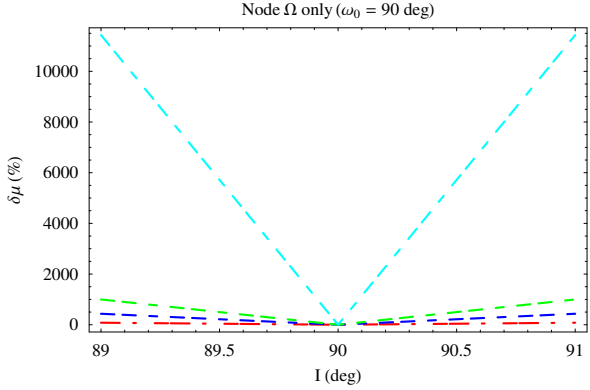}
 \caption{\footnotesize{\label{JUNO_node} Systematic percent bias on the Lense-Thirring node precession induced by the mis-modeling in the
 zonals $J_2$ (red), $J_3$ (blue), $J_4$ (green), $J_6$ (light blue) according to Table \ref{zonalsjup} for $89$ deg $\leq I\leq 91$ deg and $\omega_0=90$ deg.} }
\end{figure}

By assuming the values quoted in Table \ref{zonalsjup} for the uncertainties $\delta J_{\ell}$, $\ell=2,3,4,6$, let us see what
is the impact of \textrm{an imperfect} polar orbital geometry on the \textrm{nodal} Lense-Thirring precessions.  The results are depicted in Figure
\ref{JUNO_node} for each degree $\ell$ separately; the initial condition $\omega_0=90$ deg has been used.  It should be noted that,
in view of the likely correlations among the determined zonals, a realistic upper bound of the total bias due to them can be computed by
taking the linear sum of each mis-modeled terms.
The major source of bias is the so far poorly known $J_6$; an improvement of four orders of magnitude, which sounds rather unlikely to be
obtained even with Juno \citep{And04}, would be required to push its aliasing effect at a percent level of the Lense-Thirring effect.
The situation for the other zonals is more favorable; $J_4$ should be known better than now by a factor 1000, which is, instead, a realistic goal
according to \citet{And04}.
Thus, we conclude that a nearly-polar orbit \textrm{being} 1 deg off the ideal 90 deg case would likely \textrm{prevent a} measurement of the gravitomagnetic node precession at a decent level of accuracy.

Thanks to the high eccentricity \textrm{of Juno's} orbit, also the perijove and the mean anomaly are well defined, so that they can be used in a suitable way to remove the bias of $J_6$ and $J_2$.
\textrm{Let us write}
\begin{eqnarray}
  \delta\dot\Omega &=& \dot\Omega_{.2}\delta J_2 + \dot\Omega_{.6}\delta J_6 + \mu\dot\Omega_{\rm LT} + \Delta_{\Omega}, \\
  \nonumber\\
    \delta\dot\omega &=& \dot\omega_{.2}\delta J_2 + \dot\omega_{.6}\delta J_6 + \mu\dot\omega_{\rm LT} + \Delta_{\omega},  \\
    \nonumber\\
\delta\dot{\mathcal{M}} &=& \dot{\mathcal{M}}_{.2}\delta J_2 + \dot{\mathcal{M}}_{.6}\delta J_6 + \Delta_{\mathcal{M}},
\end{eqnarray}
\textrm{where} $\delta\dot\Psi$ denotes some  of Observed-minus-Calculated ($O-C$) quantity for the rate of the Keplerian element $\Psi$ which accounts for
every \textrm{unmodeled/mis-modeled dynamical effect. It} may be, for example, a correction to the modeled precessions to be phenomenologically estimated as a
solve-for parameter of a global fit \textrm{of Juno's} data as done by \citet{Pit05} with the planetary perihelia, or it could be a computed time-series
of\footnote{Since the Keplerian elements are not directly
measurable quantities, we use here the term ``residual'' in an improper sense.}  ``residuals'' of $\Psi$ by suitably overlapping orbital arcs.  The
gravitomagnetic force should be purposely not modeled in order to be fully present in $\delta\dot\Psi$.
The parameter $\mu$ is\footnote{It is not one of the standard PPN parameters, but it can be expressed in terms of $\gamma$ as $\mu=(1+\gamma)/2$.} 1
in GTR and 0 in Newtonian mechanics and accounts for the Lense-Thirring effect. The $\Delta$ terms include all the other systematic errors like the
precessions induced by the mis-modeled parts of the second even zonal harmonic $\delta J_4$ and the first odd zonal harmonic $\delta J_3$,
the mis-modeling due to the uncertainty in Jupiter's $GM$, etc.
By solving for $\mu$ one obtains
\eqi \delta\dot\Omega + p_1\delta\dot\omega + p_2\delta\dot{\mathcal{M}}=\dot\Omega_{\rm LT} + p_1\dot\omega_{\rm LT}+\Delta,\lb{comb}\eqf
with
\begin{eqnarray}
  p_1 &=& \rp{ {\dot{\mathcal{M}}_{.6}}\ {\dot\Omega_{.2}} - {\dot\Omega_{.6}}\ {\dot{\mathcal{M}}_{.2}}} {{\dot\omega_{.6}}\
  {\dot{\mathcal{M}}_{.2}} -{\dot{\mathcal{M}}_{.6}}\ {\dot\omega_{.2}} }, \lb{coef1}\\
  \nonumber\\
  p_2 &=& \rp{ {\dot\Omega_{.6}}\ {\dot\omega_{.2}} - {\dot\omega_{.6}}\ {\dot\Omega_{.2}}} {{\dot\omega_{.6}}\
  {\dot{\mathcal{M}}_{.2}} -{\dot{\mathcal{M}}_{.6}}\ {\dot\omega_{.2}} }. \lb{coef2}\end{eqnarray}
\textrm{\Rfr{comb} is designed}, by construction, to single out the combined Lense-Thirring
precessions and to \textrm{cancel} the combined secular\footnote{We include in them also the long-period harmonic terms for the
reasons explained before.} precessions due to $J_2$ and $J_6$ along with their mis-modeling. \textrm{However}, it is affected by $\Delta$
which acts as a systematic bias on the Lense-Thirring signal of interest. $\Delta$ globally includes the mis-modeled part of the
combined precessions induced by $J_3$ and $J_4$; the sources of uncertainty reside in $J_3$ and $J_4$ themselves and
\textrm{in Jupiter's} $GM$
through the mean motion $n$ which enters the coefficients $\dot\Omega_{.\ell},\dot\omega_{.\ell},\dot{\mathcal{M}}_{.\ell}$.

\begin{figure}[t]
\includegraphics[width=\columnwidth]{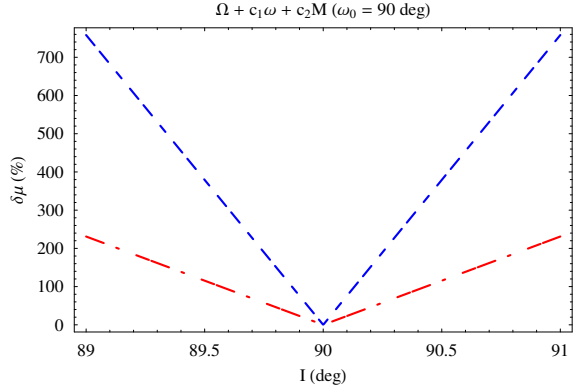}
 \caption{\footnotesize{\label{JUNO_combi_90} Systematic percent bias on the Lense-Thirring precessions,
 combined according to \rfr{comb}, induced by the mis-modeling in the uncanceled zonals $J_3$ (dash-dotted red line), $J_4$ (dotted blue line)
 according to Table \ref{zonalsjup} for $89$ deg $\leq I\leq 91$ deg and $\omega_0=90$ deg. }}
\end{figure}

\begin{figure}[t]
\includegraphics[width=\columnwidth]{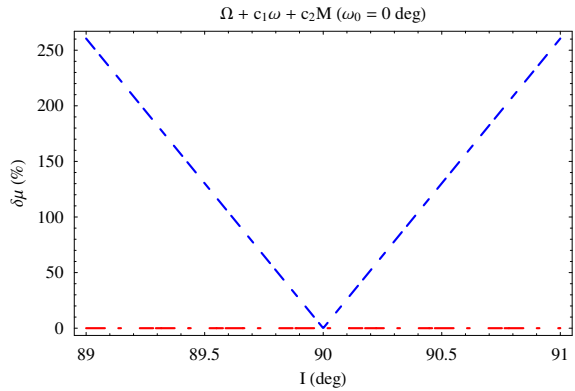}
 \caption{\footnotesize{\label{JUNO_combi_0} Systematic percent bias on the Lense-Thirring
 precessions, combined according to \rfr{comb}, induced by the mis-modeling in the uncanceled
 zonals $J_3$ (dash-dotted red line), $J_4$ (dotted blue line)  according to Table \ref{zonalsjup} for $89$ deg $\leq I\leq 91$ deg and $\omega_0=0$ deg. }}
\end{figure}

In Figure \ref{JUNO_combi_90} the impact of the mis-modeling in $J_3$ and $J_4$ for $\omega_0=90$ deg is depicted.
It corrects Fig. 2, \textrm{p.} 558 of \citet{IorioNA010}; incidentally, Figure \ref{JUNO_combi_90} yields smaller percent uncertainties.
In Figure  \ref{JUNO_combi_0}, which amends Fig. 3, \textrm{p.} 559 of \citet{IorioNA010}, we use $\omega_0 = 0$ deg.
In this case the situation is much more favorable because for a total departure of $\pm 1$ deg from $I=90$ deg, an improvement of only two orders of magnitude
in $J_3$, which is, today, still compatible with zero, and $J_4$ would be needed to reach the percent level;
let us recall that the expected improvement in $J_4$ with respect to the results by \citet{JUP230} is \textrm{about} three orders of magnitude \citep{And04}.
Note that a value of $\omega_0$ far from 90 deg is preferable to minimize
the perturbations.

Another potential source of systematic \textrm{errors is Jupiter's} $GM$ whose uncertainty $\delta(GM)$ indirectly affects \rfr{comb} through the Keplerian mean
motion $n$ entering the uncanceled $J_3$ and $J_4$ combined precessions; $\delta n$ is also present via the mean anomaly itself. However,
it turns out that it is of no concern because, according to the present-day level of relative uncertainty  \citep{JUP230}
\eqi\rp{\delta(GM)}{GM}=1.6\times 10^{-8},\eqf
its impact on the combined Lense-Thirring precessions is well below the percent level.
\subsubsection{A numerical approach}
Also in this \textrm{case we} followed an alternative approach based on preliminary numerical simulations. We investigated the impact
of the uncertainties in the first two jovian even zonals on a Juno's single six-hours pass by numerically simulating the probe's
Doppler range-rate signals due to $\delta J_2$ and $\delta J_4$. By assuming for them values as large as $2\times 10^{-10}$ and $3\times 10^{-10}$,
respectively, it turns out that the maximum Doppler shifts are roughly $1-1.5$ $\mu$m s$^{-1}$. Moreover, and more importantly, the time-dependent
patterns of the even zonals' Doppler signals are quite different from the Lense-Thirring one removing the risk of an insidious mimicking bias.
Another encouraging fact is that such simulations
indicate that an inclination of even 91 deg would not compromise the recovery of the gravitomagnetic signal of interest.

\subsection{Summary}
\textrm{We first explored} the possibility of a high accuracy measurement of the Lense-Thirring effect by performing  analytical calculations and interpreting them in a rather
conservative fashion.
A meter-level accuracy in determining the jovicentric orbit of \textrm{the Juno spacecraft} should not be an unrealistic goal to be
reached. Equivalently, the gravitomagnetic node precession of Juno amounts to 68.5 mas yr$^{-1}$, while the accuracy
in measuring its node and perijove precessions should
be of the order of $0.5-1$ mas yr$^{-1}$, given the expected improvements in our knowledge of the departure of the jovian gravitational field
from spherical symmetry. \textrm{If Juno's orbit were} perfectly polar, the long-period node precessions induced by the zonal harmonics $J_{\ell}, \ell=2,3,4,6,\textrm{\dots}$ of
the non-spherical jovian gravitational potential would vanish, thus removing a major source of systematic alias on the Lense-Thirring secular precession.
In reality, unavoidable orbit injection errors will displace the orbital plane of Juno from the ideal polar geometry; as a consequence, the mis-modeled part
of the node zonal precessions would overwhelm the relativistic signal  for just $\delta I = \pm 1$ deg, in spite of the expected  improvements
in $J_{\ell}, \ell=2,4,6$ of three orders of magnitude. A suitable linear combination of the node \textrm{$\Omega$}, the perijove $\omega$ and the mean anomaly
$\mathcal{M}$ will allow to cancel out the \textrm{effects} of $J_2$ and $J_6$; the remaining uncanceled $J_3$ and $J_4$ will have an impact on the
Lense-Thirring combined precessions which should be reduced down to the percent level or better by the improved low-degree zonals.

Instead of looking at the cumulative, secular effects over the entire duration of the mission, we also followed an alternative approach by looking at
single Doppler range-rate  measurements over time spans six hours long centered on the the probe's closest approaches to Jupiter; it turned out that,
in this way, the perspectives are even more favorable. We numerically simulated the characteristic Lense-Thirring pattern for a single science pass by
finding a maximum value of the order of hundreds $\mu$m s$^{-1}$, while the expected precision level in Juno's Doppler measurements is of the order of
one  $\mu$m s$^{-1}$. Thus, by exploiting about 25 of the planned 33 total passes of the mission it would be possible to reach a measurement accuracy
below the percent level. We repeated our numerical analysis also for the Doppler range-rate signals of $J_2$ and $J_4$ by finding quite different patterns
with respect to the gravitomagnetic one; moreover, for a level of mismodeling of the order of $2-3\times 10^{-10}$ in such zonals the maximum value of their
biasing Doppler signals is about $1-1.5$ $\mu$m s$^{-1}$. Our numerical analysis also shows that a departure from the nominal polar orbital geometry as large
as 1 deg would not compromise the successful outcome of the measurement of interest, contrary to the conservative conclusions of our analytical analysis.
Thus, this approach \textrm{suggests} that there is not a high correlation between the Lense-Thirring parameter and the jovian gravity field parameters, although a
covariance analysis would be needed to prove it. However, such a covariance analysis is outside the scope of the present paper.

In conclusion, the potential error in the proposed Juno Lense-Thirring measurement is between $0.2$ and $5$ percent. Conversely, if one assumes the
existence of gravitomagnetism as predicted by general relativity, the proposed measurement can also be considered as a direct, dynamical determination
of the jovian proper angular momentum $S$ by means of  the Lense-Thirring effect at the percent level.

\section{A note on gravitational waves: a new window into the Universe}\lb{GWs}
In the Introduction we emphasized that, as gravitomagnetic effects are usually discussed in the linearized approach of GTR, they also have  a strong connection with GWs within various theories of gravity \citep{IorCor09,IorCor10,Cafaro}.

In fact, interferometric \textrm{GW} detectors are operative at the present time ~\citep{Giazotto,IorCor10}. A direct detection will be a historic confirmation of the indirect evidence of the existence of GWs by \citet{Pulsar}. The realization of a GW astronomy, by giving a significant amount of new information, will be a cornerstone for a better understanding of gravitational physics too.  Detectors for GWs will be important for a better knowledge of the Universe \citep{Giazotto} and also because the interferometric GW detection will be the definitive test for GTR or, alternatively, a strong endorsement for Extended Theories of Gravity \citep{Essay}. In fact, if advanced projects on the detection of GWs improve their sensitivity, allowing the scientific community to perform a GW astronomy, accurate angle- and frequency-dependent response functions of interferometers for GWs arising from various theories of gravity will permit to discriminate among GTR and extended theories of gravity like Scalar Tensor Gravity, String Theory and $f(R)$ Theories. This  ultimate test will work because standard GTR admits only two polarizations for GWs, while in all extended theories the \textrm{number of polarization states is}, at least, \textrm{three; see} \citet{Essay} for details.
On the other hand,  the discovery of GW emission by the compact binary system composed \textrm{of} two neutron stars PSR1913+16 \citep{Pulsar}  has been, for  both of theoretician and experimental physicists working in this field, the fundamental thrust allowing to reach the extremely sophisticated technology needed for \textrm{investigating this} field of research \citep{IorCor10}.

GWs \textrm{arise} from Einsteinian GTR \citep{Albertino}
\textrm{and} are weak perturbations of the spacetime curvature which travel at light speed \citep{AlbertinoII,AlbertinoIII}. As GWs are quadrupolar waves, only asymmetric \textrm{astrophysical} sources can emit them \citep{IorCor10}. The most efficient ones are very dense and massive coalescing \textrm{binary} systems, like neutron stars and black holes. Instead, a single rotating pulsar, even if dense and massive,  can only rely  on spherical asymmetries, usually very small \citep{IorCor10}. Supernova explosions could have, in principle, relevant asymmetries, being potential sources \textrm{of gravitational waves} \citep{Giazotto, IorCor10}.
The most important cosmological source of GWs is  the so called  stochastic background of relic GWs which, together with the Cosmic Background Radiation (CBR), \textrm{carries} a huge amount of information on the early stages of the Universe evolution \citep{rainbow}.
This reason is a rather hopeful one \citep{rainbow, Allen, Corda}. As gravitation is the weakest of the four known fundamental interactions, the small-scale perturbations of the gravitational field decoupled from the evolution of the rest of the universe at very early times. Currently, the most detailed view of the early universe comes from the microwave background radiation, which decoupled from matter about 300.000 years after the big bang, and gives us an accurate picture of the universe at this very early time \citep{rainbow, Allen, Corda}. Some simple computations by \citet{Allen} showed that if \textrm{a background of relic GWs is observed,} they will carry a picture of the universe as it was about $10^{-22}$ s after the
\textrm{initial singularity}. This would represent a tremendous step forward in our understanding \textrm{of the primordial universe}.
The existence of a relic stochastic background of GWs is a consequence of very \textrm{general} assumptions which arise from
\textrm{an interplay} between basic principles of classical theories of gravity and of quantum field theory \citep{rainbow, Allen, Corda}. The strong variations of the gravitational field in the \textrm{primordial universe} amplified the zero-point quantum oscillations and produced relic GWs.
\textrm{This} model derives from the inflationary framework of the early universe \citep{Inflation}, which is fine tuned  with the WMAP data on the CBR. In fact, \textrm{an exponential inflation with a spectral index of $\approx1$ agrees with inflationary theories} \citep{Cosmology}. The
\textrm{inflationary} scenario \textrm{predicts} cosmological models in which the Universe undergoes a brief phase of a very rapid expansion in early times \citep{Inflation}. Such an expansion could be \textrm{either a} power-law or exponential in time \citep{Inflation} and provides solutions to the horizon and flatness problems. Inflation also contains a mechanism which creates perturbations in all fields \citep{Allen, Corda, Inflation}. This mechanism also \textrm{yields} a distinctive spectrum of relic GWs \citep{Allen, Corda}. The GW perturbations arise from the uncertainty principle and the spectrum of relic GWs is generated from the adiabatically-amplified zero-point fluctuations \citep{Allen, Corda}.

Regarding the potential GW detection, recalling some historical notes is \textrm{in order}. In 1957 F.A.E. Pirani, a member of \textrm{Hermann Bondi's} research group,  proposed the geodesic deviation equation as a theoretical foundation for designing a practical GW detector \citep{Pirani}.  In 1959, J. Weber studied a detector that, in principle, might be able to measure displacements smaller than the size of the nucleus \citep{Weber}. He realized an experiment using a large suspended bar of aluminum. Such a bar had a high resonant Q at a frequency of about 1 kHz. After this, in 1960, he tried to test the general relativistic prediction of gravitational waves from strong gravity collisions \citep{Weber2}. Weber performed further analyses in 1969, by claiming evidence for \textrm{the} observation of gravitational waves from two bars separated by 1000 km \citep{web}. Such evidence was based on coincident signals which Weber claimed to be detected by the two bars. \textrm{In order to confirm these observations, he also proposed to
detect gravitational waves} by using laser interferometers \citep{web}. In fact, all the modern detectors can be considered \textrm{to
originate} from early Weber's studies \citep{Giazotto, IorCor10}.

Currently, five cryogenic bar detectors have been built to work at very low temperatures ($<4$ K): Explorer at CERN, Nautilus at Frascati INFN National Laboratory, Auriga at Legnano National Laboratory, Allegro at Louisiana State University  and Niobe in Perth \citep{Giazotto, IorCor10}. There are also two spherical detectors, i.e. the Mario Schenberg, which operates in San Paolo (Brazil) and the Mini\-GRAIL, which is trying to detect GWs at the Kamerlingh Onnes Laboratory of Leiden University\textrm{; see} \citet{Giazotto} and \citet{IorCor10}. Spherical detectors are quite important for the potential detection of the scalar component of GWs that is admitted by Extended Theories of Gravity \citep{correlation}.
In the case of interferometric detectors,  two mirrors  are used like free falling masses. Such mirrors are separated by \textrm{by 3 km for Virgo
and 4 km for LIGO}. Thus,
\textrm{the GW tidal forces in interferometers are expected} to be several order of magnitude larger than in bar detectors. Differently from bars, interferometers like LIGO have very large bandwidth (10-10000 Hz) because mirrors are suspended to pendulums having resonance in the Hz region. In this way, above such a resonance frequency, mirrors works, in a good approximation,  like freely falling masses in the horizontal plane \citep{Giazotto, IorCor10}.

Recently, starting from the analysis in \citet{Grishchuk}, some papers have shown the importance of the gravitomagnetic effects in the framework of the \textrm{GW detection,} too; see \citet{IorCor10} and \citet{CordaI} and references \textrm{therein}. In fact, the so-called magnetic components of GWs have to be taken into account in the context of the total response functions of interferometers for GWs propagating from arbitrary directions \citep{Grishchuk,CordaI,IorCor10}. An extended analysis, which has been carefully reviewed in the recent work \citep{IorCor10}, showed that such a magnetic component becomes particularly important in the high-frequency portion of the range of ground based interferometers for GWs which arises from standard GTR.
The magnetic  component has been extended also to GWs arising from scalar-tensor gravity  \citep{Cafaro,IorCor10}. Such studies showed that
\textrm{by considering only the low-frequency approximation of the electric contribution and thereby neglecting the magnetic
contribution, a portion of about $15\%$} of the signal could be, in principle, lost in the case of Scalar Tensor Gravity too \citep{Cafaro,IorCor10}, in \textrm{close} analogy with the standard case of GTR \citep{Grishchuk,Corda}.

\textrm{It is important to give a physical/ma\-the\-ma\-ti\-cal explanation as to why the magnetic component of the gravitational field of a GW becomes important at high frequency. As interferometric GWs detection is performed in a laboratory environment on Earth, the coordinate system in which the spacetime
is locally flat is typically used \citep{IorCor10} and the distance between any two points is given simply by the difference in their coordinates in the sense of Newtonian physics. In this frame, called the frame of the local observer, GWs manifest themselves by exerting tidal forces on the masses (the mirror and the beam-splitter in the case of an interferometer). In the following, we work with $G=1$, $c=1$ and $\hbar=1$ and we call $h_{+}(t_{tt}+z_{tt})$ and $h_{\times}(t_{tt}+z_{tt})$ the weak perturbations due to the $+$ and the $\times$ polarizations which are expressed in terms of synchronous coordinates $t_{tt},x_{tt},y_{tt},z_{tt}$ in the transverse-traceless (TT) gauge \citep{IorCor10}. In this way, the most general GW propagating in the $z_{tt}$ direction can be written in terms of plane monochromatic waves \citep{IorCor10,Grishchuk}
\begin{equation}
\begin{array}{lll}
h_{\mu\nu}(t_{tt}+z_{tt}) &=& h_{+}(t_{tt}+z_{tt})e_{\mu\nu}^{(+)}+\\ \\
&+& h_{\times}(t_{tt}+z_{tt})e_{\mu\nu}^{(\times)}=\\ \\
&=& h_{+0}\exp i\omega(t_{tt}+z_{tt})e_{\mu\nu}^{(+)}+\\ \\
&+& h_{\times0}\exp i\omega(t_{tt}+z_{tt})e_{\mu\nu}^{(\times)},\end{array}\label{eq: onda generale}
\end{equation}
\textrm{with the corresponding line element}
\begin{equation}
\begin{array}{lll}
ds^{2}&=& dt_{tt}^{2}-dz_{tt}^{2}-(1+h_{+})dx_{tt}^{2}-(1-h_{+})dy_{tt}^{2}-\\ \\
&-& 2h_{\times}dx_{tt}dx_{tt}.\label{eq: metrica TT totale}
\end{array}
\end{equation}
The wordlines $x_{tt},y_{tt},z_{tt}={\rm const.}$ are timelike geodesics representing the histories of free test masses \citep{IorCor10}. The coordinate transformation $x^{\alpha}=x^{\alpha}(x_{tt}^{\beta})$ from the TT coordinates to the frame of the local observer is
\textrm{given by} \citep{IorCor10,Grishchuk}
\begin{equation}
\begin{array}{lll}
t &=& t_{tt}+\frac{1}{4}(x_{tt}^{2}-y_{tt}^{2})\dot{h}_{+}-\frac{1}{2}x_{tt}y_{tt}\dot{h}_{\times}\\ \\
x &=& x_{tt}+\frac{1}{2}x_{tt}h_{+}-\frac{1}{2}y_{tt}h_{\times}+\frac{1}{2}x_{tt}z_{tt}\dot{h}_{+}-\\ \\
&-&\frac{1}{2}y_{tt}z_{tt}\dot{h}_{\times}\\ \\
y &=& y_{tt}+\frac{1}{2}y_{tt}h_{+}-\frac{1}{2}x_{tt}h_{\times}+\frac{1}{2}y_{tt}z_{tt}\dot{h}_{+}-\\ \\
&-&\frac{1}{2}x_{tt}z_{tt}\dot{h}_{\times}\\ \\
z &=& z_{tt}-\frac{1}{4}(x_{tt}^{2}-y_{tt}^{2})\dot{h}_{+}+\frac{1}{2}x_{tt}y_{tt}\dot{h}_{\times},\end{array}\label{eq: trasf. coord.}
\end{equation}
\textrm{and}
\eqi
\begin{array}{lll}
\dot{h}_{+}&\equiv &\frac{\partial h_{+}}{\partial t}, \\ \\
\dot{h}_{\times}&\equiv &\frac{\partial h_{\times}}{\partial t}.
\end{array}
\eqf
 The coefficients of this transformation (components of the metric and its first time derivative) are taken along the central wordline of the local observer \citep{IorCor10,Grishchuk}. It is well known that the linear and quadratic terms, as powers of $x_{tt}^{\alpha}$, are unambiguously determined by the conditions of the frame of the local observer, while the cubic and higher-order corrections are not determined by these conditions \citep{IorCor10,Grishchuk}. Thus, at high-frequencies, the expansion in terms of higher-order corrections breaks down \citep{IorCor10,Grishchuk}.
Considering a free mass riding on a timelike geodesic ($x=l_{1},$ $y=l_{2},$ $z=l_{3}$ ) \citep{IorCor10,Grishchuk}, \rfr{eq: trasf. coord.} defines the motion of this mass with respect to the introduced frame of the local observer. In concrete terms one gets
\begin{equation}
\begin{array}{lll}
x(t) &=& l_{1}+\frac{1}{2}[l_{1}h_{+}(t)-l_{2}h_{\times}(t)]+\frac{1}{2}l_{1}l_{3}\dot{h}_{+}(t)+\\ \\
&+&\frac{1}{2}l_{2}l_{3}\dot{h}_{\times}(t)\\ \\
y(t) &=& l_{2}-\frac{1}{2}[l_{2}h_{+}(t)+l_{1}h_{\times}(t)]-\frac{1}{2}l_{2}l_{3}\dot{h}_{+}(t)+\\ \\
&+&\frac{1}{2}l_{1}l_{3}\dot{h}_{\times}(t)\\ \\
z(t) &=& l_{3}-\frac{1}{4[}(l_{1}^{2}-l_{2}^{2})\dot{h}_{+}(t)+2l_{1}l_{2}\dot{h}_{\times}(t).\end{array}\label{eq: Grishuk 0}
\end{equation}
In \textrm{the} absence of GWs the position of the mass is $(l_{1},l_{2},l_{3})$ \textrm{and the effect of the GW is to induce oscillations of the mass}. Thus,
in general, from \rfr{eq: Grishuk 0} all three components of motion are present \citep{IorCor10,Grishchuk}. Neglecting the terms with $\dot{h}_{+}$ and $\dot{h}_{\times}$ in \rfr{eq: Grishuk 0}, the \virg{traditional} equations for the mass motion are obtained \citep{IorCor10,Grishchuk}
\begin{equation}
\begin{array}{lll}
x(t) &=& l_{1}+\frac{1}{2}[l_{1}h_{+}(t)-l_{2}h_{\times}(t)]\\ \\
y(t) &=& l_{2}-\frac{1}{2}[l_{2}h_{+}(t)+l_{1}h_{\times}(t)]\\ \\
z(t) &=& l_{3}.\end{array}\label{eq: traditional}
\end{equation}
Clearly, this is the \textrm{analogon} of the electric component of motion in electrodynamics \citep{IorCor10,Grishchuk}, while
\textrm{the} equations
\begin{equation}
\begin{array}{lll}
x(t) &=& l_{1}+\frac{1}{2}l_{1}l_{3}\dot{h}_{+}(t)+\frac{1}{2}l_{2}l_{3}\dot{h}_{\times}(t)\\ \\
y(t) &=& l_{2}-\frac{1}{2}l_{2}l_{3}\dot{h}_{+}(t)+\frac{1}{2}l_{1}l_{3}\dot{h}_{\times}(t)\\ \\
z(t)&=& l_{3}-\frac{1}{4[}(l_{1}^{2}-l_{2}^{2})\dot{h}_{+}(t)+2l_{1}l_{2}\dot{h}_{\times}(t),
\end{array}\label{eq: news}
\end{equation}
are the \textrm{analogues} of the magnetic component of motion.
One could think that the presence of these \virg{magnetic} components is a \virg{frame artefact} due to the transformation of \rfr{eq: trasf. coord.},
but in Section 4 of \citet{Grishchuk} \rfr{eq: Grishuk 0} has been directly obtained from the geodesic deviation equation too, thus the magnetic components have a real physical significance. The fundamental point of \citet{IorCor10,Grishchuk} is that the \virg{magnetic} components become important when the frequency of the wave increases but only in the low-frequency regime. This can be understood directly from \rfr{eq: Grishuk 0}. In fact, by using \rfr{eq: onda generale} and \rfr{eq: trasf. coord.}, it turns out that \rfr{eq: Grishuk 0} becomes
\begin{equation}
\begin{array}{lll}
x(t) &=& l_{1}+\frac{1}{2}[l_{1}h_{+}(t)-l_{2}h_{\times}(t)]+\\ \\
&+&\frac{1}{2}l_{1}l_{3}\omega h_{+}(t-\frac{\pi}{2})+\frac{1}{2}l_{2}l_{3}\omega h_{\times}(t-\frac{\pi}{2})\\ \\
y(t) &=& l_{2}-\frac{1}{2}[l_{2}h_{+}(t)+l_{1}h_{\times}(t)]-\\ \\
&-&\frac{1}{2}l_{2}l_{3}\omega h_{+}(t-\frac{\pi}{2})+\frac{1}{2}l_{1}l_{3}\omega h_{\times}(t-\frac{\pi}{2})\\ \\
z(t) &=& l_{3}-\frac{1}{4[}(l_{1}^{2}-l_{2}^{2})\omega h_{+}(t-\frac{\pi}{2})+\\ \\
&+& 2l_{1}l_{2}\omega h_{\times}(t-\frac{\pi}{2}).\end{array}\label{eq: Grishuk 01}\end{equation}
Thus, the terms with $\dot{h}_{+}$ and $\dot{h}_{\times}$ in \rfr{eq: Grishuk 0} can be neglected only when the wavelength goes to infinity, while, at high-frequencies, the expansion in terms of $\omega l_{i}l_{j}$ corrections, with $i,j=1,2,3,$ breaks down \citep{IorCor10,Grishchuk}.
The attentive reader could be surprised that gravitomagnetic effects in the field of a GW have been ignored in the past. Actually, the key point is that realizing the first direct detection has been always the most important goal for scientists in this research field. To obtain such a first detection, the low frequency approximation is considered sufficient, i.e. only the \virg{electric} component is needed for ground based interferometers \citep{Grishchuk}. On the other hand, if the scientific community \textrm{aimes} to realize a concrete GW astronomy, the above discussion shows that gravitomagnetic effects in the field of a GW have to be taken in due account. }
\textrm{Finally, we recall that GM effects are important in the process of producing GWs too. Indeed, \citet{Capoz2} studied corrections to the relativistic orbits by considering high order approximations induced by GM effects for very massive binary systems (e.g. a very massive black hole on which a stellar object is spiralling in). Such corrections have been explained by taking into account \virg{magnetic} components in the weak field limit of the gravitational field. New nutation effects of order  $c^{-3}$ have been found. These effects work beside the standard periastron corrections and affect the GW emission and the gravitational waveforms due by the orbital motion of the two massive bodies \citep{Capoz2}.
The GM corrections emerge as soon as matter-current densities and vector gravitational potentials cannot be discarded into dynamics \citep{Capoz2}. GWs emitted through massive binary systems have been studied in the quadrupole approximation \citep{Capoz2}.}
\section{Conclusions}
Measuring gravitomagnetism, and, in particular, the Lense-Thirring effect is a challenging and difficult enterprize.
In this review we have shown how the detection of such a general relativistic signal is closely related to a number of other competing dynamical features of motion, so that a genuine interdisciplinary attitude is required. Looking for the prediction by Lense and Thirring may help in shedding light on different aspects of the solar system scenarios in which it is attempted to be measured. Conversely, a better knowledge of the forces {acting on dynamical systems}  may be of great help in reliably and accurately measuring the gravitomagnetic field itself.

The current tests with the LAGEOS satellites in the gravitational field of the Earth, which represent the first implemented attempts to measure the Lense-Thirring effect, should be repeated by directly modeling the gravitomagnetic force and explicitly solving for it in the data processing. Moreover, they should be complemented
by a re-analysis of the data sets of the currently ongoing spacecraft-based missions dedicated to accurately measuring the global gravitational field of the Earth like GRACE  taking into account general relativity itself; new global  solutions in which one or more parameters accounting for the Lense-Thirring effect are solved-for should be produced. For the moment, a conservative evaluation of the systematic bias in the LAGEOS-LAGEOS II tests due to the imperfect knowledge of the classical part of the Earth's gravitational field points toward a $20-30\%$ level; more optimistic evaluations yield a $10-15\%$ uncertainty.
It is unclear if the future LARES mission will be able to measure the Lense-Thirring precessions at the repeatedly claimed $\approx 1\%$ accuracy. Indeed,
its orbital configuration, different from that originally proposed, may enhance some competing dynamical effects of gravitational and non-gravitational origin at a level whose uncertainty  is difficult to be realistically evaluated.

The present-day level of accuracy in the orbit determination of the inner planets of the solar system has basically reached the magnitude of the Lense-Thirring precessions. \textrm{The most important systematic error due to mis-modeled dynamical effects is associtated with} the first even zonal harmonic of the non-spherical gravitational field of the Sun, \textrm{which particularly affects Mercury.} Anyway, an accurate measurement of
\textrm{the Sun's oblateness} is one of the goals of future spacecraft-based missions. The analysis of more accurate ranging data to present and future \textrm{spacecraft like, e.g., BepiColombo} and, especially, future interplanetary laser ranging devices should drastically improve the situation allowing for a reasonably accurate measurement of the solar gravitomagnetic field.

After the preliminary attempts with the Mars Global Surveyor probe which have paved the way, also Mars may become suitable for reliably measuring the Lense-Thirring orbital precessions in \textrm{the} near future. To this aim, the forthcoming Phobos-Grunt mission, scheduled for launch in late 2011 or early 2012, will improve our knowledge of some key physical properties of Mars and of the orbit of its satellite Phobos.

The Juno mission to Jupiter, scheduled for launch in 2011, may yield the opportunity \textrm{for} an accurate measurement of the Jovian angular momentum through the Lense-Thirring effect itself.

Finally, we emphasized the relevant connection between gravitomagnetic effects and GWs. In fact, the scientific community
{expects
the first direct detection of GWs within the next years}.
The importance of gravitomagnetic effects in the framework of GWs  increases in the high-frequency portion of the range of ground based interferometers for both of GWs {which arise from standard GTR as well as those coming} from scalar-tensor gravity. If one neglects the magnetic contribution considering only the low-frequency approximation of the electric contribution, a portion of \textrm{about $15\%$ of} the signal could be lost by interferometric detectors. Thus, a carefully analysis of the \virg{magnetic}-type contribution of GWs is needed in order to
\textrm{establish a GW astronomy within the} next years.

\section*{Acknowledgments}
L.I. thanks L.F.O. Costa for useful suggestions and critical remarks. The authors gratefully thank an anonymous referee for her/his valuable comments who contributed to improve the manuscript.

\end{document}